\documentclass{aa}  

\usepackage{changepage}
\usepackage{adjustbox}
\usepackage[utf8]{inputenc}
\usepackage{array}
\usepackage{makecell}
\usepackage{rotating}
\usepackage{lscape}
\usepackage{enumitem}
\usepackage{placeins}

\usepackage{graphicx}
\usepackage{txfonts}
\begin{document}

   \title{The Westerbork Coma Survey\thanks{The table containing the estimated physical properties of our H$\,\textsc{i}$-detected Coma galaxies and Table \ref{tab::ref_samp} are only available in electronic form at the CDS via anonymous ftp to cdsarc.u-strasbg.fr (130.79.128.5), via \protect\url{http://cdsweb.u-strasbg.fr/cgi-bin/qcat?J/A+A/}, or at \protect\url{https://github.com/molnard89/WCS-Catalog-Release}.}}
  \subtitle{A blind, deep, high-resolution H$\,\textsc{i}$ survey of the Coma cluster}

   \author{D. Cs. Moln\'{a}r\thanks{e-mail: daniel.molnar@inaf.it} \inst{1}
          \and
          P. Serra \inst{1}
          \and
          T. van der Hulst \inst{2}
          \and
          T. H. Jarrett \inst{3,4,5}
          \and
          A. Boselli \inst{6}
          \and
          L. Cortese \inst{7,8}
          \and
          J. Healy \inst{2,3}
          \and
          E. de Blok \inst{10,3,2}
          \and
          M. Cappellari \inst{9}
          \and
          K. M. Hess \inst{10,2,11}
          \and
          G. I. G. J\'{o}zsa \inst{12,13,14}
          \and
          R. M. McDermid \inst{15,16,8}
          \and
          T. A. Oosterloo \inst{10,2}
          \and
          M. A. W. Verheijen \inst{2}
          }

   \institute{INAF - Osservatorio Astronomico di Cagliari, Via della Scienza 5, I-09047 Selargius (CA), Italy
             \and
              Kapteyn Astronomical Institute, University of Groningen, Landleven 12, 9747 AV Groningen, The Netherlands
             \and
             Department of Astronomy, University of Cape Town, Private Bag X3, Rondebosch 7701, South Africa
             \and
             Inter-University Institute for Data Intensive Astronomy (IDIA), University of Cape Town, Rondebosch, Cape Town, 7701, South Africa
             \and
             Western Sydney University, Locked Bag 1797, Penrith South DC, NSW 1797, Australia
             \and
             Aix Marseille Univ, CNRS, CNES, LAM, Marseille, France
             \and
             International Centre for Radio Astronomy Research, The University of Western Australia, 35 Stirling Hw, 6009 Crawley, WA, Australia
             \and
             ARC Centre of Excellence for All Sky Astrophysics in 3 Dimensions (ASTRO 3D), Australia
             \and
             Sub-Department of Astrophysics, Department of Physics, University of Oxford, Denys Wilkinson Building, Keble Road, Oxford, OX1 3RH, UK
             \and
             ASTRON, Netherlands Institute for Radio Astronomy, Oude Hoogeveensedijk 4, 7991 PD, Dwingeloo, The Netherlands
             \and
             Instituto de Astrof\'{i}sica de Andaluc\'{i}a (CSIC), Glorieta de la Astronom\'{i}a s/n, 18008 Granada, Spain
             \and
             South African Radio Astronomy Observatory, 2 Fir Street, Black River Park, Observatory, Cape Town, 7925, South Africa
             \and
             Department of Physics and Electronics, Rhodes University, PO Box 94, Makhanda, 6140, South Africa
             \and
             Argelander-Institut f\"{u}r Astronomie, Auf dem H\"{u}gel 71, D-53121 Bonn, Germany
             \and
             Department of Physics and Astronomy, Macquarie University, Sydney NSW 2109, Australia
             \and
             Astronomy, Astrophysics and Astrophotonics Research Centre, Macquarie University, Sydney, NSW 2109, Australia
             }

   \date{Received 9 November, 2021; accepted 17 December, 2021}

\abstract{We present the blind Westerbork Coma Survey probing the H$\,\textsc{i}$ content of the Coma galaxy cluster with the Westerbork Synthesis Radio Telescope. The survey covers the inner $\sim$ 1 Mpc around the cluster centre, extending out to 1.5 Mpc towards the south-western NGC 4839 group. The survey probes the atomic gas in the entire Coma volume down to a sensitivity of $\sim$ 10$^{19}$ cm$^{-2}$ and 10$^8$ M$_{\odot}$. Combining automated source finding with source extraction at optical redshifts and visual verification, we obtained 40 H$\,\textsc{i}$ detections of which 24 are new. Over half of the sample displays perturbed H$\,\textsc{i}$ morphologies indicative of an ongoing interaction with the cluster environment. With the use of ancillary UV and mid-IR, data we measured their stellar masses and star formation rates and compared the H$\,\textsc{i}$ properties to a set of field galaxies spanning a similar stellar mass and star formation rate range. We find that $\sim$ 75 \% of H$\,\textsc{i}$-selected Coma galaxies have simultaneously enhanced star formation rates (by $\sim$ 0.2 dex) and are H$\,\textsc{i}$ deficient (by $\sim$ 0.5 dex) compared to field galaxies of the same stellar mass. According to our toy model, the simultaneous H$\,\textsc{i}$ deficiency and enhanced star formation activity can be attributed to either H$\,\textsc{i}$ stripping of already highly star forming galaxies on a very short timescale, while their H$_2$ content remains largely unaffected, or to H$\,\textsc{i}$ stripping coupled to a temporary boost of the H$\,\textsc{i}$-to-H$_2$ conversion, causing a brief starburst phase triggered by ram pressure before eventually quenching the galaxy.}


   \maketitle
%

\section{Introduction}

Galaxies in the Universe form the cosmic web, a large-scale structure built up by clusters, filaments, and voids \citep{bond96,sheth04,cautun14}. The dense clusters of galaxies reside at the intersections of an intricate network of galaxy filaments, with large, sparsely populated voids in between, as observed by mapping galaxy positions and redshifts \citep[e.g.][]{davis82,colless01,jarrett04,jones09,robotham11,huchra12}. These structures provide a range of environments with widely varying physical conditions, which affect the evolution of galaxies inhabiting them -- observations have shown that galaxies in clusters tend to be redder, passive, gas-poor, and of an earlier morphological type than field galaxies \citep[e.g.][]{hubble31,butcher78,dressler80,dressler97}.

At z $=$ 0 nearly all galaxies inside rich clusters are red and of an early morphological type. The main culprits are externally driven gas removal and the disruption of gas replenishment driven by a combination of physical processes \citep[for extensive reviews on the topic, see e.g.][]{boselli06,boselli14d,cortese21,boselli21b} whose relative importance is under debate. Starvation is one of the mechanisms thought to play an important role \citep{balogh00,bosch08}. It occurs when the hot intracluster medium (ICM) hydrodynamically strips and/or transfers heat to the halo gas surrounding cluster galaxies via mixing, thus halting its cooling and subsequent accretion onto the H$\,\textsc{i}$ disk, which in turn leads to the quenching of star formation activity and the reddening of the galaxy. Another important mechanism is ram-pressure stripping, whereby external pressure exerted by the dense ICM strips the gas reservoir itself of an infalling galaxy \citep{gunn72}. The most visually striking examples of galaxies experiencing ram-pressure stripping are the so-called jellyfish galaxies: objects trailed by tails of gas and newly formed stars \citep{cortese07,yoshida08,hester10,smith10,yagi10,fumagalli14,kenney14,poggianti16,cramer19,ramatsoku19,ramatsoku20,deb20}. During ram-pressure stripping either the external pressure is strong enough to perturb the cold molecular (H$_2$) gas component, which results in eventual rapid star formation quenching in the disk \citep[e.g.][]{lee17}, or, more commonly, it only affects the more loosely bound neutral hydrogen gas (H$\,\textsc{i}$), which in turn cuts off the supply of the H$_2$ component, leading to quenching on longer timescales. Thus, in general, ram-pressure stripped galaxies are associated with reduced  H$_2$ content \citep[][but see \citealt{moretti20}]{fumagalli09,boselli14b} and star formation activity \citep{gavazzi10,boselli14c}. However, observations of individual objects \citep[e.g.][]{cortese07,merluzzi13,kenney14,boselli21a} and, more recently, statistical samples \citep{vulcani18,roberts20,roberts22} find in some, possibly peculiar, cases an enhancement of star formation in galaxies undergoing ram pressure stripping before they are inevitably quenched. According to hydrodynamical simulations \citep[e.g.][]{tonnesen07,kapferer08,kapferer09,tonnesen08,tonnesen09,steinhauser12,lee20}, this temporary enhancement of star formation might be a result of ram pressure compressing gas into high density clouds in the disk. 

The relative importance of these and other gas removal processes strongly depend on environmental factors (such as the temperature and pressure of the surrounding gaseous medium, the number density of neighbouring galaxies, and the motion of galaxies relative to one another and to the medium), as well as the intrinsic characteristic of the galaxies themselves (stellar, neutral and molecular gas mass, orbit, etc). This suggests that in clusters of different sizes and dynamical states, the balance of external and internal gas removal processes (such as feedback from supernovae and active galactic nuclei) changes, making the task of disentangling them observationally challenging. Studying the H$\,\textsc{i}$ component provides important clues -- ongoing ram-pressure could result in tails along the orbit of the galaxy, often without stars, while extensions due to tidal interactions tend to be more aligned with the interacting partner and frequently contain a stellar component. Truncated gas disks are a sign of gas removal or starvation mechanisms that affected the galaxy at earlier epochs. Observing galaxies in the extreme environments of clusters thus gives valuable insights into a range of available galaxy evolutionary paths.

The Coma cluster (ACO 1656), situated in the massive Coma supercluster \citep{chincarini76}, is a unique laboratory for investigating the link between environment and galaxy properties mainly due to its vicinity \citep[distance of $\sim$100 Mpc;][]{carter08} and halo mass \citep[$M_{200} \approx0.5$--$1\times10^{15}$;][]{lokas03,gavazzi09}, enabling a sensitivity and a resolution that is rare among the largest clusters in the Universe. Moreover, the high mass of the Coma cluster results in more extreme cases of ram pressure and overall stronger effects of environment on galaxy evolution, than other nearby clusters of lower mass, such as the Virgo cluster.

Although initially Coma was thought to be a relaxed system \citep{kent82}, it was later found to be in the state of merger as evidenced by the distribution of its X-ray emission \citep{neumann01,neumann03,churazov21}. Its complex substructure, seen in optical spectroscopy \citep[e.g.][]{colless96,healy21}, reveals several groups of galaxies likely on their first infall. The dynamic nature of the cluster is further underlined by galaxies observed to have tails in H$_{\alpha}$ \citep{yagi10,gavazzi18,cramer19}, ultraviolet \citep[UV;][]{smith10}, radio continuum \citep{miller09,chen20,roberts21}, and H$\,\textsc{i}$ \citep{bravo00,bravo01} showing gas stripping in action. Indeed, \citet{gavazzi18} reported that 17/27 late-type galaxies in the central region of Coma exhibit extended H$_\alpha$, demonstrating the important role ram pressure stripping plays in transforming the galaxies accreted into the cluster. Coma also hosts a population of blue k+a (also called a+k or E+A) galaxies, whose spectral properties are indicative of a recent (<0.5 Gyr) SF quenching, which, in most cases, was preceded by a starburst phase \citep{bothun86,poggianti04,gavazzi10}. These lie around the edges of the X-ray structure of Coma, suggesting that ICM interaction is responsible for their sudden quenching and perhaps their starbust episodes as well.

NGC 4921, the brightest spiral galaxy of the Coma cluster, is an especially good example of the diverse physical mechanisms that ram pressure can trigger. It is a face-on galaxy whose truncated gas disk is compressed on its north-western quadrant, coincident with intense star formation activity \citep{cramer21}. High resolution optical and CO observations show filaments of dust and molecular gas on this side protruding from the main gas ring, pointing towards the projected direction of the ICM wind \citep{kenney15,cramer21}. These are likely remnants of the densest regions of the interstellar medium prior to the ICM interaction, after the sparser gas has been swept back to form the observed dust front downstream from the wind. This suggests that while in NGC 4921 ram pressure globally is quenching star formation from outside in, it can locally trigger star formation activity at the same time.

To gain a more complete overall picture of the effects ICM has on galaxies falling into the cluster, H$\,\textsc{i}$ surveys were carried out to map the distribution of atomic gas in Coma and characterize the gas content of its galaxies. It was targeted by the Very Large Array \citep[VLA;][]{bravo00,bravo01}, the Westerbork Synthesis Radio Telescope \citep[WSRT;][]{beijersbergen03}, and the Arecibo telescope \citep{gavazzi06,gavazzi13}. From these data it became clear that within 2-3 Mpc, galaxies are H$\,\textsc{i}$ deficient compared to field galaxies \citep{gavazzi06}. Most of these H$\,\textsc{i}$ deficient galaxies tend to be closer to the cluster centre, near the X-ray detected hot ICM, and without exception show very perturbed H$\,\textsc{i}$ morphologies, consistent with ram pressure stripping \citep{bravo00,bravo01}. Despite the valuable insight into the properties of Coma's H$\,\textsc{i}$ content, these observations either lack the sufficient sensitivity, angular resolution or coverage to give a general census of the H$\,\textsc{i}$ gas in Coma. For instance, the measurements of \citet{gavazzi06} targeting the late type galaxies are sensitive down to few times 10$^8$ $\rm M_\odot$ (and are complete to M$_{\rm HI} = 10^9$ M$_{\odot}$), while the more recent H$\,\textsc{i}$ Arecibo Legacy Fast ALFA Survey \citep[ALFALFA][]{giovanelli05,haynes18} data are shallower by a factor 4. Moreover, the low resolution of Arecibo (3 arcmin, $\sim$90 kpc at the distance of Coma) means that no information is available on the spatial distribution of gas around individual galaxies. On the other hand, the VLA survey by \citet{bravo00,bravo01} combined relatively high angular resolution ($\sim$ 33 arcsec) and good sensitivity (detecting several M$_{\rm HI} < 10^9$ M$_{\odot}$ sources), but was targeted only at the brightest spirals of the cluster and probed only a velocity range that corresponds to the recessional velocity dispersion of the cluster centred on the velocity of the targeted spiral, missing a large fraction of Coma galaxies in its field of view altogether. The WSRT blind survey of \citet{beijersbergen03} does not suffer from this problem because it covers the cluster central volume and extends south-west to the well-known NGC 4839 infalling group.
Over half of the H$\,\textsc{i}$ detections appeared to have central positions significantly shifted from the centres of their optical counterparts, possibly because of their infall into the cluster. However, due to the resolution and sensitivity limitations of the survey, this observation was not statistically confirmed. Furthermore, the effective integration per pointing was only 50 minutes, rendering the survey unable to detect objects with M$_{\rm HI} < 10^9$ M$_{\odot}$.

The new blind Westerbork Coma Survey (WCS) presented in this paper aims to unite the best properties of the previous H$\,\textsc{i}$ studies in order to gain a better census of the overall H$\,\textsc{i}$ properties of the Coma cluster. Our measurements have spatial and velocity resolutions that are comparable to that of \citet{bravo00,bravo01}, while mapping the central 1 Mpc of the cluster and extending out to 1.5 Mpc towards the south-western NGC 4839 group, and covering the entire velocity range of the cluster along the line of sight. In the following (Sect \ref{sect::data}) we describe our data in more detail alongside the source finding approach employed to identify H$\,\textsc{i}$ detections in our cube. We characterize our H$\,\textsc{i}$ sources by classifying them according to their H$\,\textsc{i}$ morphologies, and by estimating the star formation rate (SFR) and stellar mass (\textit{M$_\star$}) of their optical counterparts, while applying the same methods to a set of reference galaxies in the field (Sect \ref{sect::methods}). We examine the distribution of our H$\,\textsc{i}$ detections within the cluster and compare their host galaxy properties to the reference sample of field galaxies with the use of scaling relations (Sect \ref{sect::results}). Finally, in Sect \ref{sect::discussion} we attempt to find a physical interpretation for our results, summarised in Sect \ref{sect::summary}.

Throughout this paper for the Coma cluster we adopt values of R$_{200}$ = $1.8 \mathrm{Mpc}$, M$_{200}$ = $5.1\,h^{-1} \cdot 10^{14} \mathrm{M_{\odot}}$, $\sigma_{\rm cl}$ = $1045\, \mathrm{km s^{-1}} $, and $c$ = $5$ from \citet{gavazzi09}, and a distance of 100 Mpc. Star formation rates and stellar masses reported assume a \citet{chabrier03} initial mass function.

\section{Data}
\label{sect::data}

\subsection{WSRT observations}

\begin{figure}
    \centering
    \includegraphics[width=0.45\textwidth]{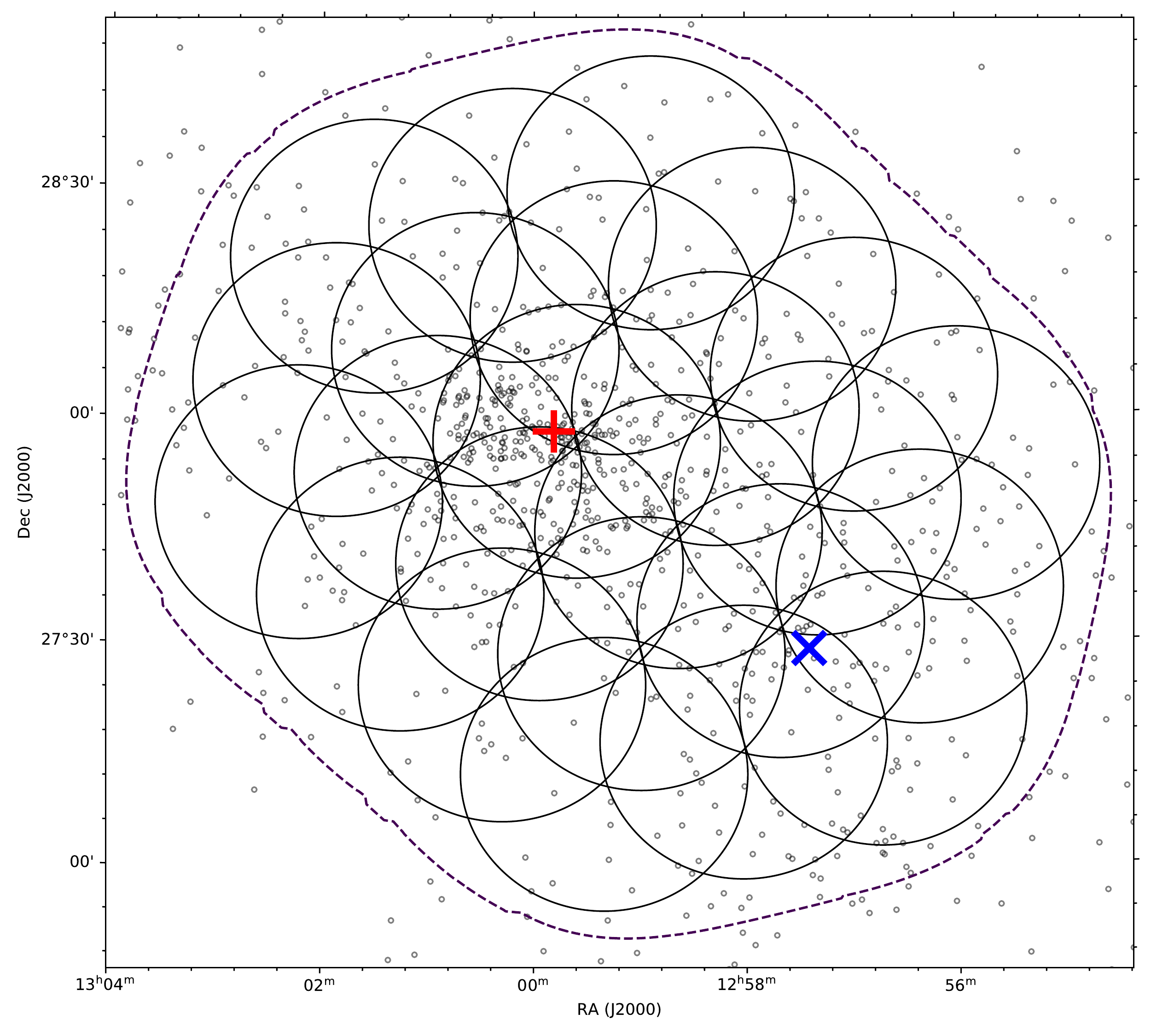}
    \caption{Positions of the 24 individual WSRT pointings that make up our WSRT H$\,\textsc{i}$ survey of the Coma cluster. Circles indicate the FWHM of each observation, while the dashed line is drawn at 25 \% peak sensitivity of the combined mosaic. The red plus sign and blue cross mark the centre of the cluster and NGC 4839, respectively. Black circles indicate the positions of Coma member galaxies from \protect\citet{healy21}.}
    \label{fig::survey_pointings}
\end{figure}

The survey consists of 24 individual WSRT pointings observed throughout June 2012 using the Multi-Frequency-Frontends. They were arranged on a hexagonal grid as shown in Fig. \ref{fig::survey_pointings}, with a spacing of 0.3 deg, which gives considerable overlap between individual observations given the WSRT primary beam (0.6 deg FWHM). We integrate 12 h per pointing which, due to the considerable overlap, results in an effective integration of 3 $\times$ 12 h per pointing in the central region of the mosaic.

\begin{figure}
    \centering
    \includegraphics[width=0.45\textwidth]{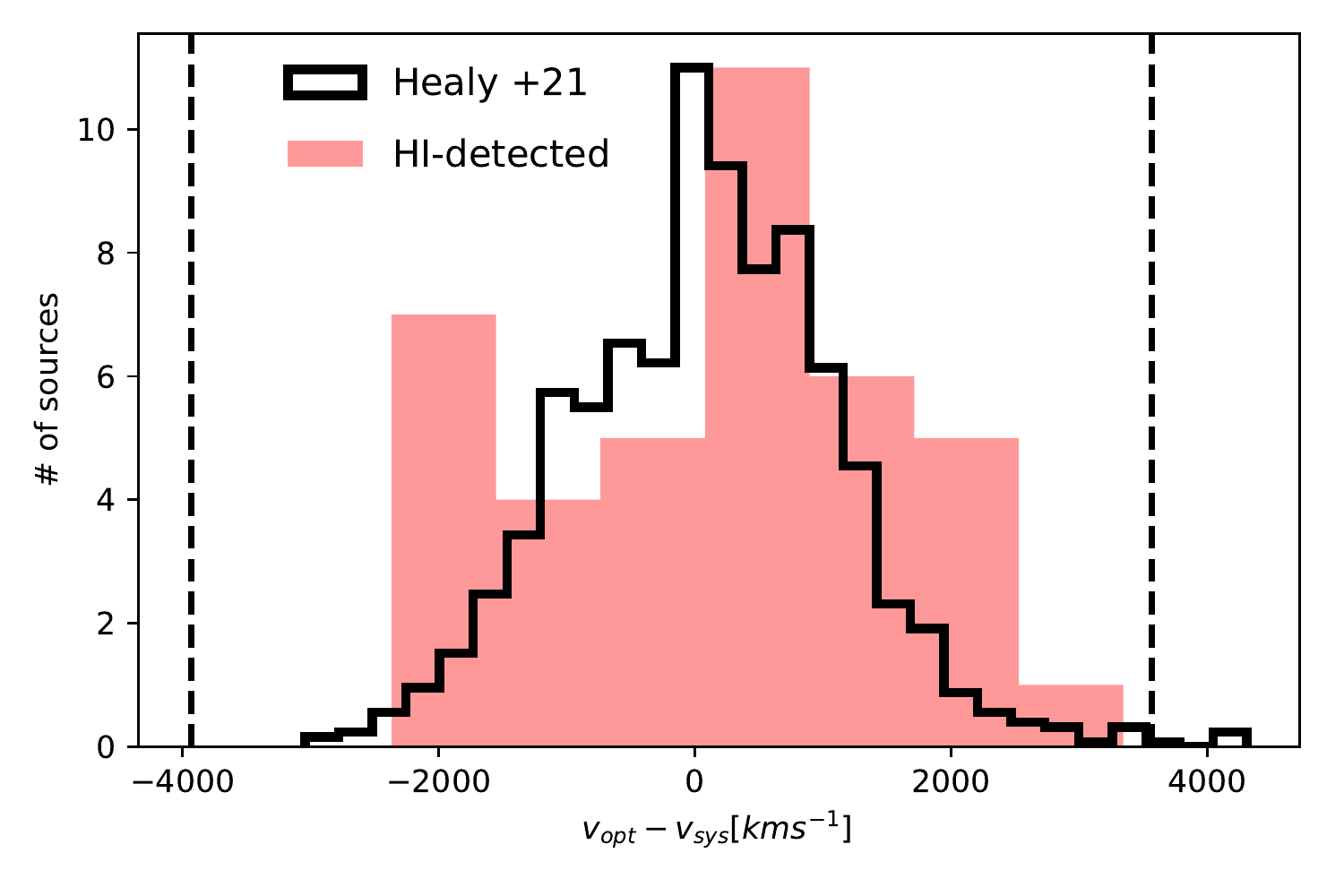}
    \caption{Distribution of line of sight velocities relative to the systemic velocity of the Coma cluster. The red shaded histogram shows H$\,\textsc{i}$-selected galaxies in our WSRT survey while the black one represents confirmed Coma members based on the optical spectroscopy by \protect\citet{healy21}. The latter has been re-scaled to ease comparison. Dashed vertical lines mark the minimum and maximum velocities at which the WCS can detect H$\,\textsc{i}$ emission. For more details on optical association see Sect. \ref{sect::opt_counter}.}
    \label{fig::v_distr}
\end{figure}

We observed with two overlapping 20-MHz-wide bands centred at 1396 and 1379 MHz -- each with 512 channels. We reduced the data independently and in a standard way for each pointing and each band using the MIRIAD package \citep{sault95}. We imaged the visibilities with robust = 0.4 weighting and 20 arcsec tapering in the $uv$ plane. For each band, we linearly mosaicked the 24 single-pointing H$\,\textsc{i}$ cubes after deconvolution. The final result consists of two mosaic cubes covering the same sky area and the velocity range $\sim$ 3,000 -- 6,700 and 6,700 -- 10,500 $\rm km\,s^{-1}$, respectively. Given the $\sim$ 6,900 $\rm km s^{-1}$ systemic velocity and $\sim$ 1,000 $\rm km s^{-1}$ velocity dispersion of Coma, this setup ensures that the WCS probes the entire volume of the cluster, as Fig. \ref{fig::v_distr} demonstrates.

The cubes have an angular resolution of 26.5 $\times$ 40.3 arcsec$^2$ and a velocity resolution of 16.5 $\rm km\,s^{-1}$ after Hanning smoothing. The minimum noise in the mosaics is 0.2 mJy beam$^{-1}$ per channel. The noise remains within a factor 1.2, 1.5, 2, and 3 of this minimum value in the inner 1.1, 1.7, 2.2, and 2.7 deg$^2$ of the mosaic, respectively, and increases rapidly further out. Accordingly, the 3$\sigma$ H$\,\textsc{i}$ column density sensitivity at the centre of the is 1.3$\times$10$^{19}$ cm$^{-2}$ in a 25 $\rm km\,s^{-1}$ channel. At the distance of Coma, assuming a 100 $\rm km\,s^{-1}$ line-width, corresponds to a 5$\sigma$ point-source H$\,\textsc{i}$ mass of 9.6$\times$10$^{7}$ M$_{\odot}$.

In order to assess the accuracy of the flux scale in our final image, we compared the flux density of 20 continuum sources sampling both the centre and the outskirts of our mosaic to the flux densities measured by the NRAO VLA Sky Survey \citep[NVSS;][]{condon98}. We find a mean flux density ratio of 0.99 with a scatter of 0.08, suggesting no systematic shift between our flux scales and that of the NVSS, with an accuracy of better than 10 \%. We thus adopt a conservative 10 \% flux scale uncertainty in our H$\,\textsc{i}$ line intensity measurements.

\subsection{H$\,\textsc{i}$ source detection}

We use SoFiA \citep{serra15,westmeier21} to find H$\,\textsc{i}$ sources in the two mosaic cubes. Within SoFiA, we remove noise variations in the mosaic cubes based on \textit{i)} the sensitivity images returned by the mosaicking algorithm and \textit{ii)} a measurement of the noise as a function of velocity. We then run SoFiA’s smooth+clip detection algorithm with a threshold of 4$\sigma$ using 15 3D smoothing kernels. Each kernel consists of a 2D circular Gaussian on the sky and a 1D box filter in velocity. The FWHM of the 2D Gaussian takes the values 0 (no smoothing), 30, and 60 arcsec. The width of the 1D box filter is 0 (no smoothing), 25, 41, 74, 140, and 272 $\rm km\,s^{-1}$. The 15 filters are obtained using all possible combinations of the above 2D Gaussian and 1D box filters with the exclusion of filters with the largest size both on the sky and in velocity (60 arcsec with 272 and 140 $\rm km s^{-1}$, and 30 arcsec with 272 $\rm km\,s^{-1}$). We merge the detected voxels into individual sources by using a merging radius of 3 10-arcsec pixels along both RA and Dec axis, and 8 8.25 $\rm km\,s^{-1}$ channels in velocity. We reject all sources smaller than 2 pixels along RA or Dec, and 3 channels in velocity. We also reject sources at the edge of the mosaic cubes ($\sigma$ > 3$\sigma_{\rm min}$  = 0.6 mJy beam$^{-1}$), where the noise gradient is large, except one detected at a $> 6\sigma$ level. We make use of the resulting masks in the two following ways.

First, we calculate the reliability of all sources as described by \citet{serra12} and retain only the reliable ones. For this purpose we work in the parameter space defined by source total flux, peak flux (both normalised by the local noise), and number of voxels. The optimum settings of the reliability algorithm are still under investigation, and in this case we tune them to maximise the agreement with our visual assessment of which detections are true. Specifically, we smooth the distribution of sources in this parameter space using a kernel whose shape is that of the covariance matrix and whose size is scaled down to a Skellam convergence of $\sim$0.6. We retain only sources with a \textsc{reliability.fMin} $\geq$ 70 and \textsc{reliability.threshold} $>$ 0.89\footnote{For a detailed description of these parameters see \protect\url{https://github.com/SoFiA-Admin/SoFiA/wiki/SoFiA-Control-Parameters}.}. We then dilate the masks of all reliable detections in order to include faint emission at the objects’ edge. The dilation is by one 10-arcsec pixel. This choice is based on the visual inspection of the size of the mask relative to the size of the emission deemed to be reliable in the cube. Our selection criteria above result in 37 H$\,\textsc{i}$ detections. Two of these formally reliable sources were at the edge of detectability due to their low signal-to-noise ratios and at the same time both had no robust optical counterpart (for our matching criteria see Sect \ref{sect::opt_counter}). After careful visual inspection we decided to consider them as likely spurious detections and hence were discarded from our sample.

The second use of the initial masks is to select objects with a spectroscopic optical counterpart regardless of their reliability. We use the spectroscopic catalogues from SDSS DR12 \citep{alam15}, GOLDMine \citep{gavazzi03} and \citet{healy21}. The first two catalogues include $\sim$650 galaxies each and have nearly 100 \% overlap in the volume covered by the WCS (30 GOLDMine objects are not in SDSS DR12, and 25 SDSS DR12 objects are not in GOLDMine). The spectroscopic sample of \citet{healy21} consists of 953 galaxies within the volume probed by our WSRT data -- 630 of them were also included in the GOLDMine (SDSS DR12) sample, thus the addition of the \citet{healy21} catalogue further increased our spectroscopic prior sample by $\sim$ 50 \%. For each spectroscopic object in the union of these three catalogues we look for H$\,\textsc{i}$ detections whose geometric centre is within 30 arcsec and 150\,$\rm km s^{-1}$ of the optical position and velocity, respectively. This results in 17 new detections not included in the blind initial catalogue described above. As above, we dilate the mask of the new detections to properly capture the surrounding faint flux components missed by the automatic SoFiA masks. Adding these sources to the initial catalogue we obtain a sample of 52 H$\,\textsc{i}$ detection candidates.

\subsection{Associated optical counterparts and detection verification}
\label{sect::opt_counter}

\begin{figure*}
   \centering
   \includegraphics[width=0.85\textwidth]{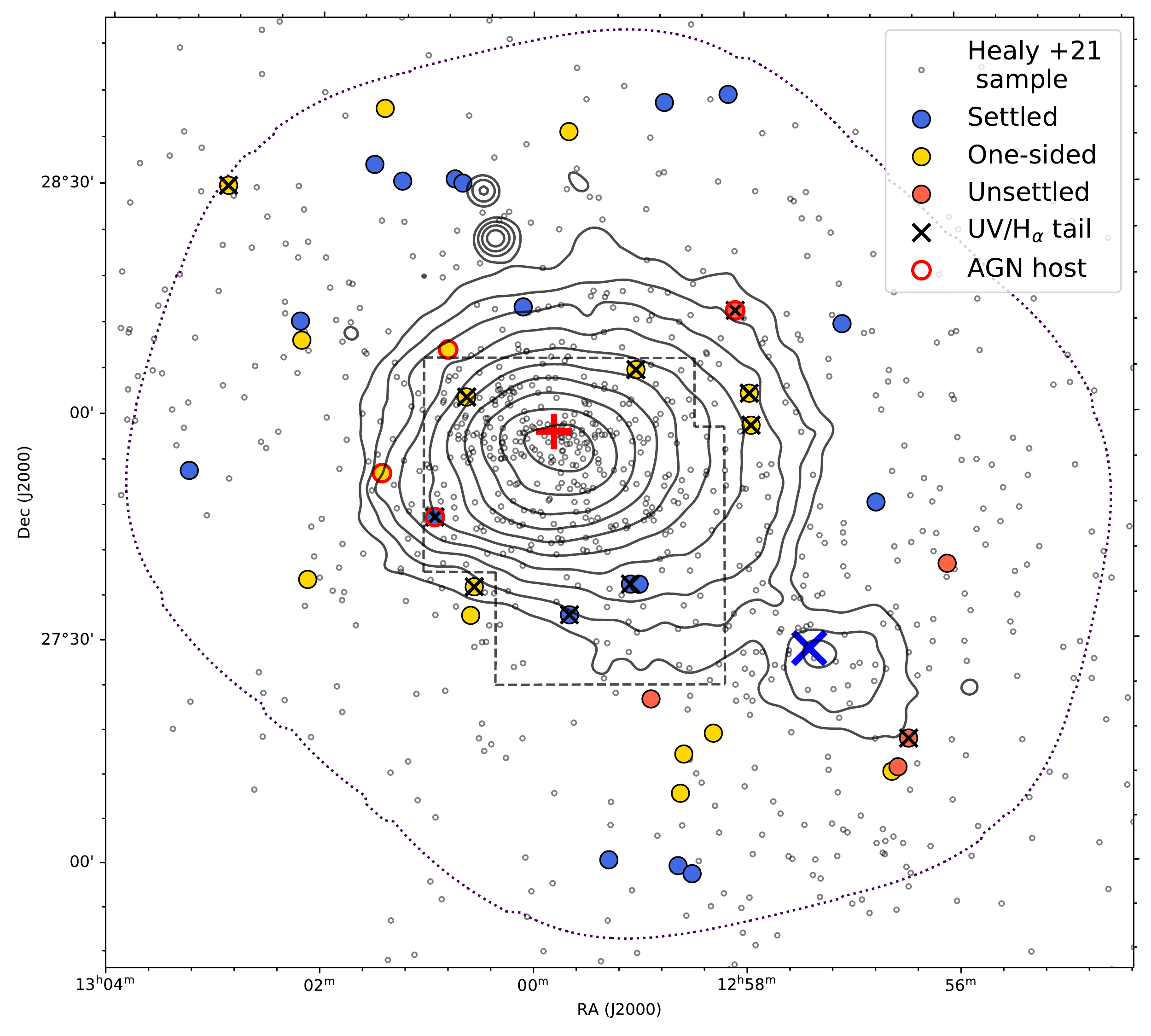}
   \caption{Position of H$\,\textsc{i}$-selected galaxies (coloured circles) in the plane of sky. Spectroscopically confirmed Coma members from \citet{healy21} are shown as grey circles. The red plus sign represents the centre of the Coma cluster, while the blue cross is placed at the position of NGC 4839. Solid black contours represent 2 - 7.2 keV X-ray distribution observed by XMM-Newton \protect\citep{neumann03}. The lowest contour was drawn at 5 c s$^{-1}$ deg$^{-2}$, while subsequent contours are $\sqrt{2^n}$ multiples of it. The dotted contour is drawn at the 25 \% of the peak primary beam sensitivity of our WSRT mosaic. Black crosses highlight sources with UV \protect\citep{smith10} and/or H$_\alpha$ \protect\citep{yagi10} tails. Dashed lines mark the H$_\alpha$ coverage of \protect\citet{yagi10}. All but one of our H$\,\textsc{i}$ detections were found to have H$_\alpha$ tails within this region. Finally, red circles indicate galaxies identified as AGN in previous literature \protect\citep{mahajan10,gavazzi11,toba14}.}
   \label{fig::comaMap}
\end{figure*}

As a final step in assessing the robustness of the 52 H$\,\textsc{i}$ detection candidates, we visually inspected their H$\,\textsc{i}$ cubes and source masks in combination with the optical maps and counterpart velocities. Due to having only low-significance detections in 1 or 2 channels and a lack of clear optical counterpart, we labelled 12 H$\,\textsc{i}$ detection candidates as likely spurious. These were removed from our catalogue and all subsequent analysis. Thus our final H$\,\textsc{i}$-selected sample of Coma galaxies consists of 40 H$\,\textsc{i}$ sources with robust optical counterparts. Their distribution in the plane of the sky is shown in Fig. \ref{fig::comaMap}.

\begin{figure}
    \centering
    \includegraphics[width=0.45\textwidth]{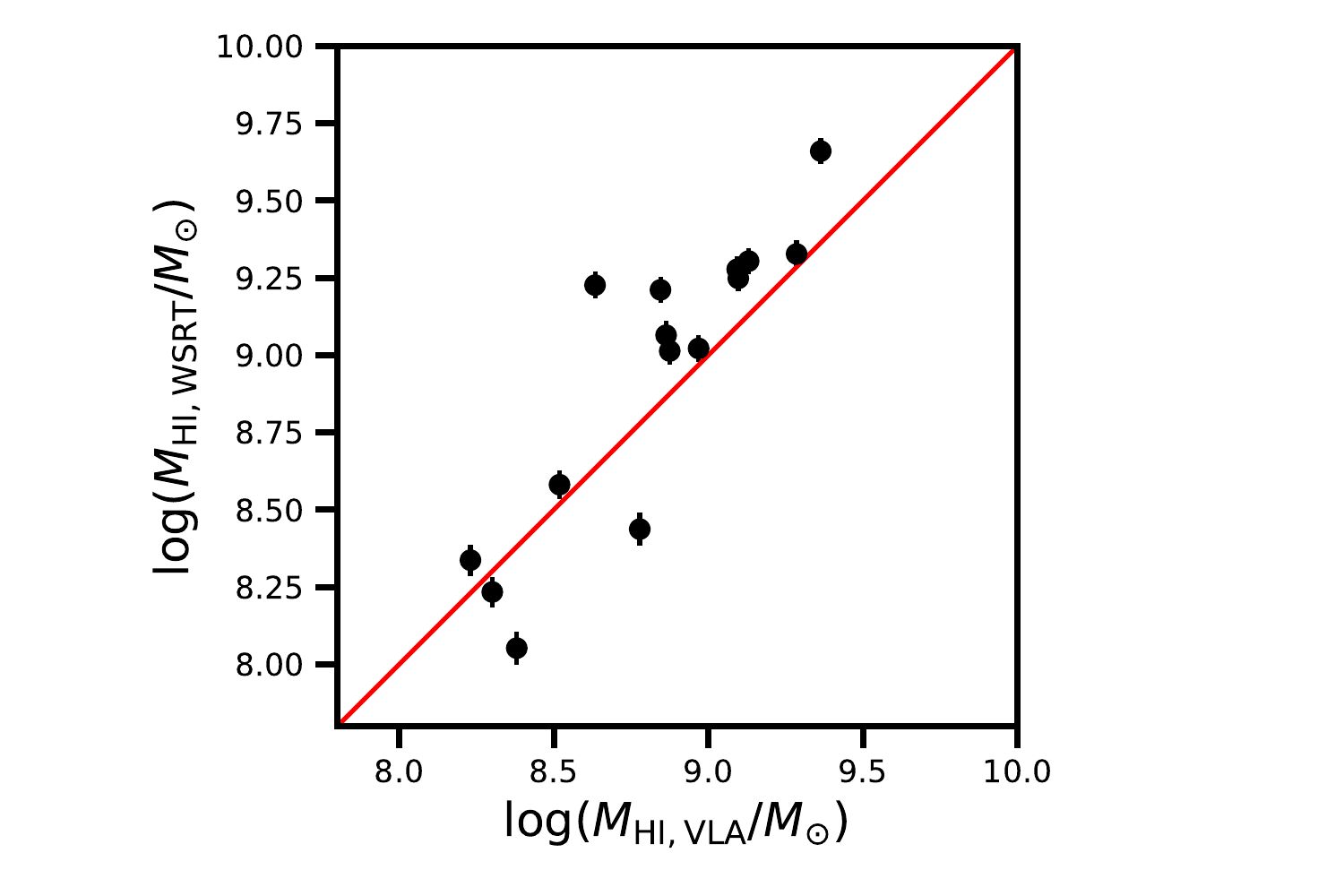}
    \includegraphics[width=0.45\textwidth]{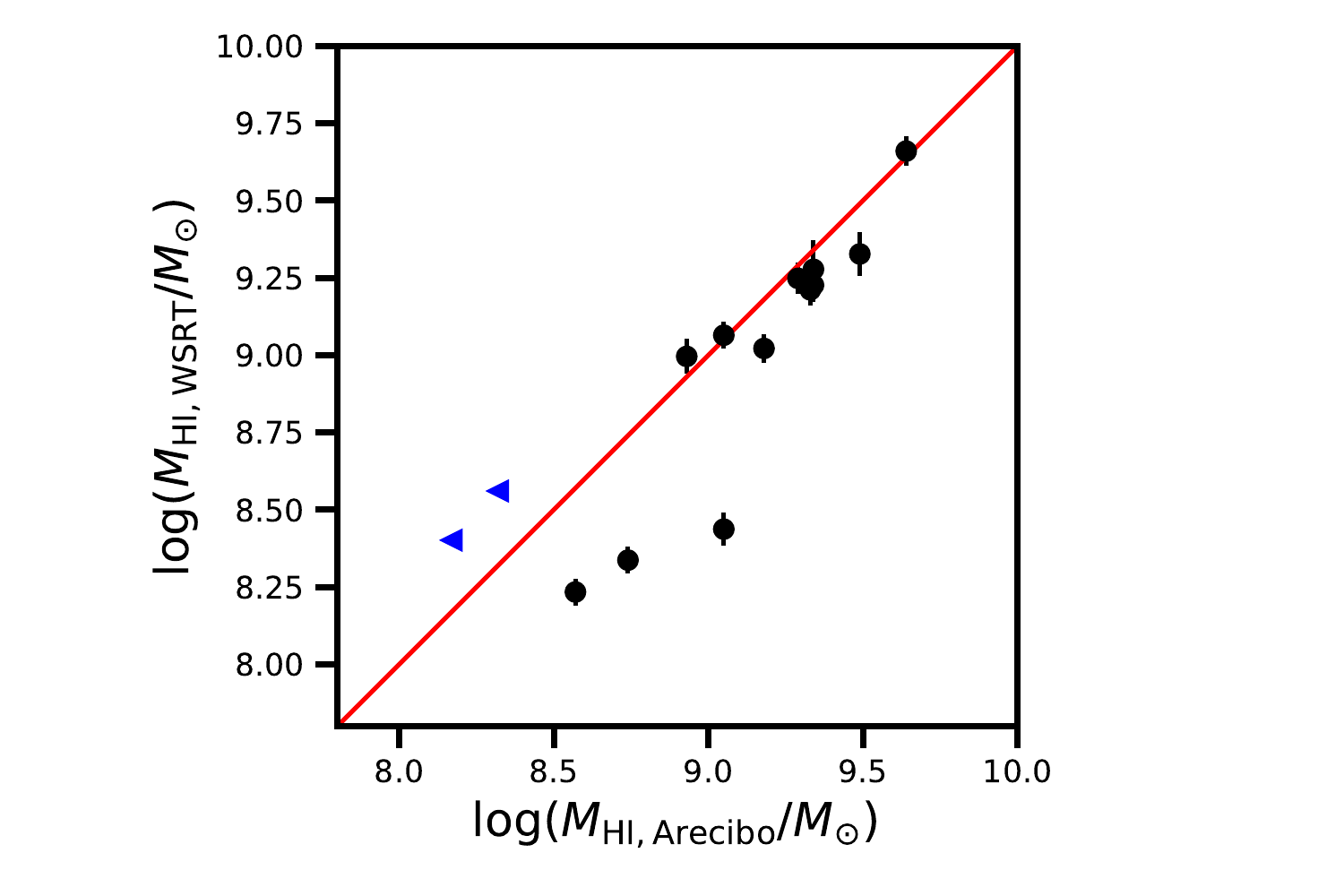}
    \caption{A comparison of H$\,\textsc{i}$ masses recovered by our WSRT survey and the VLA measurements of \protect\citet{bravo00,bravo01}, and the Arecibo survey of \protect\citet{gavazzi06} in the top and bottom panels, respectively. Red lines in both figure represent a 1:1 relation. Blue triangles in the bottom panel indicate \textit{M$_{\rm HI}$} upper limits in the \protect\citet{gavazzi06} catalogue.}
    \label{fig::BA_WSRT_comp}
\end{figure}

We estimated the H$\,\textsc{i}$ mass (\textit{M$_{\rm HI}$}) with the use of Eq. 50 in \citet{meyer17}. We have detected all 16 galaxies in our WSRT footprint previously observed by \citet{bravo00,bravo01}. We find that our detections, on average, recover 0.11 dex more flux (with a scatter of 0.23 dex), as seen in Fig. \ref{fig::BA_WSRT_comp}. \citet{chen20} found 0.3 dex higher H$\,\textsc{i}$ fluxes for their three H$\,\textsc{i}$ detections (NGC 4848, NGC 4911, and IC 4040) than that of \citet{bravo00}. A comparison of H$\,\textsc{i}$ fluxes in \citet{chen20} to ours reveals a median flux ratio of $\sim$ 1. \citet{chen20} proposed that \citet{bravo00} may have missed the diffuse H$\,\textsc{i}$ component due signal-to-noise constraints. Fourteen of our H$\,\textsc{i}$ detections also have unambiguous counterparts in the Arecibo survey of \citet{gavazzi06}, that is we only found one source in the $\sim$ 3.8 arcmin Arecibo beam. Two of these were tabulated as upper limits. Examining the remaining 12 direct detections we find that, on average, our \textit{M$_{\rm HI}$} measurements are 0.16 dex lower (with a scatter of 0.19 dex), than those of the \citet{gavazzi06} sample. In summary, our flux measurements are broadly consistent with publicly available data found in the literature.

\section{Analysis}
\label{sect::methods}

\subsection{H$\,\textsc{i}$ morphological classification}
\label{sect::hi_morph}

The spatial distribution of neutral hydrogen around galaxies is linked to its assembly history. Therefore, in order to facilitate a more physically motivated investigation into the link of ongoing star formation activity, large-scale environment, and gas content in our sample, we divided all of our H$\,\textsc{i}$ detections into three distinct morphological types. This classification was carried out through visual inspection of all 40 H$\,\textsc{i}$ sources (by two adjudicators), considering both their projected 2D distribution in integrated H$\,\textsc{i}$ intensity maps as well as the individual channel maps extracted from the data cubes relative to the associated stellar bodies. Using these information we established three morphological types based on the following criteria:

\begin{itemize}
    \item \textbf{Settled sources} -- characterized by symmetric H$\,\textsc{i}$ distribution centred on the stellar body and a velocity gradient consistent with rotation. We note that unresolved sources were added to this category as well.
    \item \textbf{One-sided asymmetry} -- having either a regular disk-like component, similar to settled sources, with an additional tail-like extension in one particular direction or a regular component but with significant excess H$\,\textsc{i}$ flux on one side of the stellar body.
    \item \textbf{Unsettled} -- showing H$\,\textsc{i}$ flux extensions in multiple directions from the stellar body, or severe 3D asymmetries/displacement relative to the optical light with no clear disk-like component, resulting in a generally irregular H$\,\textsc{i}$ distribution and/or kinematics.
\end{itemize}{}

\noindent
Appendix \ref{app::stamps} contains the H$\,\textsc{i}$ and optical maps of all 40 H$\,\textsc{i}$-selected sources organised according to their H$\,\textsc{i}$ morphologies. In these figures our assigned H$\,\textsc{i}$ morphological classes appear as blue, yellow or red circles in the corner of each image referring to the settled, one-sided asymmetry, and unsettled categories, respectively. In the same order these groups correspond to a numerical flag of 0, 1, and 2 in our table available as an online supplement\footnote{At the CDS via anonymous ftp to cdsarc.u-strasbg.fr (130.79.128.5), via \protect\url{http://cdsweb.u-strasbg.fr/cgi-bin/qcat?J/A+A/} or at \url{https://github.com/molnard89/WCS-Catalog-Release}.} containing all main source parameters. We note that while other methods exist that quantitatively characterise the magnitude of kinematic and spatial \citep[e.g.][]{haynes98,espada11,giese16,scott18,reynolds20} asymmetries present in the H$\,\textsc{i}$ gas, we decided not to employ these due to the insufficient velocity and angular resolution of our data, as well as our relatively small sample size. Given the scope of our observations, the visual approach described in this section proved adequate to reliably distinguish between perturbed and unperturbed systems.

\subsection{Field galaxy reference sample}
\label{sect::refsamp}

In order to study the effect environment has on the star formation properties and the gas content of our Coma cluster H$\,\textsc{i}$ detections, we compiled a set of galaxies found in less dense environments as a comparison sample. The bulk of these field galaxies (the reference sample henceforth) came from the Herschel Reference Survey \citep[HRS;][]{boselli10}, a K-band selected, volume-limited compilation of nearby galaxies spanning a wide range of stellar mass and morphological types. We select objects in HRS that have robust H$\,\textsc{i}$ detections from \citet{boselli14a}. These H$\,\textsc{i}$ measurements mainly consist of ALFALFA data from \citet{haynes11} and the collection of single-dish detections by \citet{springob05}. To target sources in less dense environments, we omitted HRS galaxies residing within the densest regions of the Virgo cluster based on the environmental flags of the HRS team (from priv. comm. with Luca Cortese). These selection criteria yielded 155 sources from HRS. To extend the stellar mass range of the reference sample below $\log(M_{\star} / \mathrm{M_{\odot}})$ $\sim$ 9, and thus to better probe the stellar mass range of our Coma H$\,\textsc{i}$ detections, we added sources from the Void Galaxy Survey \citep[VGS;][]{kreckel11,kreckel12} to the 155 HRS sources. The VGS is a geometrically selected\footnote{The VGS sample selection was carried out via employing a unique void identifying algorithm. A galaxy density field was reconstructed using the SDSS DR7 data, followed by a purely geometric void finding procedure. In the end, 60 galaxies were selected in the centre of clearly defined void regions.} sample of 60 galaxies in the least dense environments of the nearby Universe. Among these, 41 had a secure H$\,\textsc{i}$ detection. Of the H$\,\textsc{i}$-selected galaxies we managed to obtain stellar mass and SFR estimates for 37, using the same as methods described in Sect \ref{sect::mstar_sfr_calc}. Therefore, the final reference sample consists of 192 H$\,\textsc{i}$-selected field galaxies with stellar mass and SFR measurements.

\subsection{Measuring stellar masses and star formation rates}
\label{sect::mstar_sfr_calc}

Any given stellar mass and SFR calculation technique may introduce biases to estimated \textit{M$_{\star}$} and SFR values compared to other such methods \citep[see e.g.][]{kennicutt12, madau14}. Moreover, the variation of the point spread function (PSF) characteristics and flux extraction approaches between different instruments and surveys could result in additional, more subtle biases in the derived galaxy properties. Since our primary goal is to look for differences between galaxies residing in the Coma cluster and in the field, we attempted to minimise the impact of potential systematic errors on our comparison by applying the same \textit{M$_{\star}$} and SFR estimation methods using the same data products in both the H$\,\textsc{i}$-selected Coma sample and the reference sample.

Firstly, Wide-field Infrared Survey Explorer \citep[WISE;][]{wright10} 3.4\,$\mu$m, 4.6\,$\mu$m, 12\,$\mu$m, and 22\,$\mu$m flux measurements were carried out at the positions of the optical counterparts of H$\,\textsc{i}$-selected Coma and reference sample galaxies with the custom pipeline of \citet{jarrett13,jarrett19}. Mass-to-light ratios and subsequent stellar masses were derived using 3.4\,$\mu$m and 4.6\,$\mu$m photometry, following the calibration of \citet{cluver14}:

\begin{equation}
    \log(M_{\star} / L_{\rm W1}) = -1.96 (\mathrm{W1} - \mathrm{W2}) - 0.03,
\end{equation}

\noindent
with 

\begin{equation}
    L_{\rm W1} = 10^{-0.4(M-M_{\rm Sun})},
\end{equation}

\noindent
where $L_{\rm W1}$ is the rest-frame 3.4\,$\mu$m luminosity of a given galaxy in solar luminosity units, $M$ is its absolute magnitude in the same band, $M_{\rm Sun} = 3.24$, and $(\mathrm{W1} - \mathrm{W2})$ is the rest-frame (4.6\,$\mu$m - 3.4\,$\mu$m) colour measured by WISE.

\begin{figure}
    \centering
    \includegraphics[width=0.45\textwidth]{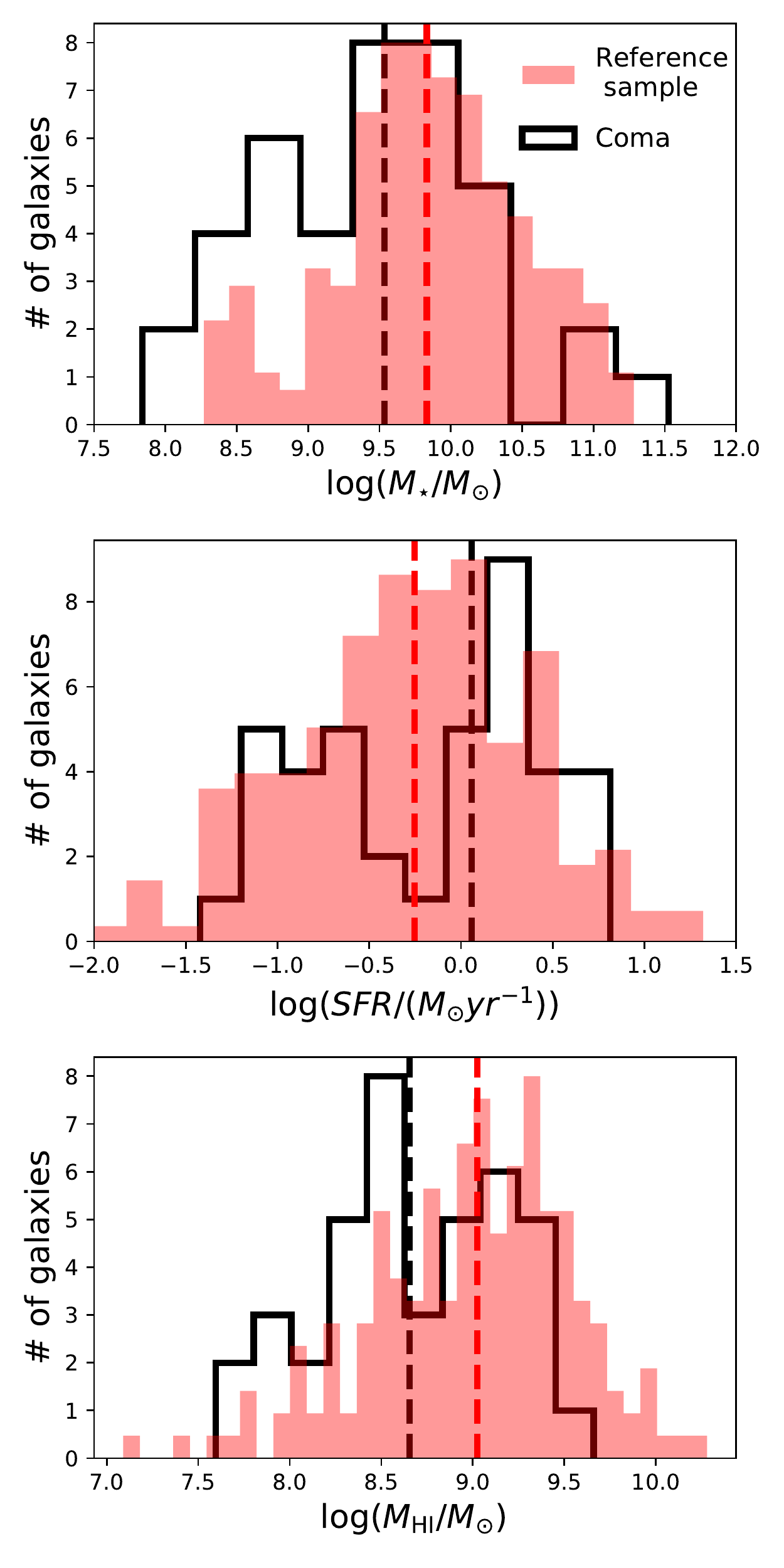}
    \caption{Distribution of stellar mass (top), SFR (middle), and $\rm M_{\rm HI}$ of H$\,\textsc{i}$-selected Coma galaxies (black) and the reference sample (red). The latter has been re-scaled to ease comparison. In each panel, black and red vertical dashed lines are drawn at the median values of the H$\,\textsc{i}$-selected Coma sample and the reference sample, respectively.}
    \label{fig::sfrmstar_hist}
\end{figure}

Since our reference and Coma samples cover $\sim$ 3.5 dex in stellar mass (Fig. \ref{fig::sfrmstar_hist}, top panel), they are expected to show substantial variation in the balance of obscured and unobscured star formation as a function of their stellar mass \citep[e.g.][]{whitaker17}. In order to mitigate any potential biases arising from this effect, and to derive consistent SFR estimates across the whole \textit{M$_{\star}$} range, we adopted the hybrid SFR estimation approach described in \cite{boquien16}.

This method requires UV photometry. Thus we added near ultriaviolet (NUV) measurements from the Galaxy Evolution Explorer (GALEX) All-Sky Survey \citep[GASC;][]{seibert12} to the optical counterparts of our H$\,\textsc{i}$-selected Coma and VGS galaxies via spatial cross-matching with a radius of 20 arcsec. For HRS sources we used the UV photometry of \citet{cortese12}. Near ultriaviolet magnitudes were corrected for Galactic extinction using the coefficient in Table 2 of \citet{yuan13} and tabulated E(B-V) values of GASC and \citet{cortese12}. The total SFR is then computed as

\begin{equation}
    \left(\frac{\mathrm{SFR}}{\mathrm{M_\odot} \mathrm{yr}^{-1}}\right) = c_{\rm NUV} \times \left(\frac{L_{\rm NUV}}{L_\odot} + k_i \times \frac{L_{22 \mu m}}{L_\odot}\right),
    \label{eq::sfr_hybrid}
\end{equation}

\noindent
where $L_{\rm NUV}$ is the observed luminosity in the NUV band, $L_{22 \mu m}$ is the 22\,$\mu$m luminosity, $k_i$ = 6.17 is the luminosity-weighted correction factor of \citet{boquien16} (from their Table 3), and we use the conversion factor $c_{\rm UV} \approx 2.58 \times 10^{-10} \mathrm{M_\odot} \mathrm{yr}^{-1} L_\odot^{-1}$. For galaxies lacking either a secure $L_{\rm NUV}$ or $L_{22 \mu m}$ measurement we still calculated SFRs using Equation \ref{eq::sfr_hybrid} with the un-detected luminosity set to zero. In the H$\,\textsc{i}$-selected Coma sample only one source had no significant 22\,$\mu$m detection and all of them were measured in the NUV band, while in the reference sample 21 and 18 of the 192 sources lacked $L_{22 \mu m}$ and $L_{\rm NUV}$ estimates, respectively.

Figure \ref{fig::sfrmstar_hist} shows the distributions of \textit{M$_{\star}$}, SFR, and \textit{M$_{\rm HI}$} values. The tables containing H$\,\textsc{i}$ and counterpart galaxy properties of galaxies both in the Coma and the reference samples are available as an online supplement\footnote{At the CDS via anonymous ftp to cdsarc.u-strasbg.fr (130.79.128.5), via \protect\url{http://cdsweb.u-strasbg.fr/cgi-bin/qcat?J/A+A/} or at \url{https://github.com/molnard89/WCS-Catalog-Release}.}. The reference sample catalogue is partially shown in Table \ref{tab::ref_samp}.

\section{Results}
\label{sect::results}

\subsection{Spatial distribution of H$\,\textsc{i}$-selected galaxies within the Coma cluster}

\begin{figure*}
   \centering
   \includegraphics[width=0.85\textwidth]{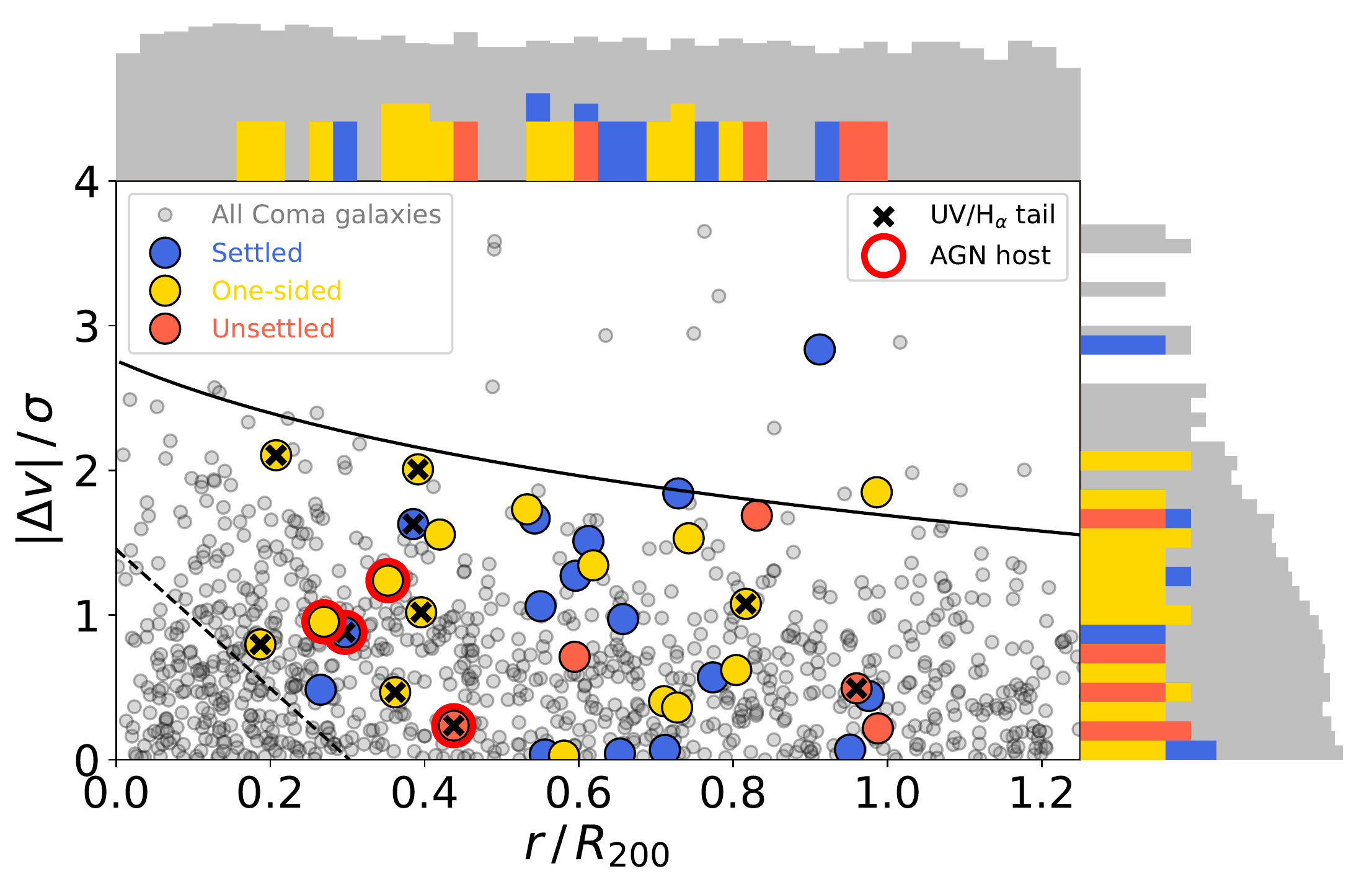}
   \caption{Phase-space diagram of the Coma cluster. Red, yellow, and blue circles mark our H$\,\textsc{i}$ detections, while grey circles represent the underlying distribution of all spectroscopically confirmed Coma sources from \protect\citet{healy21}. Logarithmic histograms of appropriate colour along the horizontal and vertical axes show the marginalised distributions of cluster centre distance and line-of-sight velocities, respectively. The dashed line is drawn at the upper boundary of the statistically most virialised galaxy population from Equation 3 and 4 of \citet{pasquali19} using $p=1$. The black curve is the cluster escape velocity as a function of distance from the centre, calculated via Equation 2 of \protect\citet{jaffe15}. The black crosses highlight sources with UV \protect\citep{smith10} and/or H$_\alpha$ \protect\citep{yagi10} tails. Finally, the red circles indicate galaxies identified as AGN in previous literature
 \protect\citep{mahajan10,gavazzi11,toba14}.}

\label{fig::pps}
\end{figure*}

As seen in Fig. \ref{fig::comaMap}, our H$\,\textsc{i}$ detections envelop the densest, X-ray emitting region of the cluster core with disturbed morphologies with tails becoming slightly more common towards the centre. Similarly, H$\,\textsc{i}$ detections are scarcer towards the south-west NGC 4839 group. In accordance with the spatial distribution, H$\,\textsc{i}$-selected galaxies have a flatter and broader line-of-sight velocity distribution with a dispersion of $\sim$ 1420 km/s (as opposed to the $\sim$ 1045 km/s dispersion found for entire Coma sample, as seen in Fig. \ref{fig::v_distr}), already marking the H$\,\textsc{i}$-selected population kinematically distinct from the overall Coma population \citep[as found for the star forming, late-type, and therefore likely gas-rich galaxies by][]{colless96,boselli06}. Indeed, combining spatial and velocity information by placing galaxies on the phase-space diagram (i.e. projected cluster centre distance versus line-of-sight velocity offset with respect to the cluster systemic velocity, seen in Fig. \ref{fig::pps}) reveals that our H$\,\textsc{i}$ detections appear in a regime similar to the phase-space region found in \citet{jaffe15}, for example. Specifically, H$\,\textsc{i}$ galaxies avoid the virialised region as well as high line-of-sight velocities with low projected distances. The majority of galaxies in the so-called virialised cone have been members of the cluster for several Gyrs \citep{oman13}, more than the crossing time of the cluster, and therefore had at least one pericentric passage and likely lost a significant portion of their H$\,\textsc{i}$ content as a result.

\subsection{Star formation and gas content}

\begin{figure}
   \centering
   \includegraphics[width=0.45\textwidth]{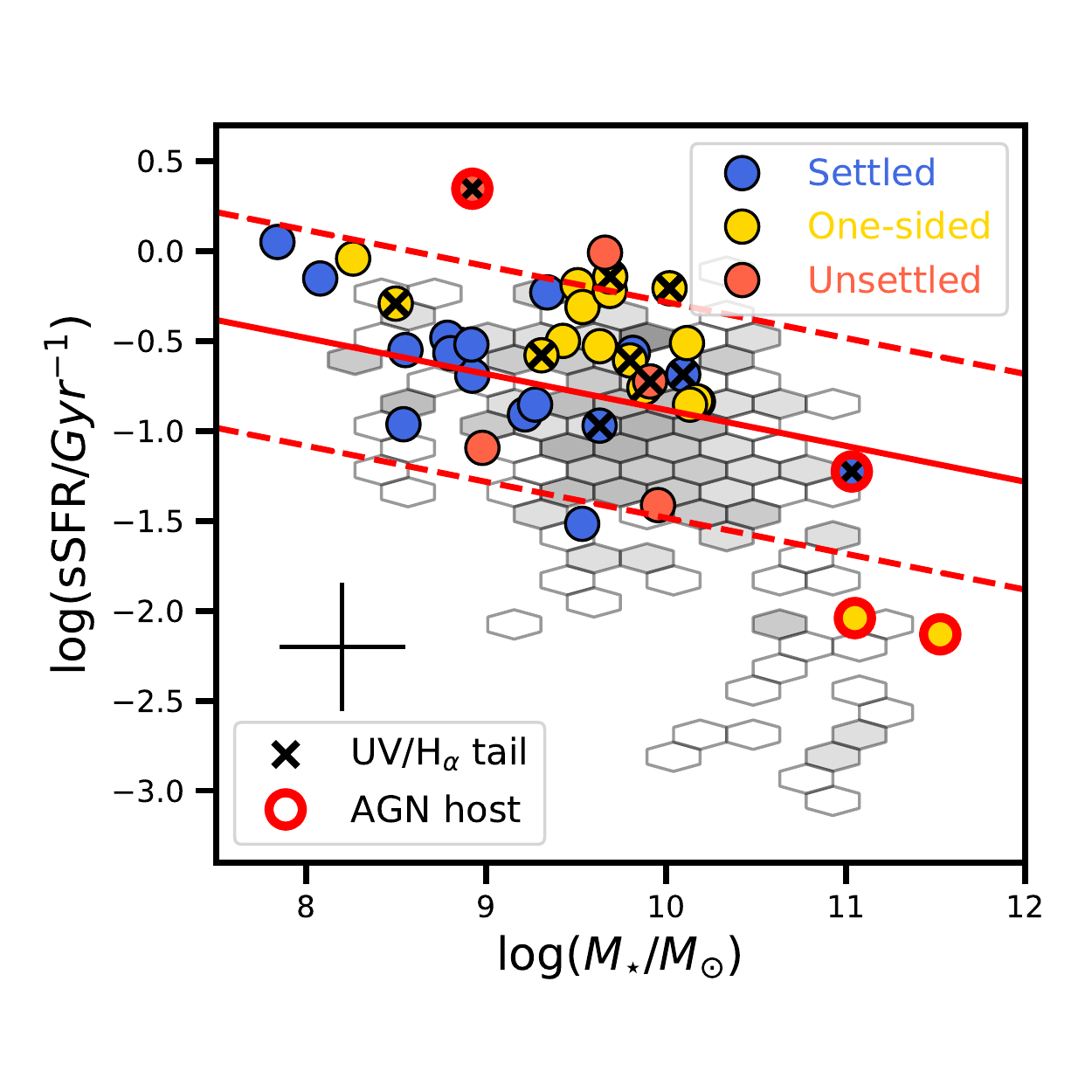}
   \caption{Distribution of H$\,\textsc{i}$-selected Coma galaxies (coloured circles) and the field galaxy reference sample (hexagonal bins indicating the number of sources in each bin with a logarithmic colourscale) in the stellar mass -- specific star formation rate plane. The red line represents the z $=$ 0 star forming main sequence relation of \protect\cite{sargent14}, while dashed lines above and below indicate the $\pm$ 0.6 dex offset from SFMS. Black crosses mark Coma galaxies with tails at H$_{\alpha}$ \protect\citep{yagi10} or UV \protect\citep{smith10} wavelengths, while red circles denote ongoing AGN activity. Black errorbars are the mean uncertainties of the Coma measurements. Finally, red circles indicate galaxies identified as AGN in the literature
 \protect\citep{mahajan10,gavazzi11,toba14}.}
   \label{fig::sfms}%
\end{figure}

Figure \ref{fig::sfms} shows the H$\,\textsc{i}$-selected Coma sample against the reference sample in the $M_{\star}$ -- specific star formation rate (sSFR\footnote{The specific star formation rate, sSFR, is defined as $\mathrm{sSFR} \equiv \mathrm{SFR}/M_{\star}$.}) plane. Out of our 40 galaxies, 34 fall on the star forming main sequence (SFMS), meaning they are found within the typical 0.6 dex scatter of the relation (shown as dashed red lines in Fig. \ref{fig::sfms}). However, on average, H$\,\textsc{i}$-selected Coma galaxies show an enhancement in SF since they lie $\sim$ 0.2 $\pm$ 0.1\footnote{The uncertainties of the median measurements throughout the paper are estimated with the standard error on the median ($\sigma_{\mathrm{med}}$), that is $\sigma_{\mathrm{med}} \equiv 1.2533 \, \sigma /\sqrt{n}$, where $\sigma$ is the standard deviation of the data, and $n$ is the number of data points.} dex above the main ridge of the SFMS. In particular, the median sSFR offset from the SFMS of sources with one-sided asymmetry is $\sim$ 0.3 $\pm$ 0.1 dex \citep[which is in an excellent quantitative agreement with the 0.3 dex offset of the Coma cluster stripping candidates of ][]{roberts20}, while settled sources have, on average, only a 0.1 $\pm$ 0.1 dex excess in sSFR, indicating that galaxies with disturbed H$\,\textsc{i}$ morphologies are the main reason for the overall enhanced star formation activity in our H$\,\textsc{i}$-selected Coma sample. This is consistent with studies finding enhanced star formation activity in ram pressure stripped galaxies \citep[e.g.][]{vulcani18,roberts22}. We note that neither here nor in later stages of our analysis we could obtain statistically meaningful median estimates for the 5 unsettled sources, because their high scatters and small number resulted in large errors on the median, rendering them consistent with all other median measurements.

Other notable outliers are the 4 starbursts (i.e. more than 0.6 dex higher sSFR than predicted by the SFMS relation at their stellar mass) in the $\log(M_{\star}/\mathrm{M_{\odot}})$ < 10 regime. Interestingly all of them show disturbance in their H$\,\textsc{i}$ distribution, and 3/4 were found to have tails at UV and/or H$_{\alpha}$ wavelengths. However, most disturbed galaxies can be found on the SFMS, and two of them are even below it, suggesting that morphological disturbances might be necessary but not sufficient to generate high star formation activity (assuming that there is a causal relation between the two in the first place). Another possibility is that disturbances remain observable longer than any perturbations of the SF activity associated with them. If ram-pressure is linked to elevated SF activity, it is also worth noting that galaxies with enhanced SFR mostly fall into the 9.3 < $\log(M_{\star}/\mathrm{M_\odot})$ < 10.3 regime, and are less common amongst the lower mass H$\,\textsc{i}$ detections. This could be due to the fact that lower stellar mass galaxies lose their H$\,\textsc{i}$ gas faster, biasing our H$\,\textsc{i}$-selected sample against $\log(M_{\star}/\mathrm{M_\odot})$ < 9 galaxies with disturbed gas morphologies.

In Fig. \ref{fig::sfms} we marked 4 sources, NGC 4921, NGC 4907, NGC 4911, and NGC 4848, with red circles, indicating they have been found to harbour an AGN in previous studies \citep{mahajan10,gavazzi11,toba14}. The two most massive ones ($\log(M_{\star}/\mathrm{M_\odot}) \approx 11.0 -- 11.5$), NGC 4921 and NGC 4907, are quenched late-type galaxies. NGC 4911 is of similar $M_{\star}$, but resides on the SFMS, while NGC 4848 ($\log(M_{\star}/\mathrm{M_\odot}) \approx 9$) has a sSFR consistent with that of a starburst galaxy. NGC 4921, NGC 4907, and NGC 4848 have disturbed H$\,\textsc{i}$ components. We note that, according to their MIR photometry, none of them have W1-W2 colours greater than 0.8, where most AGN lie \citep{stern12}. They are most likely Seyferts, whose star formation can be comparable with their AGN activity \citep[e.g. in Circinus][]{for12}, and as a result can have MIR colours similar to regular SF galaxies. Thus, we consider their $\rm SFR$ estimates to be reliable within the uncertainties. Nevertheless, it is important to keep in mind that their internal, AGN-related processes might play an important role in shaping their SF and gas properties. Disentangling the influence of their AGN from the role of their environment is beyond the scope of this work. However, we note that recent literature has suggested a link between the onset of AGN activity and the presence of ram pressure stripping \citep{poggianti17,ricarte20,peluso21}, which could possibly occur when the external pressure funnels gas inwards \citep{tonnesen09,ramos-martinez18}. In this context, it is interesting that 3/4 AGN in our Coma sample were observed to have disturbed H$\,\textsc{i}$ morphologies.

\begin{figure}
   \centering
   \includegraphics[width=0.45\textwidth]{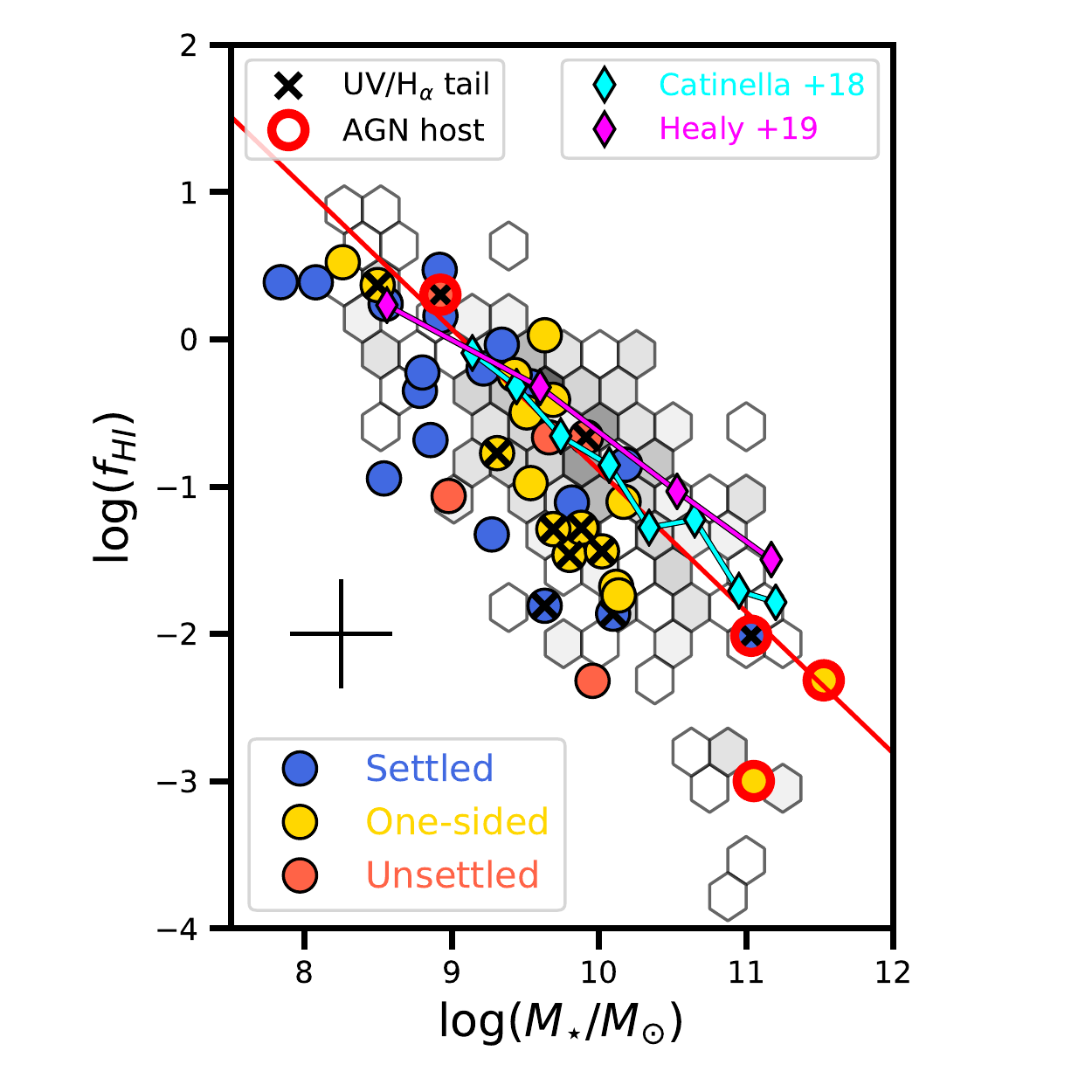}
   \caption{Distribution of H$\,\textsc{i}$-selected Coma galaxies (coloured circles) and the field galaxy reference sample (hexagonal bins) in the stellar mass -- atomic gas fraction plane. Red line shows our best-fit linear relation to the SFMS galaxies (defined using Fig \ref{fig::sfms}) in the reference sample. Black crosses mark Coma galaxies with tails at H$_{\alpha}$ or UV wavelengths, while red circles denote ongoing AGN activity. Black errorbars are the mean uncertainties of the Coma measurements. The cyan and magenta diamonds indicate the median gas fractions measured by \citet{catinella18} and \citet{healy19}, respectively.}
   \label{fig::gasfrac}%
\end{figure}

In Fig. \ref{fig::gasfrac} we plot the gas fraction (defined as $f_{\rm HI} \equiv M_{\rm HI}/M_{\star}$) of our H$\,\textsc{i}$-selected samples as a function of their stellar mass. Coma galaxies, on average, lie below the typical gas fraction across the entire observed stellar mass range. This is in accordance with the findings of \citet{healy21}, who probed the overall H$\,\textsc{i}$ content of Coma using the same WCS data by stacking non-detections, and found that Coma galaxies in general are at least 10 times more H$\,\textsc{i}$ deficient than field galaxies. In order to characterize the typical atomic gas fraction of the field galaxy sample and to create a reference scaling relation for the H$\,\textsc{i}$-selected Coma galaxies, we fitted a linear model to $M_{\star}$ -- $f_{\rm HI}$ of the star forming main sequence galaxies\footnote{We consider galaxies to be on the SFMS if their offset from the main sequence of \cite{sargent14} is < 0.6 dex. These sources lie between the dashed lines in Fig. \ref{fig::sfms}. In the reference sample 150/192 sources fulfil this criterion.} of the reference sample. Since (i) \textit{M$_{\star}$} appears in both axes of this relation, and therefore the stellar mass and gas fraction errors are correlated, and (ii) the uncertainties are comparable in magnitude (in other words neither are negligible relative to the other), we performed the fit with the bivariate correlated errors and intrinsic scatter \citep[BCES;][]{akritas96,nemmen12} method by minimising the squared orthogonal distances of the data points to the modelled relation. The resulting best-fit gas fraction scaling was found to be

\begin{equation}
\label{eq::gasfrac_scaling}
   \log(f_{\mathrm{HI}}) = - (0.96 \pm 0.06) \cdot \log(M_{\star}) + (8.71 \pm 0.6).
\end{equation}

As seen in Fig. \ref{fig::gasfrac} our best-fit model is broadly consistent with two recent characterization of the correlation by \citet{catinella18} and \citet{healy19}. Between our fit and the stacked $\log(f_{\rm HI})$ of \citet{healy19} we find a mean offset of $\sim$ 0.2 dex, with a 0.5 dex maximum at $\log(M_{\star}/\mathrm{M_\odot})$ $\sim$ 11, resulting in a slightly steeper correlation than that of \citet{healy19}. On the other hand, the measurement of \citet{catinella18} has only a mean offset of 20 \% compared to our predictions given the stellar masses of their bins. We note that considering the reference sample galaxies not within 0.6 dex of the star forming main sequence when fitting the $M_{\star}$ -- $f_{\rm HI}$ correlation would result in a steeper slope of $-1.26 \pm 0.07$, and hence a higher discrepancy compared to these measurements from the literature, but it would not impact our main results qualitatively.

When compared to our best-fit gas fraction scaling relation (Eq \ref{eq::gasfrac_scaling}), H$\,\textsc{i}$-selected Coma galaxies have 0.4 $\pm$ 0.1 dex lower $f_{\mathrm{HI}}$ than typical field galaxies of the same $M_{\star}$. As opposed to the SF properties, there is no statistically significant difference in the H$\,\textsc{i}$-deficiencies of sources with different H$\,\textsc{i}$ morphological classes.

\begin{figure}
   \centering
   \includegraphics[width=0.45\textwidth]{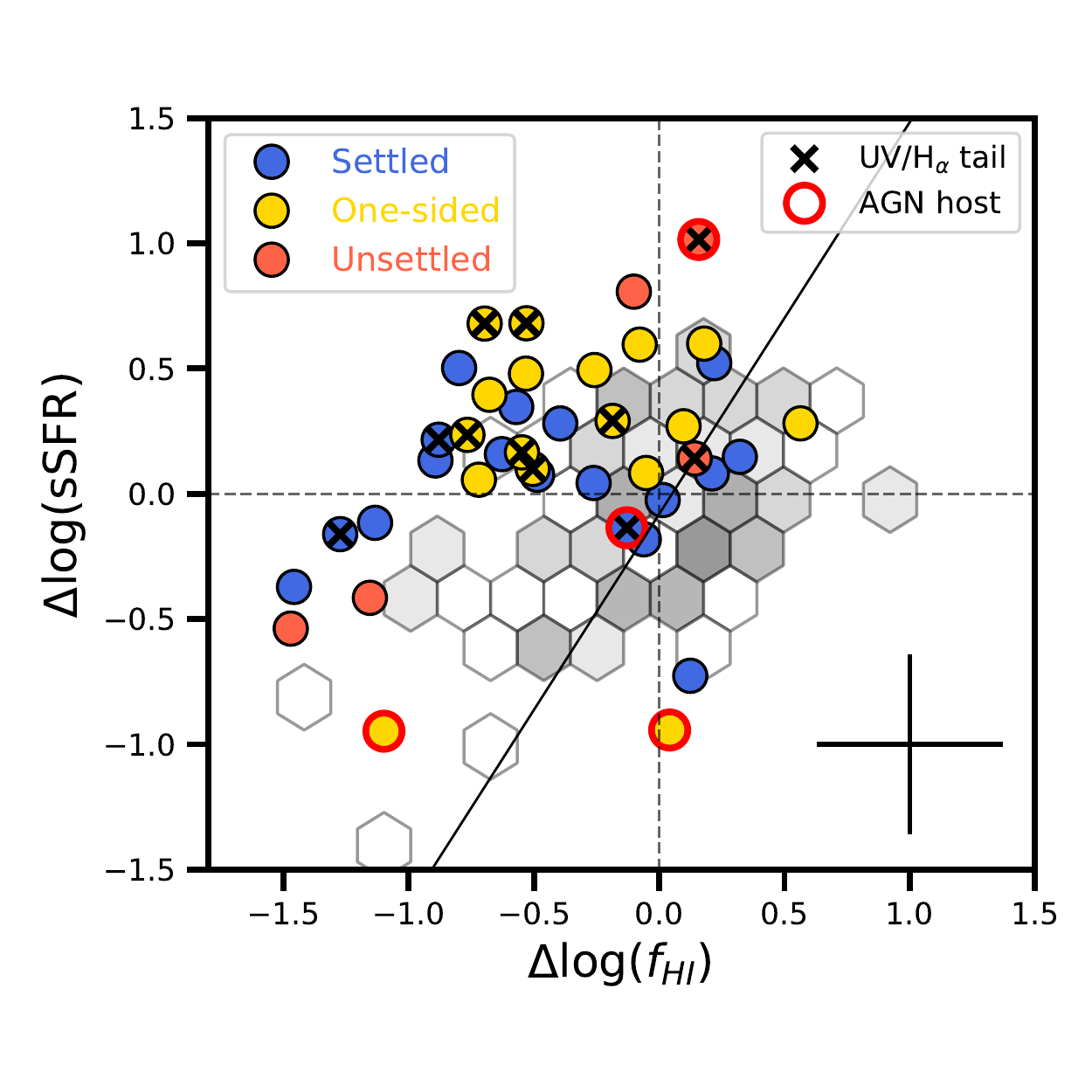}
   \caption{Offsets from the star forming main sequence (as seen in Fig. \protect\ref{fig::sfms}) versus the gas fraction -- stellar mass scaling relation (Fig. \ref{fig::gasfrac}). The black line shows the best-fit linear relation for reference sample galaxies (Eq \ref{eq::df_dssfr_bestfit}). Black errorbars are the mean uncertainties of the Coma measurements.}
   \label{fig::sf_frac_off}%
\end{figure}

Since, to the first order, stellar mass is the main parameter determining star formation and gas properties, it is important to control for it when investigating galaxy samples covering a wide range of stellar masses. Thus, in Fig. \ref{fig::sf_frac_off} we show the ratio of the measured SFR and the expected value based on the SFMS of \citet{sargent14} for each galaxy as a function of the ratio of the observed and expected gas fraction (the latter calculated via equation \ref{eq::gasfrac_scaling}). In this $\Delta$sSFR -- $\Delta f_{\rm HI}$ parameter space reference sample galaxies follow a positive correlation, as seen in Fig. \ref{fig::sf_frac_off}. In other words, field galaxies with excess sSFR at their stellar mass tend to have an excess amount of H$\,\textsc{i}$ and vice versa \citep[as already found in other samples, see e.g.][]{schiminovich10,saintonge16,janowiecki20}. However, over half of the H$\,\textsc{i}$-selected Coma sources are concentrated in the upper left quadrant of the parameter space, where galaxies simultaneously have an H$\,\textsc{i}$ gas deficit and enhanced SF activity for their stellar mass. This combination is much rarer among field galaxies (<15 \%). While individual sources might have uncertainties large enough to question the significance of their classification as an object with high star formation and H$\,\textsc{i}$ deficit, the H$\,\textsc{i}$-selected Coma galaxies as a whole display a striking offset in this direction in comparison to the reference sample. To statistically prove this impression, we first fitted the reference sample galaxies with a linear model in the $\Delta$sSFR -- $\Delta f_{\rm HI}$ plane:

\begin{equation}
    \Delta \log\left(\mathrm{sSFR}\right) = 1.56 \cdot \Delta \log\left(f_{\mathrm{HI}}\right) - 0.08.
\label{eq::df_dssfr_bestfit}
\end{equation}

\begin{figure}
   \includegraphics[width=0.45\textwidth]{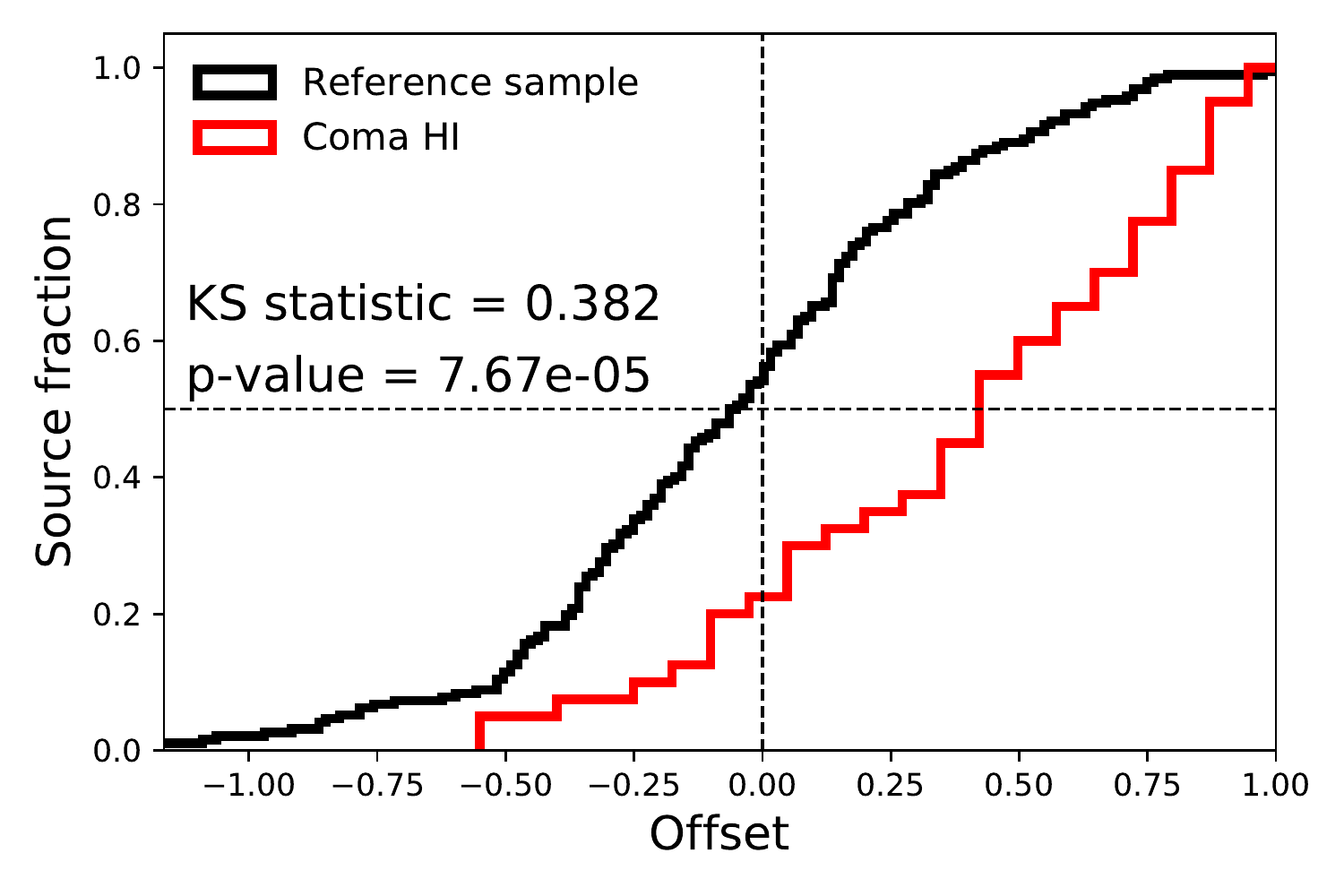}
   \caption{Cumulative distributions of the orthogonal offsets of the reference sample (black) and the Coma galaxies (red) relative to the linear best-fit relation in Fig. \protect\ref{fig::sf_frac_off}.}
   \label{fig::cumul_ref_coma}
\end{figure}

\noindent
We then examined the orthogonal distances relative to this best-fit model for both the field galaxies and the H$\,\textsc{i}$-selected Coma galaxies in Fig. \ref{fig::sf_frac_off}. The resulting cumulative distributions are presented in Fig. \ref{fig::cumul_ref_coma}. The median orthogonal distance of H$\,\textsc{i}$-selected Coma galaxies from the best-fit line is 0.5 $\pm$ 0.1 dex, with no statistical differences found between the H$\,\textsc{i}$ morphological classes. We carried out a two-sided Kolmogorov-Smirnov test on the data, and disproved the null hypothesis (i.e. that the two samples are drawn from the same underlying distribution) at a 99.999 \% significance. This provides evidence that the H$\,\textsc{i}$-selected galaxies of Coma have fundamentally different joint star formation and H$\,\textsc{i}$ gas properties than that of the galaxies found in less dense environments.

\section{Discussion}
\label{sect::discussion}

To briefly summarise Sect. \ref{sect::results}, our key results are that

\begin{enumerate}
    \item H$\,\textsc{i}$-selected Coma galaxies are on average associated with $\sim$ 0.5 dex less atomic gas mass than typical field galaxies of the same stellar mass,
    \item the H$\,\textsc{i}$-selected Coma population has, on average, a $\sim$ 0.2 dex enhancement in SFR relative to the SFMS and thus relative to the field reference sample as well,
    \item specifically, Coma galaxies with H$\,\textsc{i}$ distributions showing one-sided asymmetries have the highest median SFR enhancement ($\sim$ 0.3 dex above the SFMS).
\end{enumerate}

\noindent
These observations are broadly suggestive of the physical processes regulating the evolution of H$\,\textsc{i}$-selected Coma galaxies. Specifically, \textit{1)} suggests that the H$\,\textsc{i}$ reservoirs of infalling galaxies are likely being actively removed via for example, ram pressure or tidal interactions induced by galaxy fly-bys, while \textit{2)} and \textit{3)} in conjunction might be a clue that even though the removal of their H$\,\textsc{i}$ supply is eventually going to stop star formation, the external pressure from the dense ICM might boost H$\,\textsc{i}$-to-H$_2$ conversion, leading to a temporary enhancement of SF activity in some cases. Elevated H$_2$-to-H$\,\textsc{i}$ fractions have already been observed in particular galaxies affected by ram pressure \citep{moretti20,ramatsoku20}, and some recent studies have found enhanced amounts of molecular gas on the leading side of ram pressure stripped galaxies, coincident with areas of increased star formation \citep{lee17,cramer20,roberts20,cramer21}, lending further credibility to such a scenario occurring in some galaxies of our H$\,\textsc{i}$-selected sample.

\begin{figure}
   \centering
   \includegraphics[width=0.45\textwidth]{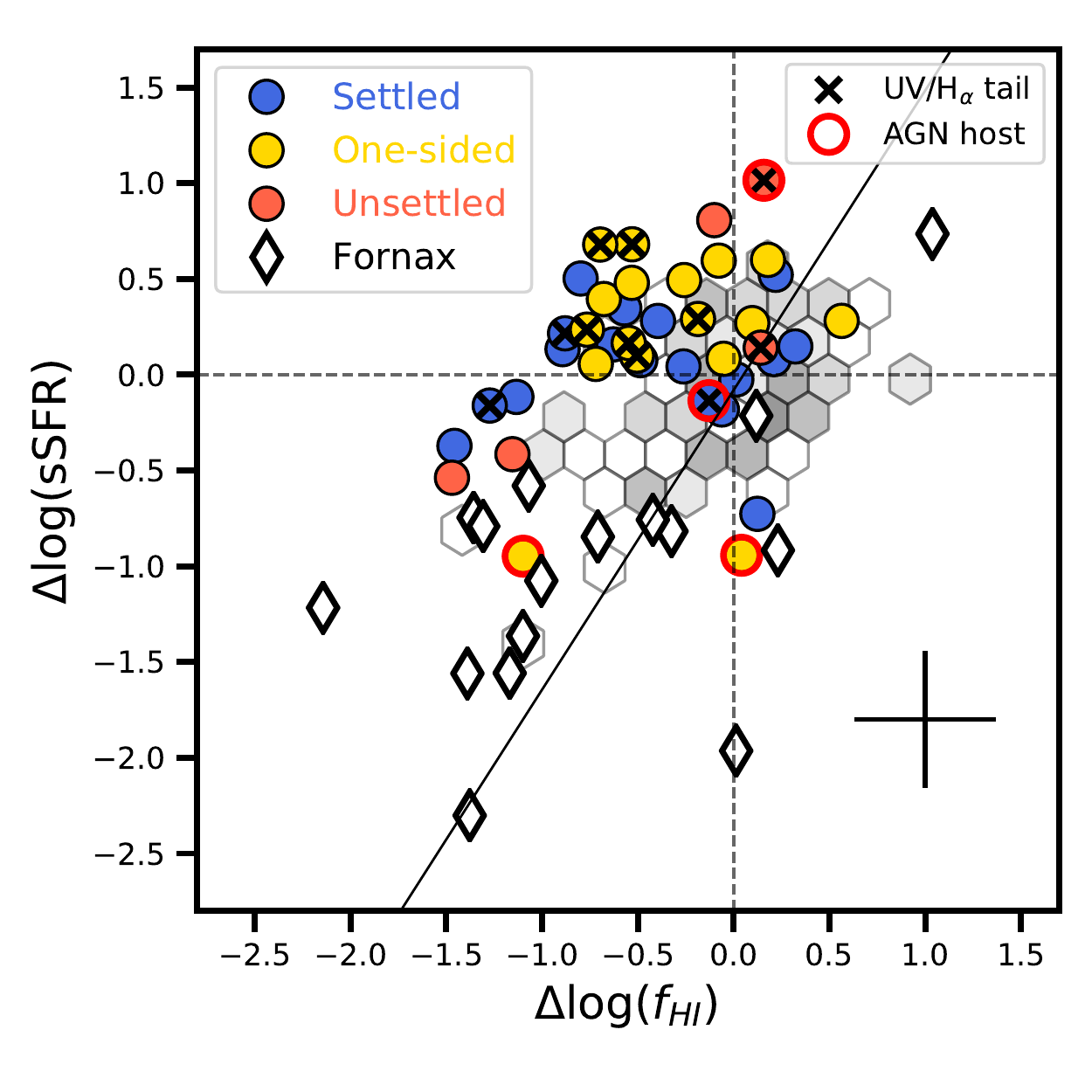}
   \caption{Offsets from the star forming main sequence versus the gas fraction -- stellar mass scaling relation for H$\,\textsc{i}$-selected Coma (circles) and Fornax (diamonds) cluster galaxies, and the reference sample (hexagonal bins). All notations are identical to that of Fig. \ref{fig::sf_frac_off}.}
   \label{fig::sf_frac_off_wFornax}%
\end{figure}

 Interestingly, H$\,\textsc{i}$-selected sources in the $\sim$ 10 times less massive Fornax cluster paint a very different picture based on data taken from \cite{loni21}. As seen in Fig. \ref{fig::sf_frac_off_wFornax}, these generally appear as both gas-poor and low SF activity sources, and thus lie at the extension of the $\Delta$sSFR -- $\Delta f_{\rm HI}$ correlation. This is likely the result of an H$\,\textsc{i}$ removal timescale slower than the H$_2$-to-stars conversion timescale, allowing the SF activity to `adjust' itself to the atomic gas content of Fornax galaxies in similar manner as field galaxies \citep{loni21}. The different behaviour of H$\,\textsc{i}$-detected Coma galaxies implies more efficient H$\,\textsc{i}$ removal or the presence of other physical mechanisms, probably linked to the fact that Coma is about 10 times richer in galaxies and more massive than the Fornax cluster. Thus, ram pressure is expected to be 10-100 times weaker in the Fornax cluster. We note that both our study and \citet{loni21} used the same methods for measuring \textit{M$_{\star}$} and SFR, and an identical reference sample for comparison, thus the difference between our results are minimally impacted by any potential systematics.

To test our aforementioned hypothesis explaining observations \textit{1)} -- \textit{3)}, we constructed a simple model that simulates trajectories in the parameter space of Fig. \ref{fig::sf_frac_off}. Furthermore, we explored the dependency of offsets from the scaling relations on H$\,\textsc{i}$ disk size to gain insights into the physical mechanisms that drive the differences between the cluster and field galaxies.

\subsection{Potential trajectories in the $\Delta$sSFR -- $\Delta f_{\rm HI}$ parameter space}
\label{sect::toymod}

In this section we present a toy model capable of simulating galaxies evolving

\begin{enumerate}[label=\Roman*.]
\item in isolation, accreting cold gas;
\item in a cluster, where accretion of cold gas onto the disk has stopped;
\item in a cluster, where in addition to halted accretion, H$\,\textsc{i}$ is actively being removed by interactions with the ICM.
\end{enumerate}

\noindent
We attempt to keep the free parameters of this toy model to a minimum and understand what combination of parameter values populates the region of Fig \ref{fig::sf_frac_off} occupied by our Coma H$\,\textsc{i}$ detections.

\begin{figure*}
    \centering
    \includegraphics[width=0.95\textwidth]{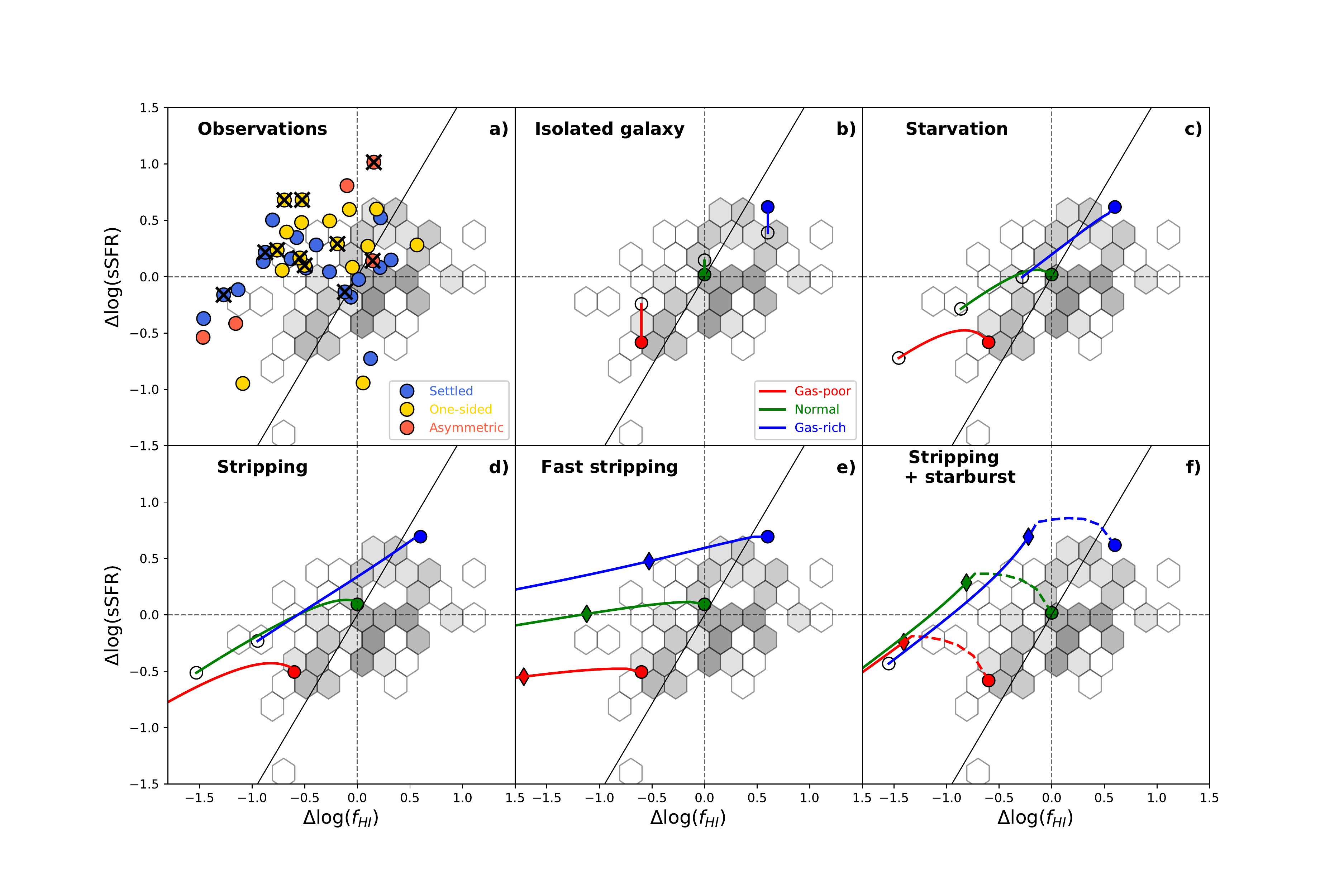}
    \caption{Simulated trajectories of galaxies in the $\Delta$sSFR -- $\Delta f_{\rm HI}$ space represented by red, green, and blue curves with filled circles marking the gas-poor, normal, and gas-rich starting points, respectively, while empty circles are the end states. Hexagonal bins show the reference sample with the black line as its best-fit correlation. The top left panel (\textit{a)} shows our H$\,\textsc{i}$-selected sample as seen in Fig. \ref{fig::sf_frac_off}. Each subsequent panel demonstrates a different scenario of galaxy evolution as described in Sect. \ref{sect::toymod}. Specifically, \textit{b)} represents a field galaxy with atomic gas accretion, \textit{c)} has no H$\,\textsc{i}$ replenishment or removal, \textit{d)} shows H$\,\textsc{i}$ stripping on timescales comparable with the H$\,\textsc{i}$-to-H$_2$ conversion timescale, \textit{e)} shortens this timescale by a factor of 10, while \textit{f)} adds a 300 Myr starburst episode to the models of \textit{e)} with a factor 8 enhancement to the H$\,\textsc{i}$-to-H$_2$ conversion efficiency. In panels \textit{e)} and \textit{f)} diamonds indicate the model states after 300 Myr, and in panel \textit{f)} dashed lines highlight the starburst phases.
    }
    \label{fig::toymod}
\end{figure*}

Simulated galaxies in our model at any given epoch are primarily characterized by the mass of three components: their atomic gas, molecular gas, and stars. There are four possible pathways for mass to transfer between them as well as between the external medium, and the model galaxies. These are, H$\,\textsc{i}$ replenishment from the IGM and/or the halo, H$\,\textsc{i}$ loss to the IGM, H$\,\textsc{i}$ conversion into H$_2$, and H$_2$ conversion into stars. The three aforementioned galaxy evolution scenarios \textit{I)}, \textit{II)} and \textit{III)} can be modelled by allowing different combinations of these four mass transfer channels --- specifically: an isolated galaxy (scenario \textit{I}) is modelled by enabling H$\,\textsc{i}$ replenishment, H$\,\textsc{i}$-to-H$_2$, and H$_2$-to-stars conversion; for starvation (scenario \textit{II}) H$\,\textsc{i}$ replenishment is turned off; and in the H$\,\textsc{i}$ stripping case (scenario \textit{III}) we enable H$\,\textsc{i}$ removal. For a quick reference, we gathered these information in Table \ref{tab::model_scenarios} -- it shows exactly which mass transfer channels were active during each of the three scenarios, and on which panel of Fig. \ref{fig::toymod} they appear.

\begin{table}
    \centering
    \caption{Mass transfer channels active in our three toy model scenarios. The final row lists the figures where the appropriate scenarios are shown.}
    \label{tab::model_scenarios}
    \begin{tabular}{c|ccc}
          & \makecell{Isolated galaxy \\ (I)} & \makecell{Starvation \\ (II)} & \makecell{Stripping \\ (III)} \\ \hline
          \makecell{H$\,\textsc{i}$ \\ replenishment} & \checkmark & & \\
         \makecell{H$\,\textsc{i}$-to-H$_2$ \\ conversion} & \checkmark & \checkmark & \checkmark \\ 
         \makecell{H$_2$-to-stars \\conversion} & \checkmark & \checkmark & \checkmark \\
         \makecell{H$\,\textsc{i}$ \\ removal} &  & & \checkmark \\ \hline
         Figure & 12b) & 12c) & 12d),e),f)
    \end{tabular}
\end{table}

The rate at which mass flows in these channels is regulated by timescales defined in the following.
We initialise the model by setting an initial stellar mass, an optional offset from the stellar mass dependent H$\,\textsc{i}$ fraction relation in Eq. \ref{eq::gasfrac_scaling}, and an initial H$_2$-to-H$\,\textsc{i}$ ratio, $r_{\rm mol, 0}$. Together these values define the starting mass of our three modelled components. Then we re-evaluate the mass of each component every 10 Myr based on a series of timescales, which stay constant during each model run, and which regulate the conversions between each component. Specifically, at each time step we first calculate the amount of H$\,\textsc{i}$ removed  from the galaxy and incorporated into the ICM with the use of the H$\,\textsc{i}$ removal rate (HIRR), defined as

\begin{equation}
    \mathrm{HIRR} = \frac{M_{\rm H_I}}{\tau_{\rm remov, HI}},
    \label{eq::hirr}
\end{equation}

\noindent
where $\tau_{\rm remov, HI}$ is the timescale over which all of the current H$\,\textsc{i}$ would be removed assuming the HIRR remains constant. Once a fraction of $M_{HI}$ is subtracted accordingly, we calculate the fraction of the remaining H$\,\textsc{i}$ converted to H$_2$ in a similar fashion via the H$_2$ formation rate, H$_2$FR,

\begin{equation}
    \mathrm{H_2FR} = \frac{M_{\rm HI}}{\tau_{\rm conv, HI}},
    \label{eq::h2fr}
\end{equation}

\noindent
where $\tau_{\rm conv, HI}$ is the timescale over which all current H$\,\textsc{i}$ gas would be converted into H$_2$ assuming a constant $\mathrm{H_2FR}$. Finally, taking the sum of $M_{H_2}$ from the previous time step and the amount of just formed H$_2$, we compute the SFR as

\begin{equation}
    \mathrm{SFR} = \frac{M_{\rm H_2}}{\tau_{\rm dep, H_2}},
    \label{eq::sfr}
\end{equation}

\noindent
where $\tau_{\rm dep, H_2}$, analogous to the previously described two timescales, is the amount of time over which all H$_2$ would be converted to stars with the current SFR. For the purpose of our simple model we neglected processes that turn some stellar mass back into gas phase material (e.g. stellar winds and supernovae). These three parameters, $\tau_{\rm remov, HI}$ (when used), $\tau_{\rm conv, HI}$, and $\tau_{\rm dep, H_2}$ were kept constant (to values which we detail further down in this section) for 3 Gyr time intervals. We note, that due to the time-dependent $M_{\rm HI}$ and $M_{\rm H_2}$ values, HIRR, H$_2$FR, and SFR are varying with time as well, despite the fixed timescales. In other words, the H$\,\textsc{i}$ and H$_2$ components are not actually depleted over these times, since our model always assumes that at any given time the currently available reservoirs will be converted, and updates each conversion rate accordingly. For example, if there were no H$_2$ replenishment from H$\,\textsc{i}$, with a fixed $\tau_{\rm dep, H_2}$ the SFR would steadily decrease at each epoch. At every time step the H$\,\textsc{i}$ mass was compared to the expected H$\,\textsc{i}$ mass of Eq. \ref{eq::gasfrac_scaling}, assuming the scaling relation itself does not change with time, while the SFR values were compared to the redshift-dependent SFMS prescription of \citet{sargent14} to infer the offsets and draw trajectories in the parameter space shown in Fig. \ref{fig::sf_frac_off}.

In order to set the baseline values of the timescales $\tau_{\rm conv, HI}$ and $\tau_{\rm dep, H_2}$ in our models, we first examined scenario \textit{A)} with a gas replenishment rate that keeps the amount of H$\,\textsc{i}$ gas in the modelled galaxy fixed to the \textit{f$_{\rm HI}$} -- \textit{M$_\star$} relation at all time steps. We chose an initial stellar mass of $\log(M_\star/\mathrm{M}_{\odot}) = 9.1$, since we found, in general, by the end of our simulated 3 Gyr time spans, this starting value, depending on various model parameters, leads to a final $\log(M_\star/\mathrm{M}_{\odot}) \approx$ 9.3 -- 9.6. This puts simulated galaxies within $<$ 50 \% of the median stellar mass value, $\log(\overline{M}_\star/\mathrm{M_\odot}) = 9.48$, of our H$\,\textsc{i}$-selected Coma sample. We tuned $r_{\rm mol, 0}$, $\tau_{\rm dep, H_2}$, and $\tau_{\rm conv, HI}$ to keep our simulated galaxy roughly on the main sequence for the full 3 Gyr duration, while retaining realistic parameter values (e.g. maximum timescales of the order of 1 Gyr, and an initial molecular gas fraction below 1). We found that $r_{\rm mol, 0} = 0.4$, $\tau_{\rm dep, H_2} = 1$ Gyr, and $\tau_{\rm conv, HI} = 1.5$ Gyr result in a model that starts out on the SFMS and also roughly remains a main sequence galaxy during the simulated time period with a change in $\Delta \mathrm{sSFR}$ of < 0.3 dex between the first and last time steps, as shown in Fig. \ref{fig::toymod}b) for three cases: a galaxy starting on-,  0.6 dex above-, and 0.6 dex below the \textit{f$_{\rm HI}$} -- \textit{M$_\star$} relation, respectively.

As a next step, we turned off gas replenishment and kept the initial conditions unchanged to investigate galaxy starvation-like tracks in the $\Delta$sSFR -- $\Delta f_{\rm HI}$ parameter space using scenario \textit{B)}. Figure \ref{fig::toymod}c) shows examples of such evolutionary paths for galaxies starting on the gas fraction scaling relation, 0.6 dex below, and 0.6 dex above, representing initially normal, gas-poor, and gas-rich sources, respectively. These modelled galaxies stayed broadly consistent with measurements from the reference sample throughout the whole 3 Gyr.  Adding H$\,\textsc{i}$ removal (scenario \textit{C}) with $\tau_{\rm remov, HI} = 2$ Gyr, which loosely corresponds to the crossing time ($\tau_{\rm cross} \sim R_{200}/\sigma_{\rm cl} \approx 1.7$ Gyr), our test sources fall further below the SFMS and become more gas poor, as expected, resembling less the reference sample in their end state, as seen in Fig. \ref{fig::toymod}d).

So far evolutionary tracks of both the starvation and the slow H$\,\textsc{i}$ removal scenarios avoided the gas-poor and highly star forming quadrant of the $\Delta$sSFR -- $\Delta f_{\rm HI}$ space, where the majority of H$\,\textsc{i}$-selected Coma galaxies reside. As shown in Fig. \ref{fig::toymod}e), a ten times faster H$\,\textsc{i}$ removal (with $\tau_{\rm remov, HI} = 0.2$ Gyr) can, for a few 100 Myr, put initially gas-rich galaxies in this regime, due to the combined effect of sudden H$\,\textsc{i}$ removal and the comparatively slower wind down of star formation activity, since in our model the H$_2$ gas is unaffected by stripping. Galaxies that do not start being very H$\,\textsc{i}$-rich, though, do not populate the upper-left quadrant of the figure in this case. Another pathway into the upper left quadrant is via a quick initial 300 Myr starburst episode, during which the H$\,\textsc{i}$-to-H$_2$ conversion rate is boosted by a factor of 8, simulating the consequences of the increased external pressure on the central, more strongly bound H$\,\textsc{i}$ gas due to ram pressure in the cluster environment. Adding this phase to the slow H$\,\textsc{i}$ removal case ($\tau_{\rm remov, HI} = 2$ Gyr) leads to a brief raise above the SFMS, as seen in Fig. \ref{fig::toymod}f). This causes not only already gas-rich, but also more normal galaxies to enter the $\Delta$sSFR -- $\Delta f_{\rm HI}$ regime occupied by most H$\,\textsc{i}$-selected Coma sources, making it a somewhat more likely scenario than very sudden H$\,\textsc{i}$ removal. In all cases, the end result is the quenching of the stripped galaxy, but the paths leading to quenching can be diverse.

In summary, we found that slow, secular gas removal could not prompt galaxies initially similar to the reference sample to become gas-poor and highly star forming at the same time -- these processes are more suitable to explain the SF and gas properties of Fornax cluster galaxies (see Fig. \ref{fig::sf_frac_off_wFornax}). However, more violent and quick processes could result in sources more similar to our H$\,\textsc{i}$-selected Coma galaxy sample. Specifically, both the effective, almost instantaneous H$\,\textsc{i}$ removal in gas-rich (and hence highly star forming) galaxies and the short starburst phase represented by a very effective H$\,\textsc{i}$-to-H$_2$ conversion in our model resulted in evolutionary tracks that could reproduce the gas-poor and vigorously star forming Coma galaxies, starting from the reference field population. Since the starburst scenario can also elevate galaxies with normal gas content and star formation rate (and not just gas-rich and highly star forming ones) to the region coinciding with most Coma H$\,\textsc{i}$ sources, it appears to be a more common and likely evolutionary path for our H$\,\textsc{i}$-selected sample. This scenario is further supported by the relatively high incidence of post-starburst galaxies in massive clusters, whose fraction with respect to the total galaxy population in clusters at Coma's halo mass of $10^{15}\,\mathrm{M_\odot}$ is measured to be $\sim$ 5 \%, as opposed to the typical 1 \% of field galaxies \citep{paccagnella19}. Indeed, in the Coma cluster post-starburst galaxies are clustered around the densest regions appearing in higher fractions than in the less dense regions \citep{gavazzi10}.

\subsection{Star formation and gas properties as a function of H$\,\textsc{i}$ and optical size ratios}

Gas removal begins in the outskirts of galaxies where their gravitational pull is not strong enough to hold onto their gas supplies \citep[see e.g.][and references therein]{cortese21,boselli21b}. Therefore, a reduced extent of the H$\,\textsc{i}$ disk relative to the size of the stellar body is indicative of past interactions with for instance, the ICM. In order to measure the ratio of H$\,\textsc{i}$ and optical size, and investigate possible links to SF properties and H$\,\textsc{i}$ deficiency, we first estimated the H$\,\textsc{i}$ size of galaxies with an H$\,\textsc{i}$ morphology we classified as settled. We followed the method described in \citet{wang16}, that is, by placing elliptic annuli centred on each detection on the H$\,\textsc{i}$ intensity maps, we obtained H$\,\textsc{i}$ surface brightness profiles, which in turn were used to measure the radii where the profile reached 3.75 $\cdot 10^{19}\,\mathrm{cm}^{-1}$ (or 0.3 $\rm M_{\odot}\,\mathrm{pc}^{-2}$). We chose a threshold lower than the 1 $\rm M_{\odot}\,\mathrm{pc}^{-2}$ of \citet{wang16}, due to the poor angular resolution of our data, several galaxies in our sample have not reached this value even at their centres. Finally, the H$\,\textsc{i}$ diameters, $D_{\rm HI}$, were calculated after the radii were deconvolved with the beam via

\begin{equation}
    D_{\rm HI} = \sqrt{\left(2r_{\rm HI}^{\prime}\right)^2 - B_{\rm maj}B_{\rm min}},
    \label{d_deconv}
\end{equation}

\noindent
where $r_{\rm HI}^{\prime}$ is the radius measured from the observed surface brightness profile as described above, and $B_{\rm maj}$ and $B_{\rm min}$ are the major and minor axes of the beam, respectively. Fifteen of our 17 settled sources were found to be resolved with this method. Optical sizes were taken from \citet{healy21}, and were measured by an analogous approach by taking the diameter at 25 mag/arcsec$^{2}$ in a pseudo B-band image\footnote{\citet{healy21} created so-called pseudo B-band images by combining $\it g$- and $\it i$-band data as prescribed by their Equation 1, taken from Table 1 in \citet{cook14}.}.

\begin{figure}
    \centering
    \includegraphics[width=0.45\textwidth]{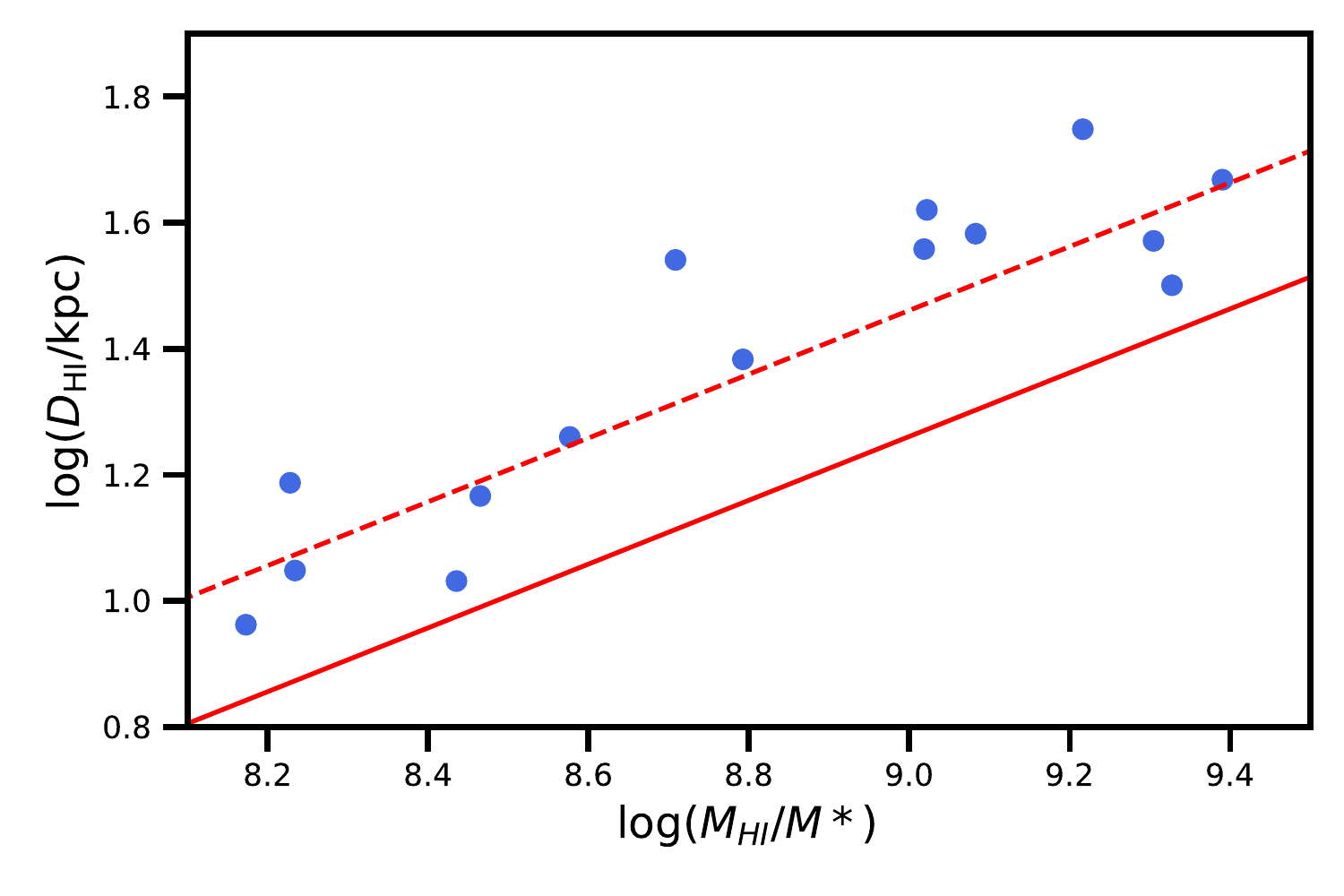}
    \caption{H$\,\textsc{i}$ size of sources with a settled H$\,\textsc{i}$ morphology, measured at 0.3 $M_{\odot} \mathrm{pc}^{-2}$ surface brightness as a function of H$\,\textsc{i}$ mass. The solid red line indicates the scaling relation of \protect\citet{wang16} for sizes measured at 1 $M_{\odot} \mathrm{pc}^{-2}$. Dashed red line is the same scaling relation shifted 0.2 dex higher.}
    \label{fig::HIsize_vs_mass}
\end{figure}

Figure \ref{fig::HIsize_vs_mass} shows the H$\,\textsc{i}$ size as a function of H$\,\textsc{i}$ mass alongside the scaling relation of \citet{wang16}. Our sources lie scattered around a similar trend but $\sim$ 0.2 dex higher, due to the lower surface brightness threshold applied in our measurement. Nevertheless, they still follow a similar correlation with a scatter of 0.1 dex \citep[slightly larger than the 0.06 dex scatter of][]{wang16}, suggesting our lower than usual threshold is a robust tracer of the H$\,\textsc{i}$ disk size.

\begin{figure}
   \includegraphics[width=0.45\textwidth]{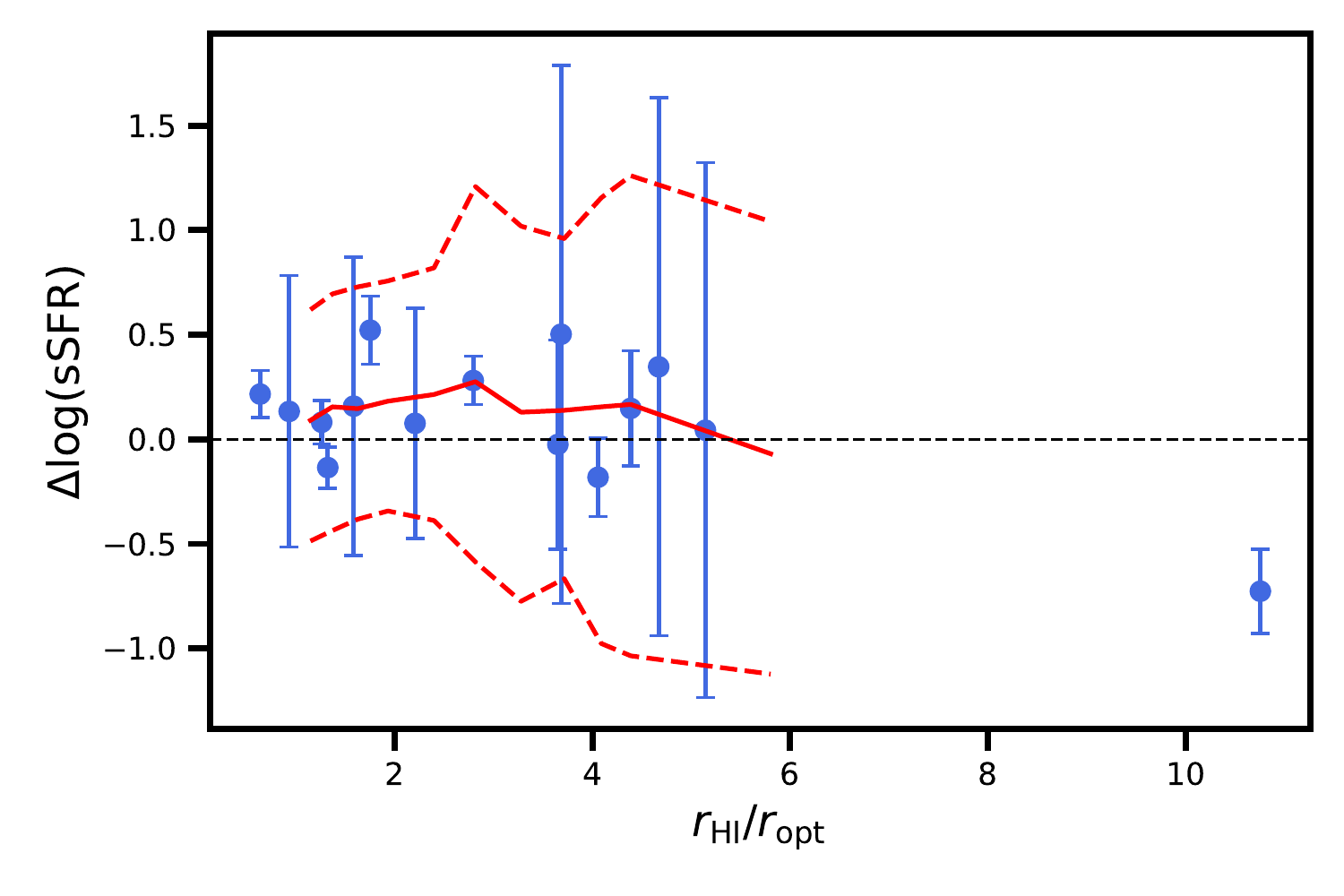}
   \includegraphics[width=0.45\textwidth]{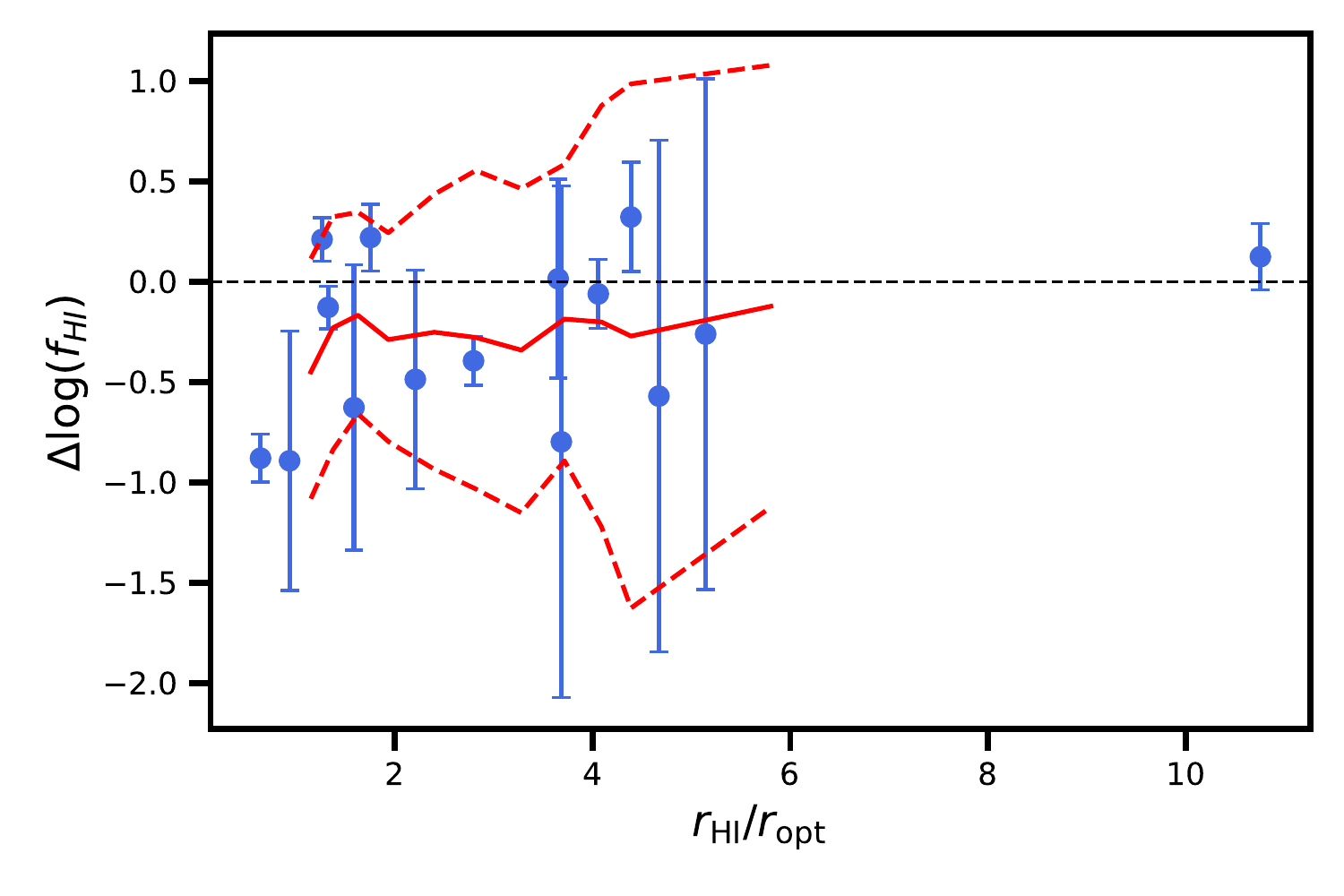}
   \caption{Offset our fitted H$\,\textsc{i}$ fraction -- stellar mass relation (top) and from the star forming main sequence (bottom) as a function of H$\,\textsc{i}$-to-optical size ratio for our H$\,\textsc{i}$-selected galaxies in Coma. Solid red lines are the running median curves, dashed red lines are the estimated $\pm$ 3\,$\sigma$ confidence intervals via bootstrapping.}
   \label{fig::offsets_vs_size}%
\end{figure}

In Fig. \ref{fig::offsets_vs_size} we present the offset from the SFMS (top panel) and the offset from the gas fraction -- stellar mass scaling relation as a function of H$\,\textsc{i}$-to-optical size ratio for the 15 resolved Coma galaxies with a settled H$\,\textsc{i}$ morphology. Both the gas fraction offset and the SFMS offset appears largely independent of the size ratio, although the running medians (indicated by the solid red line in Fig. \ref{fig::offsets_vs_size}) give a very weak impression of a possibly increasing and decreasing trend, respectively. If these indeed proved to be the underlying correlations, they could be indicative of a scenario whereby ram pressure stripping removes the atomic gas in the outskirts of the galaxies, while at the same time the pressure exerted on the inner disk component boosts star formation efficiency. We emphasise, however, that 3\,$\sigma$ uncertainties we calculated by bootstrapping our measurements 1000 times (marked as dashed red lines in both panels of Fig. \ref{fig::offsets_vs_size}) show the running medians to be consistent with no evolution, thus the accuracy of our measurements and our relatively small sample size do not permit such an assessment. Nevertheless, it might prompt follow-up investigations in future studies.

\section{Summary}
\label{sect::summary}

With a new sensitive and blind H$\,\textsc{i}$ survey, WCS, covering a 1 Mpc radius around the cluster core and out to 1.5 Mpc towards the infalling NGC 4839 group, we have mapped the H$\,\textsc{i}$ gas of the Coma cluster down to M$_{\rm HI} \approx 10^8 M_{\odot}$ (N$_{\rm HI} \approx 10^{19} \mathrm{cm}^2$) with an angular resolution of $\sim$ 33 arcsec ($\sim$ 15 kpc at the distance of Coma). After visually inspecting blind and optical prior-based H$\,\textsc{i}$ detections in our cube, as well as considering optical counterparts, we obtained 40 secure H$\,\textsc{i}$ detections of which 24 are new. A little over half of them display disturbed H$\,\textsc{i}$ morphologies, and nearly a third had been observed to have tails at H$_{\alpha}$ and/or UV wavelengths and/or in the radio continuum in the literature. They are distributed around the cusp of the hot, X-ray emitting central region of the cluster or further out, reaffirming the results of previous H$\,\textsc{i}$ surveys.

With the use of UV and mid-IR data, we estimated star formation rate and stellar mass in a self-consistent way both for our H$\,\textsc{i}$-detected Coma galaxies and a reference sample of field galaxies. We found that, statistically, H$\,\textsc{i}$-selected, gas-rich Coma galaxies are simultaneously more star forming (by $\sim$ 0.2 dex) and have less H$\,\textsc{i}$ gas (by $\sim$ 0.5 dex) than field galaxies of the same stellar mass. In an attempt to pinpoint the physical processes underpinning this difference, we tested whether the star formation and gas content properties depend on the ratio of H$\,\textsc{i}$-to-optical size considering resolved H$\,\textsc{i}$ detections of a settled H$\,\textsc{i}$ morphology. We found no statistically significant correlations. However, with the use of a simple toy model we showed that Coma galaxies with enhanced star formation and high H$\,\textsc{i}$-deficiency could have been either \textit{a)} highly star forming and gas-rich galaxies experiencing very short timescale ($<$300 Myr) H$\,\textsc{i}$-removal due to ram pressure stripping that left their molecular gas content mostly intact, or \textit{b)} galaxies of more regular gas and star formation properties experiencing H$\,\textsc{i}$ removal and a short ($<$ 300 Myr) temporary boost to their H$\,\textsc{i}$-to-H$_2$ conversion efficiency and hence their star formation activity.

Overall, the most gas-rich Coma galaxies are statistically still strongly H$\,\textsc{i}$ deficient, commonly display signs of ongoing H$\,\textsc{i}$ removal and are likely on their first infall. Since there is a wealth of evidence for gaseous and star forming trails in the late-type galaxies of Coma indicating ongoing ram pressure stripping  \citep{yagi10,smith10,gavazzi18,cramer19,chen20,roberts19,roberts21}, it is plausible that in the cluster outskirts newly accreted galaxies are first cut off from their H$\,\textsc{i}$ supply by the hot ICM, then upon entering the denser inner regions of the cluster (within $\sim$ 0.5 Mpc) ram pressure increases sharply and strips galaxies of their H$\,\textsc{i}$ rapidly \citep{roberts19}, corresponding to our toy model scenario \textit{a)}. In some galaxies this interaction also might trigger a short, few 100 Myr starburst episode, resulting in the detected enhanced SF activity, according to toy model scenario \textit{b}). This picture is consistent with the presence of a post-starburst population that spatially coincides with our innermost H$\,\textsc{i}$ detections \citep{bothun86,poggianti04,gavazzi10}. We emphasise, however, that even if ram pressure temporarily boosts star formation in some particular cases, eventually it quenches all infalling galaxies, and thus our findings and conclusions are not at odds with the clear consensus that star formation activity is eventually suppressed across all stellar masses in the cluster. Indeed, contrasting our results, \citet{loni21} found mostly H$\,\textsc{i}$-poor galaxies with low levels of star formation among the H$\,\textsc{i}$-selected sources of a much less massive cluster, Fornax, suggesting significantly different H$\,\textsc{i}$-removal and star formation quenching scenarios occurring in smaller clusters.

For a deeper understanding of the physical processes at play, some more information is still missing. For example, acquiring H$_2$ data towards both the H$\,\textsc{i}$ sources as well as the H$\,\textsc{i}$-undetected but star forming Coma population would provide important clues as to whether ram pressure is enhancing the conversion of atomic to molecular gas, or perhaps the H$_2$-to-star conversion, while the kinematics and spatial distribution of H$_2$ would inform us whether ram pressure widely affects the molecular disk of infalling galaxies. Furthermore, high sensitivity integral field unit maps could result in detailed star formation histories for the main stellar body as well as the tail components, unveiling the potential positive star formation feedback of ram pressure.

\begin{acknowledgements}
We thank the anonymous reviewer for the insightful and constructive report on our paper.
This project has received funding from the European Research Council (ERC) under the European Union’s Horizon 2020 research and innovation programme (grant agreement no. 679627; project name FORNAX). JMvdH acknowledges funding from the European ResearchCouncil under the European Union’s Seventh Framework Programme (FP/2007-2013)/ERC Grant Agreement No. 291531 (‘HIStoryNU’). LC acknowledges support from the Australian Research Councils Discovery Project and Future Fellowship funding schemes (DP210100337,FT180100066). Parts of this research were conducted by the Australian Research Council Centre of Excellence for All Sky Astrophysics in 3 Dimensions (ASTRO 3D), through project number CE170100013. JH acknowledges research funding from the South African Radio Astronomy Observatory. MV acknowledges support by the Netherlands Foundation for Scientific Research (NWO) through VICI grant 016.130.338. 
\end{acknowledgements}

\bibliographystyle{aa} 
\bibliography{ref}

\begin{appendix}

\section{Reference sample}

In Table \ref{tab::ref_samp} we show a portion of the catalogue containing the properties of our field galaxy reference sample. For details on sample selection and the methods used for estimating galaxy properties, see Sect. \ref{sect::refsamp}. Source IDs were adopted from \citet{boselli10} and \citet{kreckel11}, respectively, with a prefix indicating which catalogue they were taken from.

\begin{table*}[h!]
    \caption{Part of the field galaxy reference sample described in Sect. \ref{sect::refsamp}. The full table containing all 192 galaxies is available as an online supplement at the CDS via anonymous ftp to cdsarc.u-strasbg.fr (130.79.128.5), via \protect\url{http://cdsweb.u-strasbg.fr/cgi-bin/qcat?J/A+A/} or at \protect\url{https://github.com/molnard89/WCS-Catalog-Release}.}
    \label{tab::ref_samp}
    \centering
    \begin{tabular}{ccccccccc}
    ID & RA & Dec & $\log(M_{\star})$ & $\Delta\log(M_{\star})$ & $\log(\rm SFR)$ & $\Delta\log(\rm SFR)$ & $\log(M_{HI})$ & $\Delta\log(M_{HI})$\\
    - & [deg] & [deg] & [$\log(M_{\odot})$] & [$\log(M_{\odot})$] & [$\log(M_{\odot} \mathrm{yr^{-1}})$] & [$\log(M_{\odot} \mathrm{yr^{-1}})$] & [$\log(M_{\odot})$] & [$\log(M_{\odot})$] \\
HRS238 & 190.636 & -1.351 & 9.10 & 0.11 & -2.02 & 0.11 & 9.24 & 0.07 \\ 
HRS83 & 180.917 & 2.642 & 8.54 & 0.10 & -1.44 & 0.02 & 8.09 & 0.07 \\ 
HRS32 & 163.703 & 17.621 & 10.06 & 0.10 & -1.68 & 0.04 & 7.92 & 0.07 \\ 
HRS305 & 208.910 & 4.985 & 9.49 & 0.10 & -1.26 & 0.03 & 7.60 & 0.06 \\ 
HRS76 & 179.207 & 53.160 & 9.26 & 0.10 & -1.15 & 0.04 & 8.63 & 0.07 \\
VGS51 & 228.048 & 24.562 & 8.71 & 0.31 & -0.52 & 0.03 & 9.28 & 0.06 \\ 
VGS14 & 158.777 & 55.147 & 8.38 & 0.31 & -1.27 & 0.02 & 8.81 & 0.07 \\ 
VGS58 & 236.217 & 36.479 & 9.15 & 0.14 & -0.96 & 0.01 & 8.85 & 0.04 \\ 
VGS13 & 157.970 & 31.843 & 9.51 & 0.17 & -1.42 & 0.17 & 9.09 & 0.09 \\ 
VGS46 & 220.910 & 32.334 & 8.91 & 0.23 & -1.11 & 0.02 & 8.74 & 0.13 \\
    \end{tabular}
\end{table*}


\section{Stamps of H$\,\textsc{i}$ detections with their optical counterparts}
\label{app::stamps}

In order to facilitate a more in-depth examination of our detections, we present all our 40 measurements individually in combination with their optical counterparts in Figs \ref{fig::stamp1} - \ref{fig::stamp6} organised according to their H$\,\textsc{i}$ morphologies (see Sect \ref{sect::hi_morph} for a description of our classification scheme). The table containing their estimated physical properties is available as an online supplement\footnote{At the CDS via anonymous ftp to cdsarc.u-strasbg.fr (130.79.128.5), via \protect\url{http://cdsweb.u-strasbg.fr/cgi-bin/qcat?J/A+A/} or at \url{https://github.com/molnard89/WCS-Catalog-Release}.}.

\begin{figure*}[h!] 
   \centering
   \includegraphics[width=0.45\textwidth]{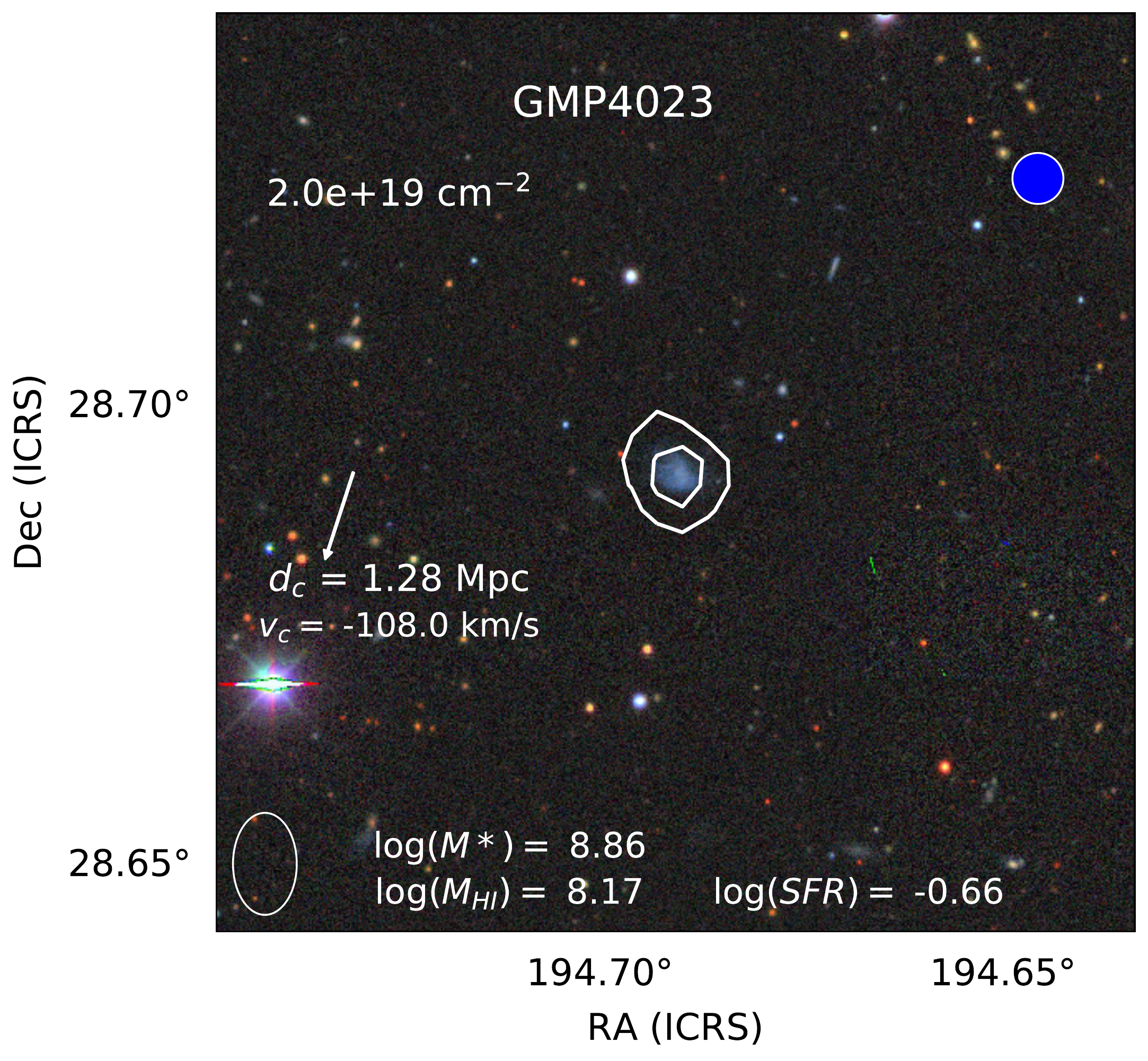} 
   \includegraphics[width=0.45\textwidth]{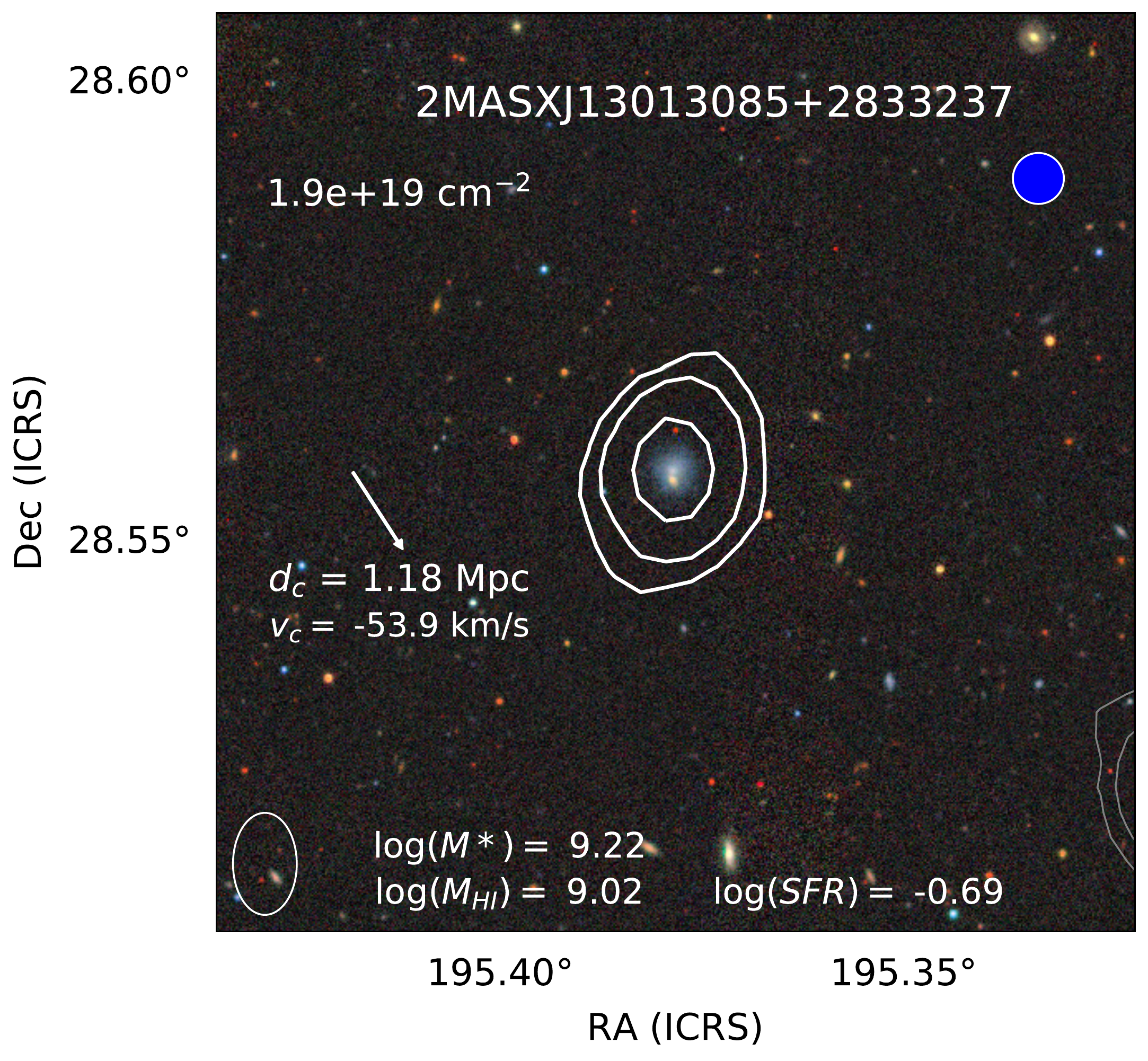} 
   \includegraphics[width=0.45\textwidth]{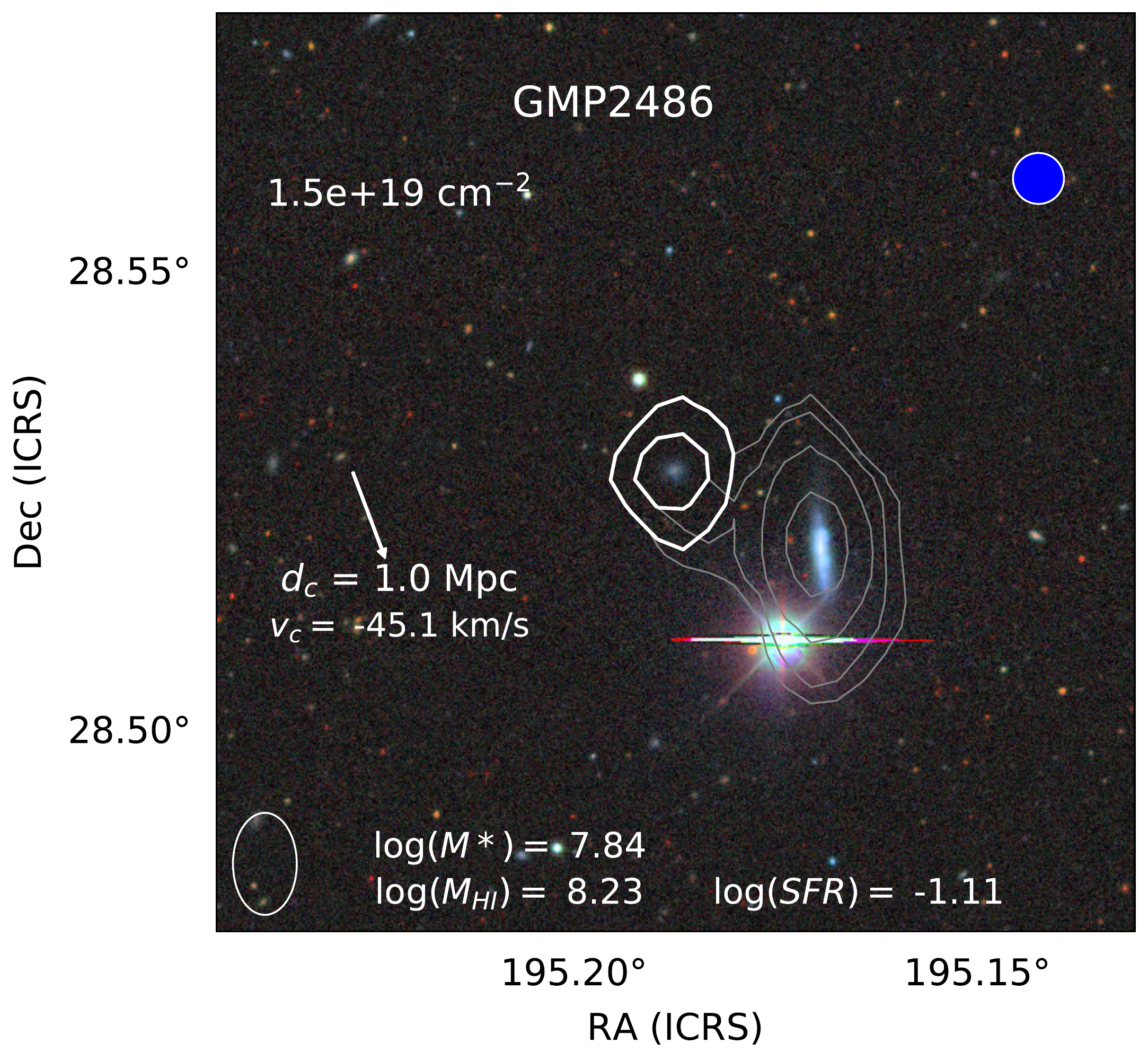} 
   \includegraphics[width=0.45\textwidth]{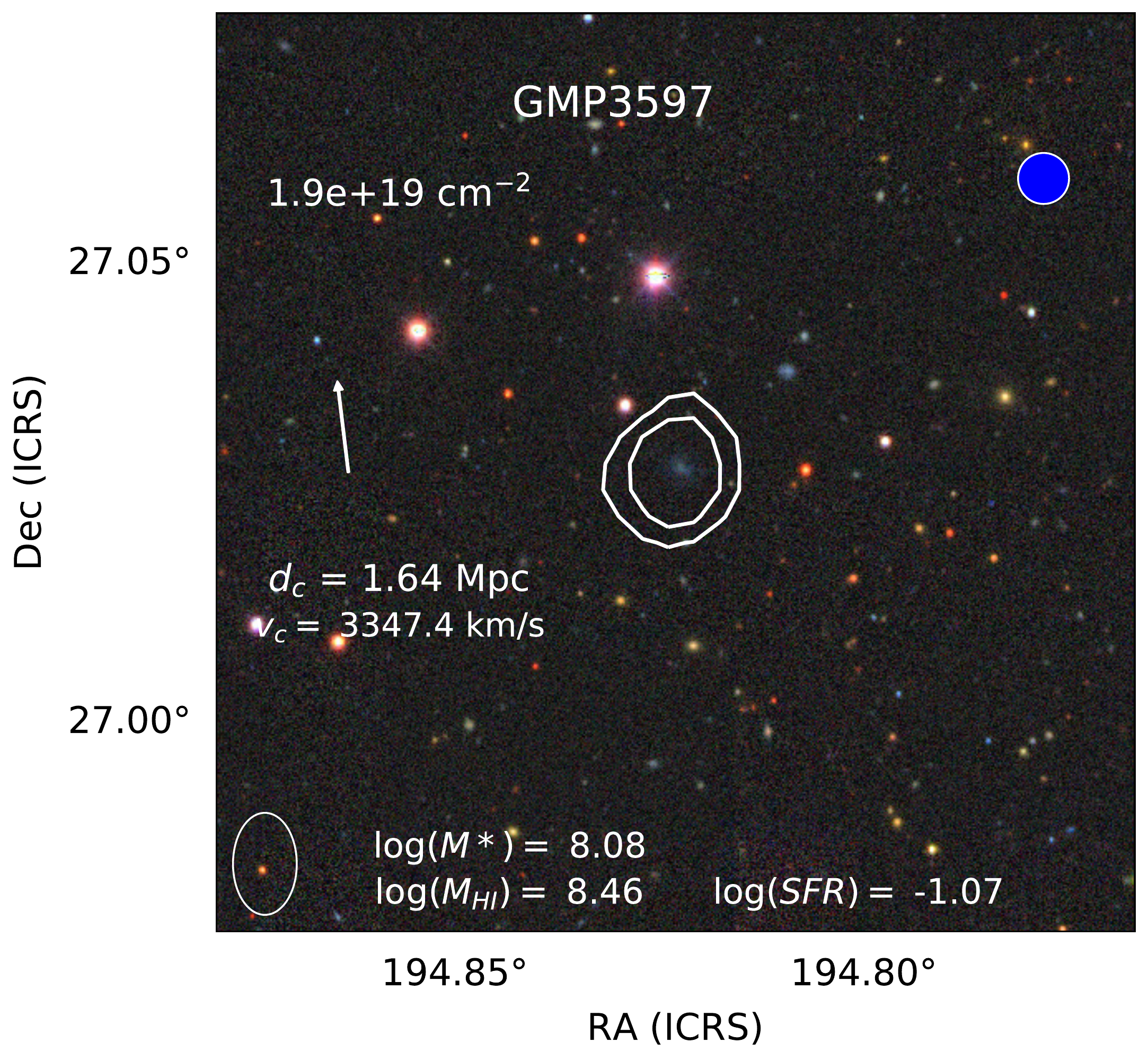} 
   \includegraphics[width=0.45\textwidth]{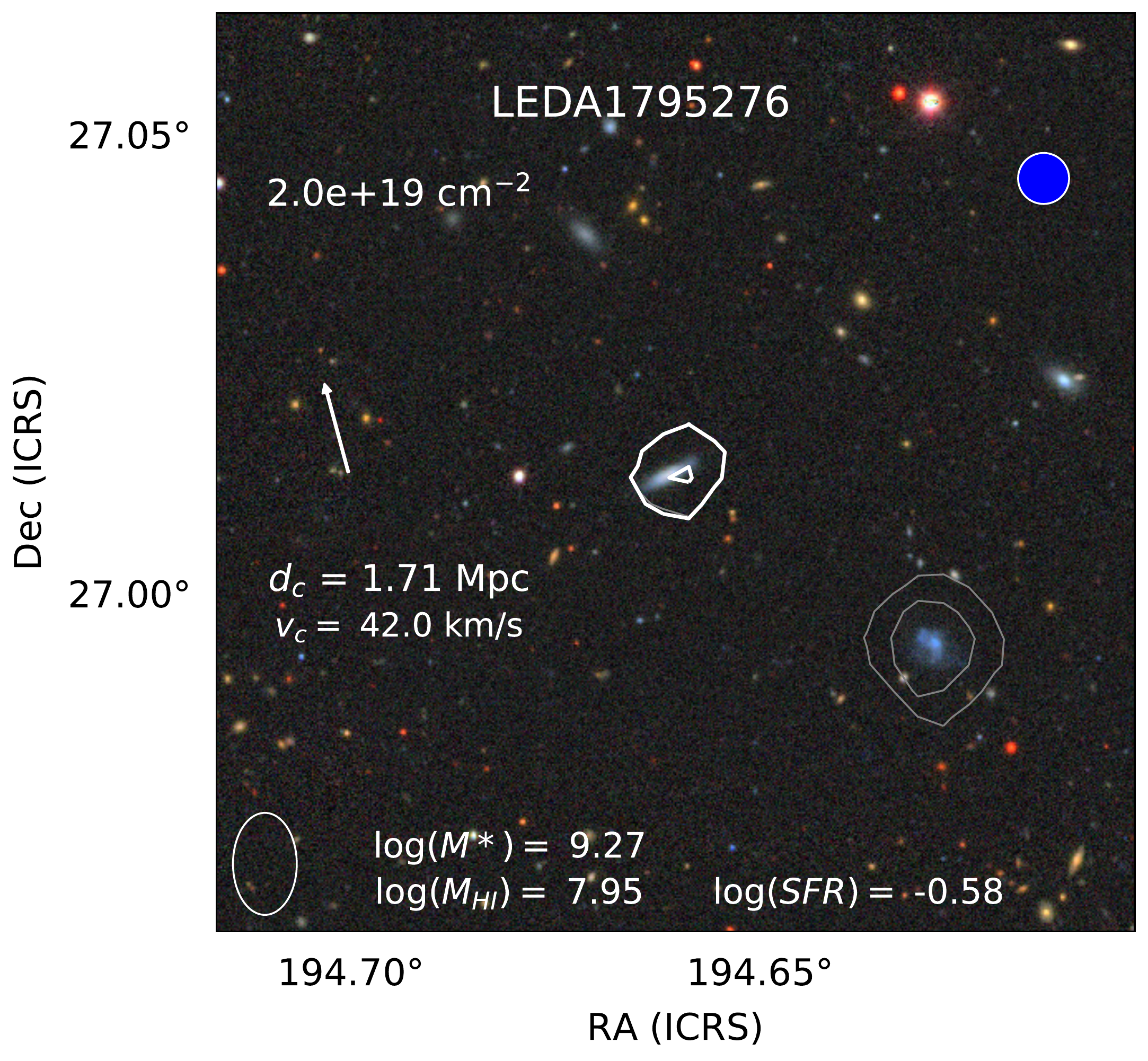} 
   \includegraphics[width=0.45\textwidth]{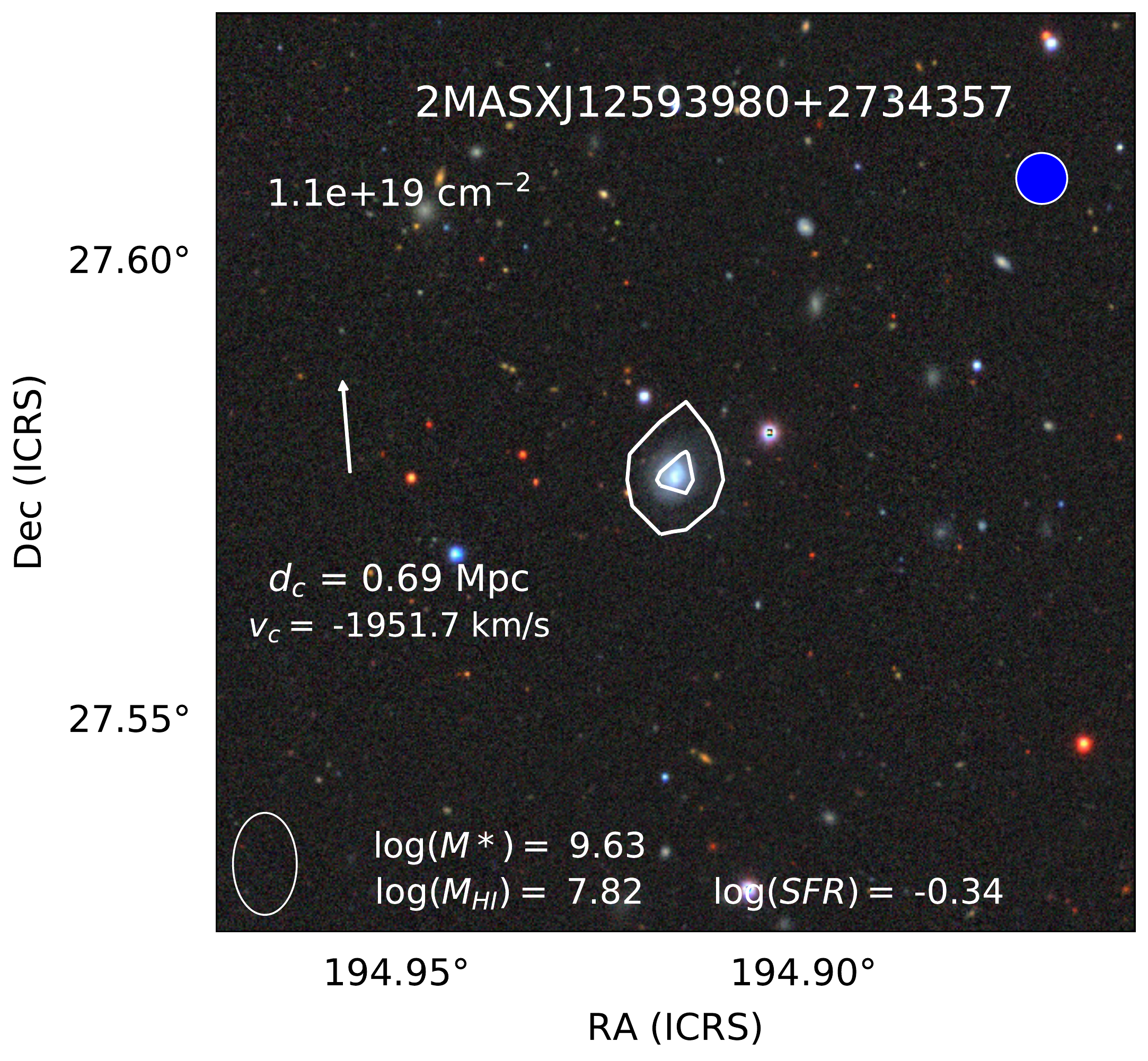} 
   \caption{Integrated intensity distribution of H$\,\textsc{i}$ overlaid on DECalS $g,r,z$  image. White column density contours outline the detection showcased by the panel. Grey contours are background or foreground H$\,\textsc{i}$ sources. Lowest contour is shown in the top left. All other contours are $3^n$ multiples of it. The SFR, stellar, and H$\,\textsc{i}$ masses are in the bottom of each image. The arrow points towards the centre of the Coma cluster, with its projected distance in Mpc ($d_c$) and the H$\,\textsc{i}$ source's velocity offset relative to the systemic velocity of the cluster ($v_c$). The WSRT synthesised beam is in the lower left corner. Fully coloured circles indicate the H$\,\textsc{i}$ morphological classes: settled (blue), one-sided asymmetry (yellow), or unsettled (red).}
   \label{fig::stamp1}%
              
\end{figure*}

\begin{figure*}[h!] 
   \centering
   \includegraphics[width=0.45\textwidth]{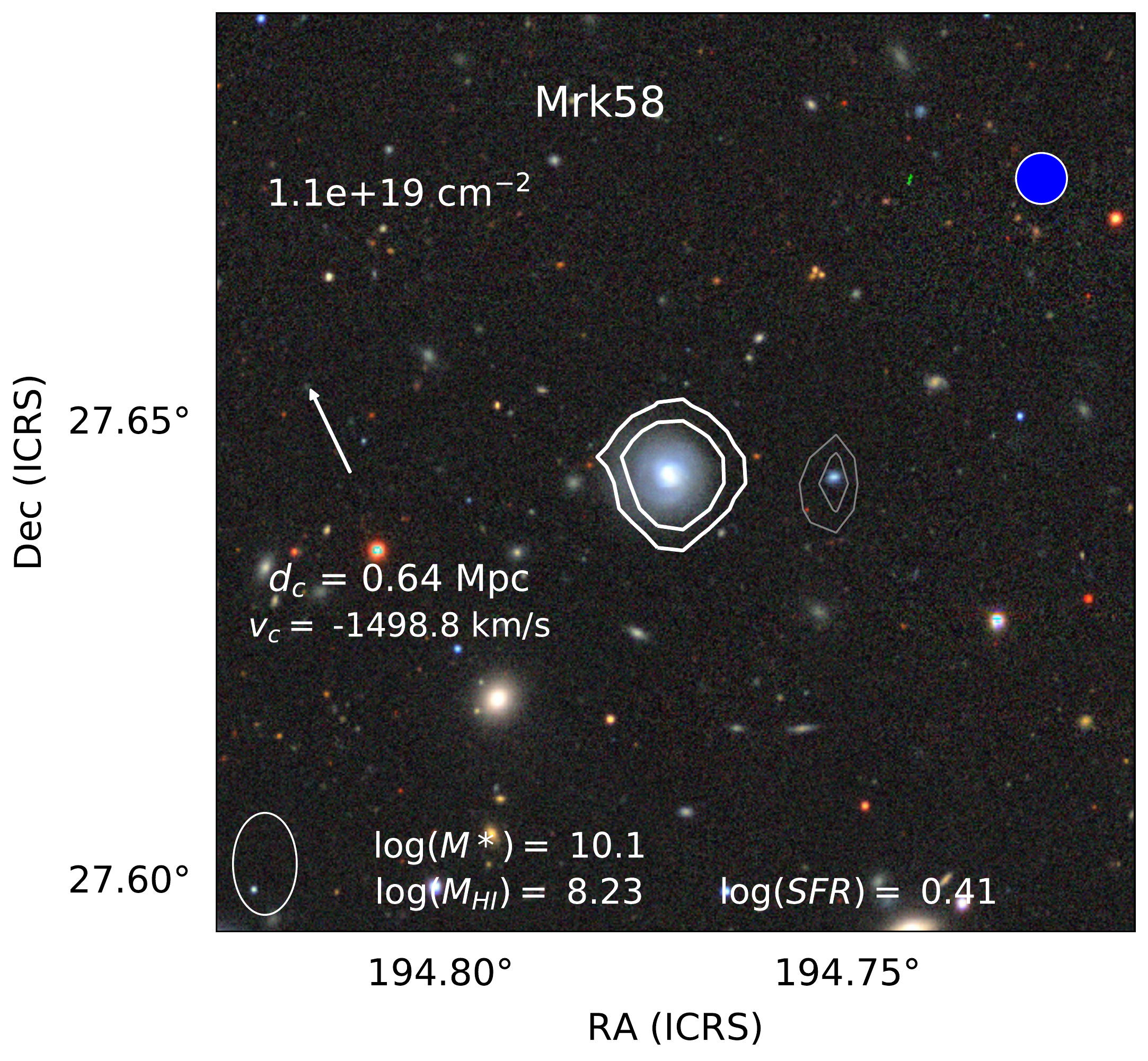} 
   \includegraphics[width=0.45\textwidth]{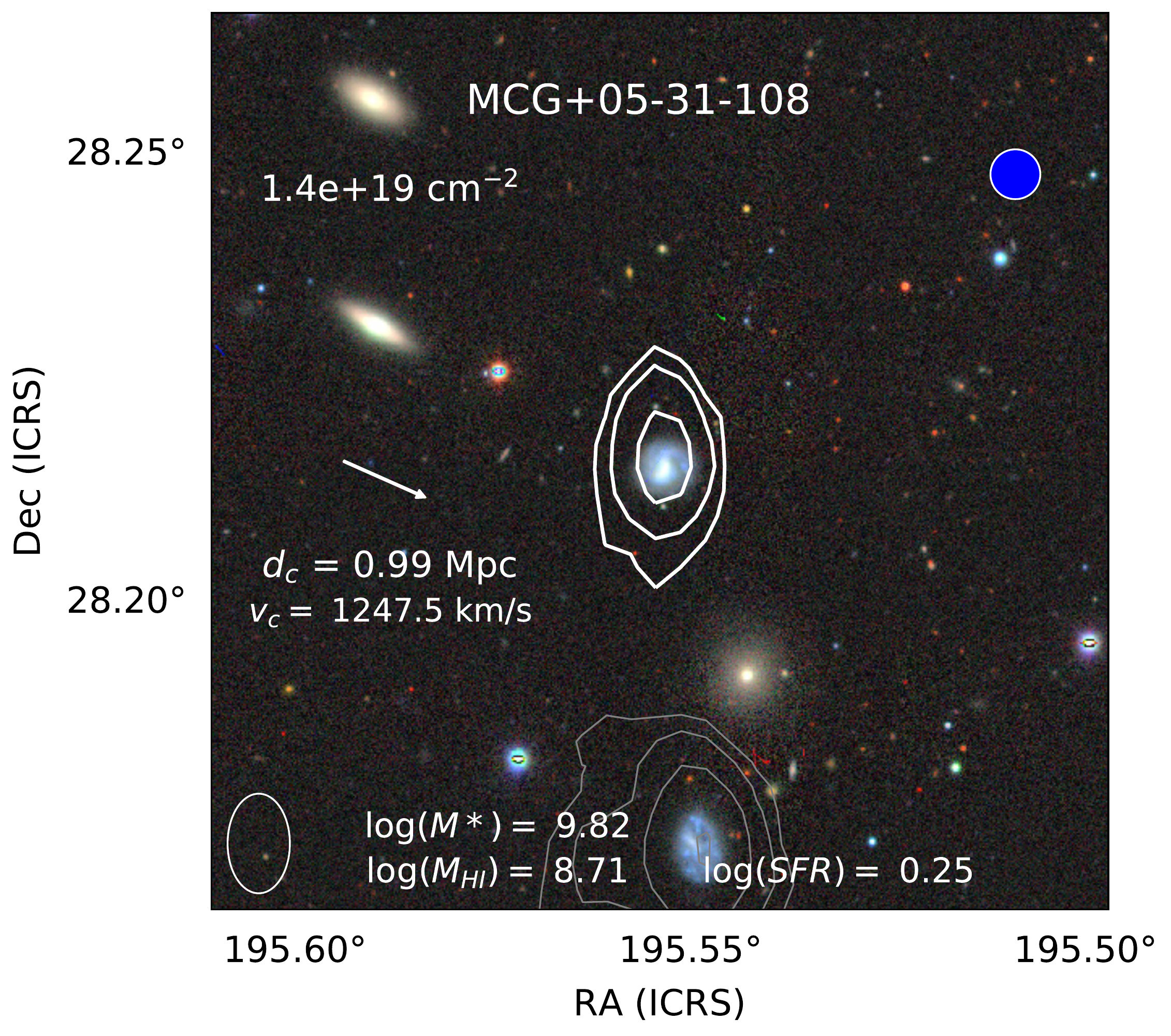} 
   \includegraphics[width=0.45\textwidth]{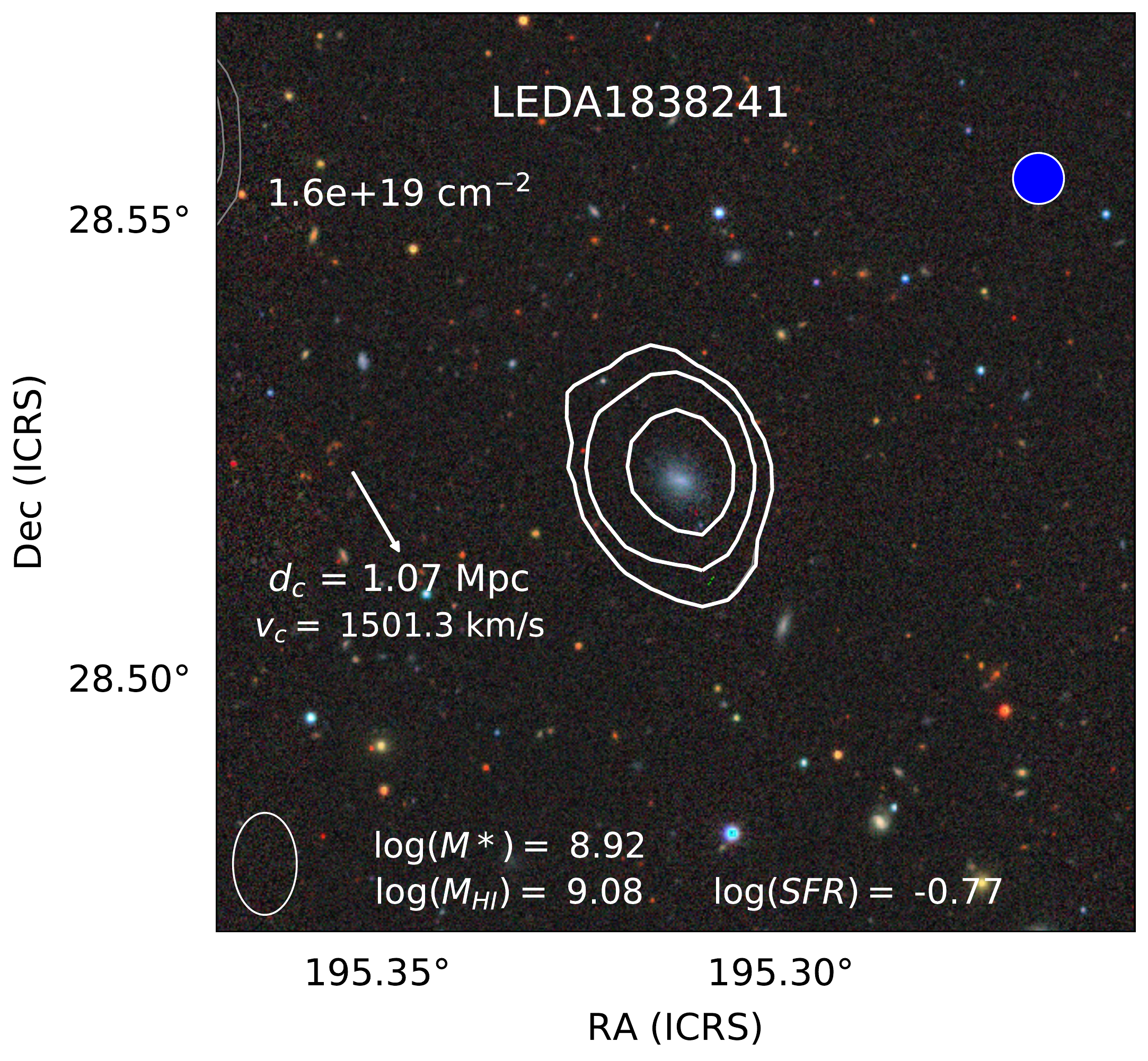} 
   \includegraphics[width=0.45\textwidth]{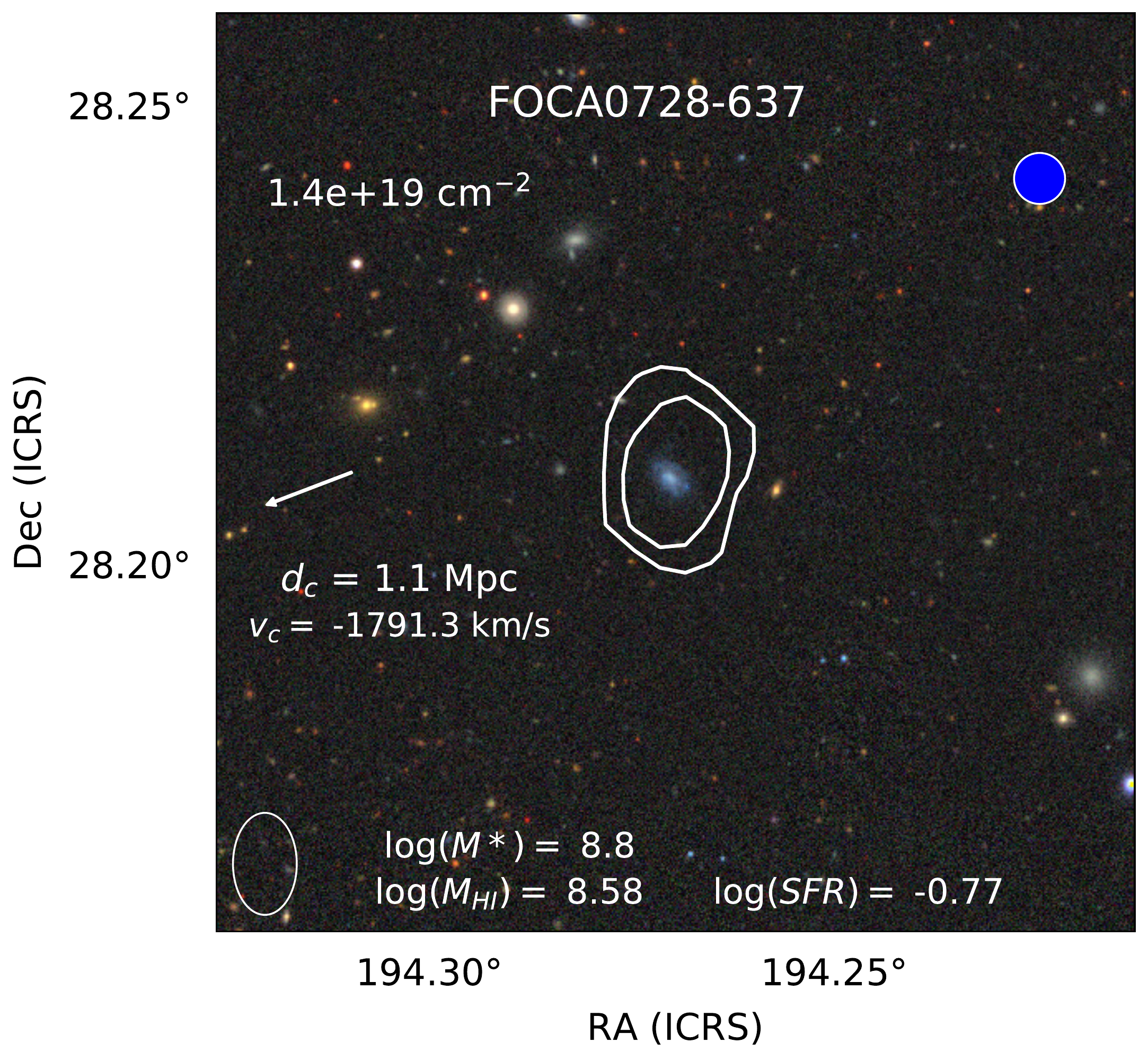} 
   \includegraphics[width=0.45\textwidth]{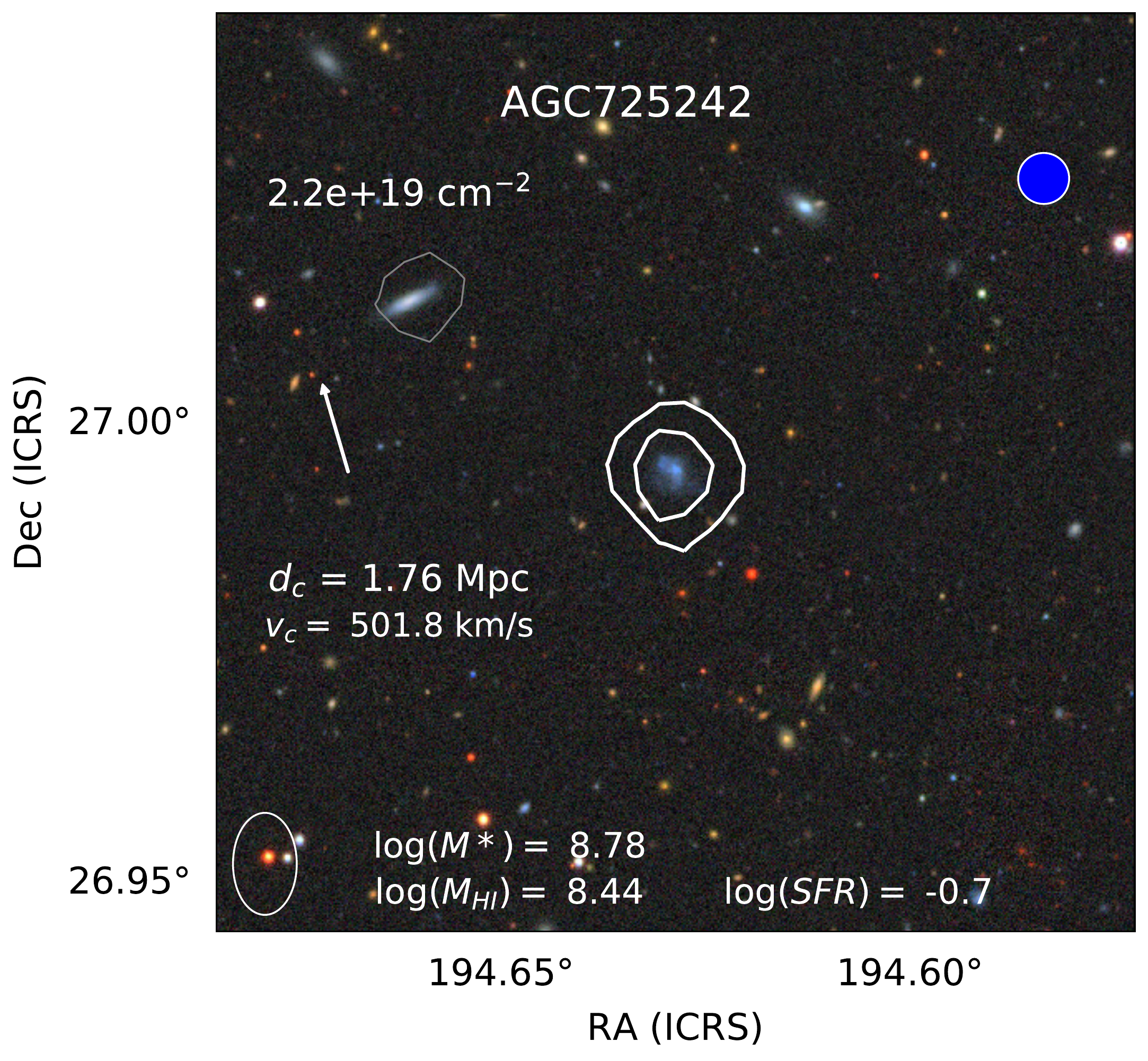} 
   \includegraphics[width=0.45\textwidth]{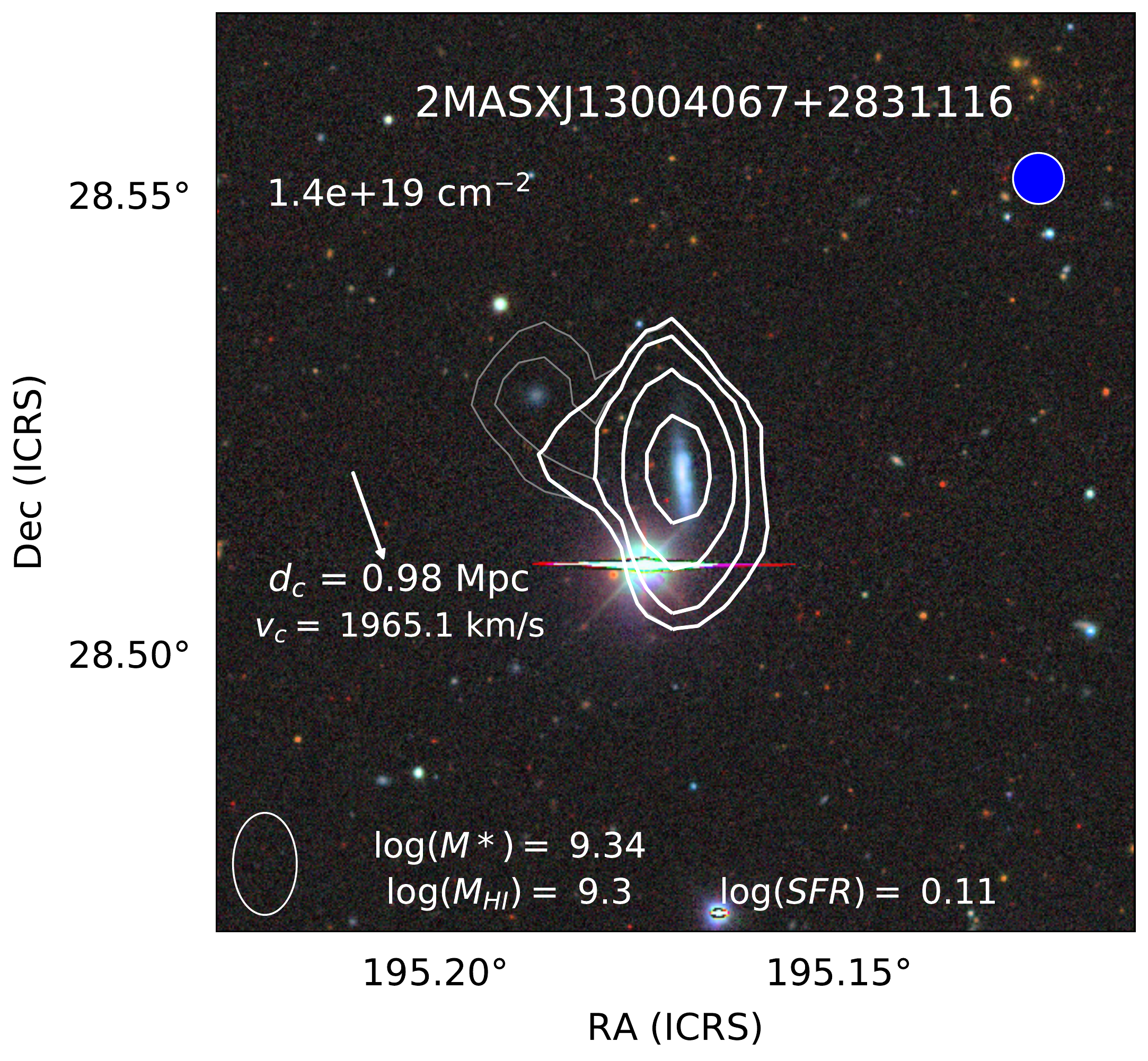} 
   \addtocounter{figure}{-1}
   \caption{continued.}
              \label{fig::stamp2}%
\end{figure*}

\begin{figure*}[h!] 
   \centering
   \includegraphics[width=0.45\textwidth]{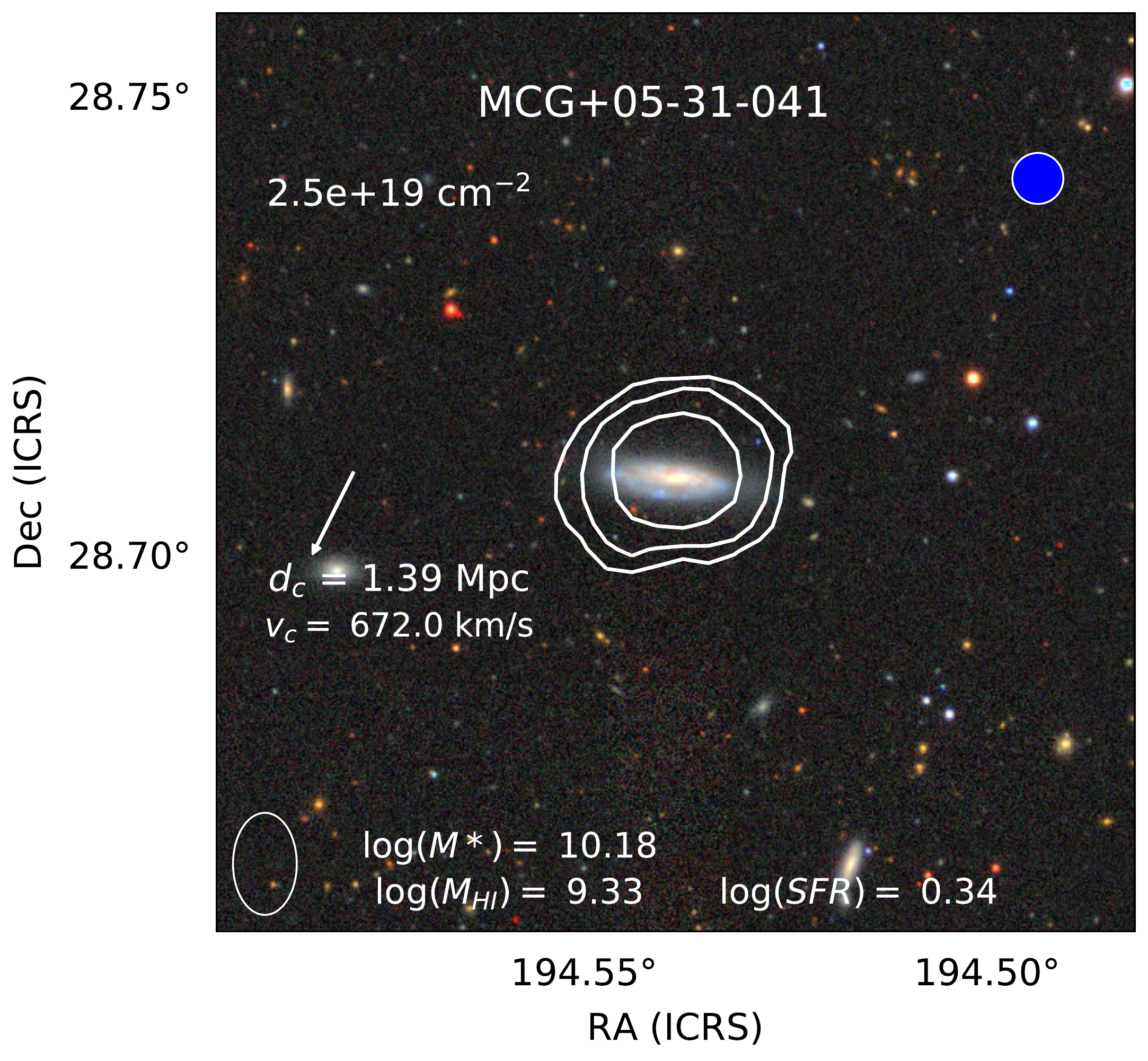} 
   \includegraphics[width=0.45\textwidth]{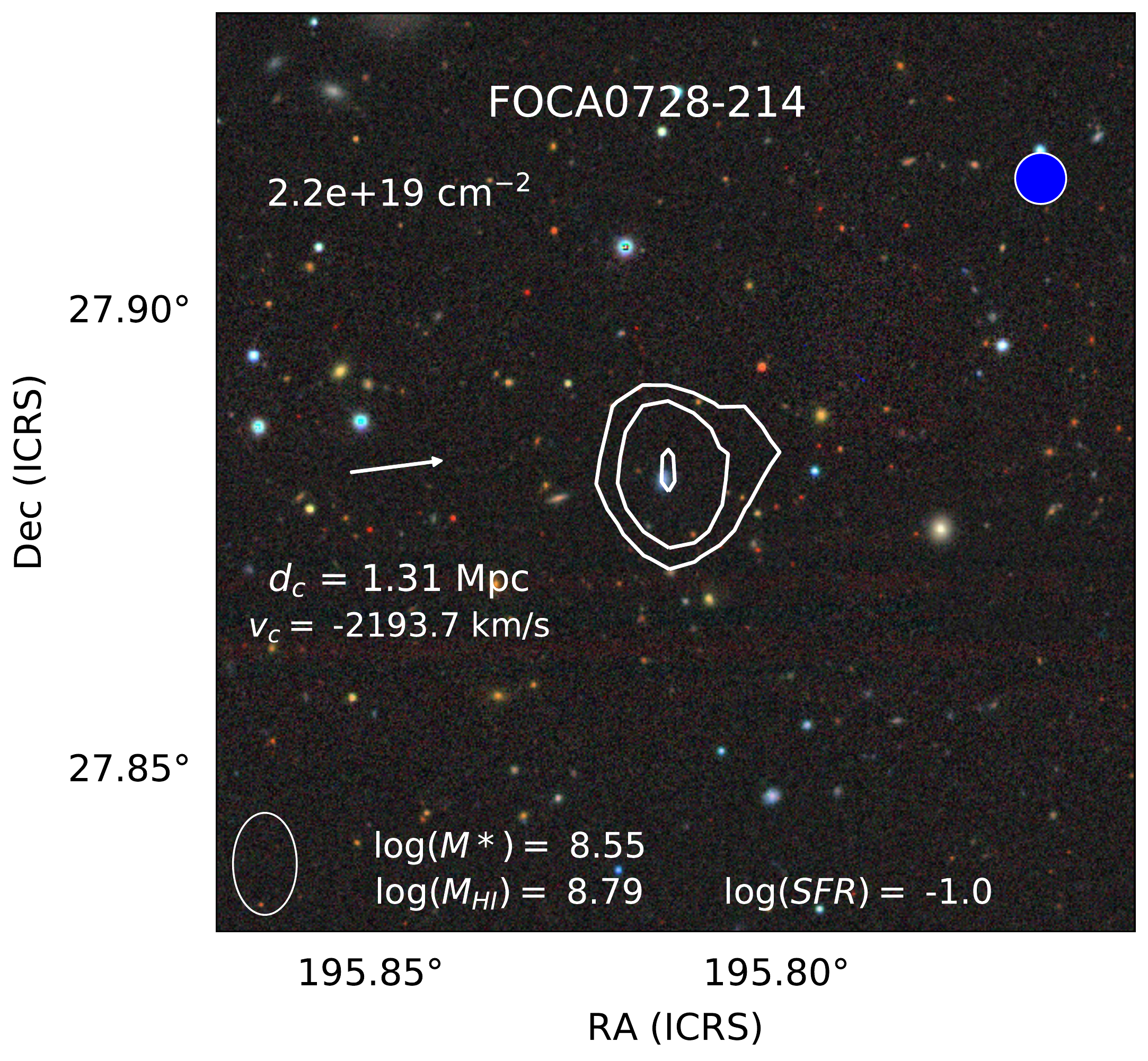} 
  \includegraphics[width=0.45\textwidth]{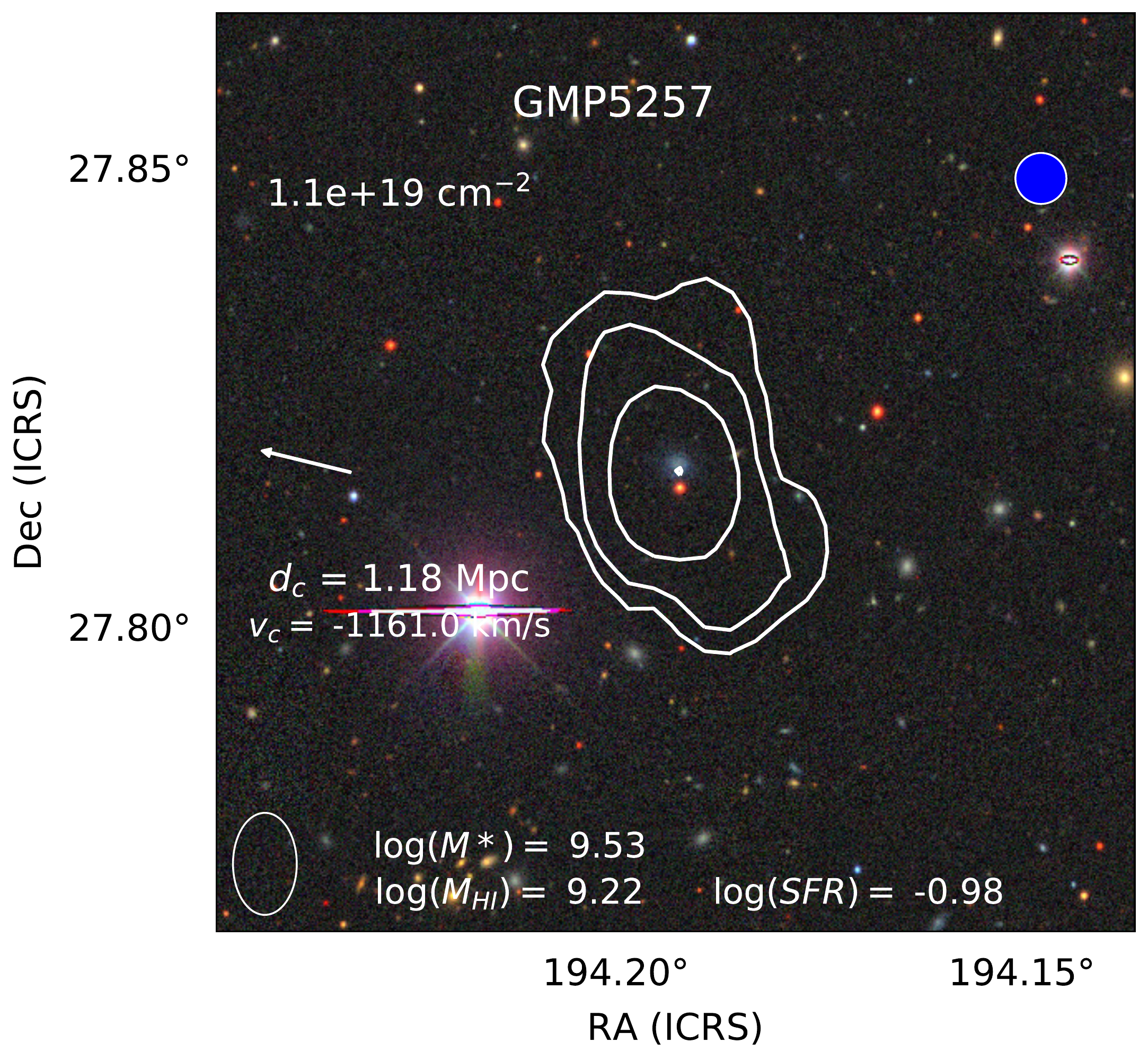} 
   \includegraphics[width=0.45\textwidth]{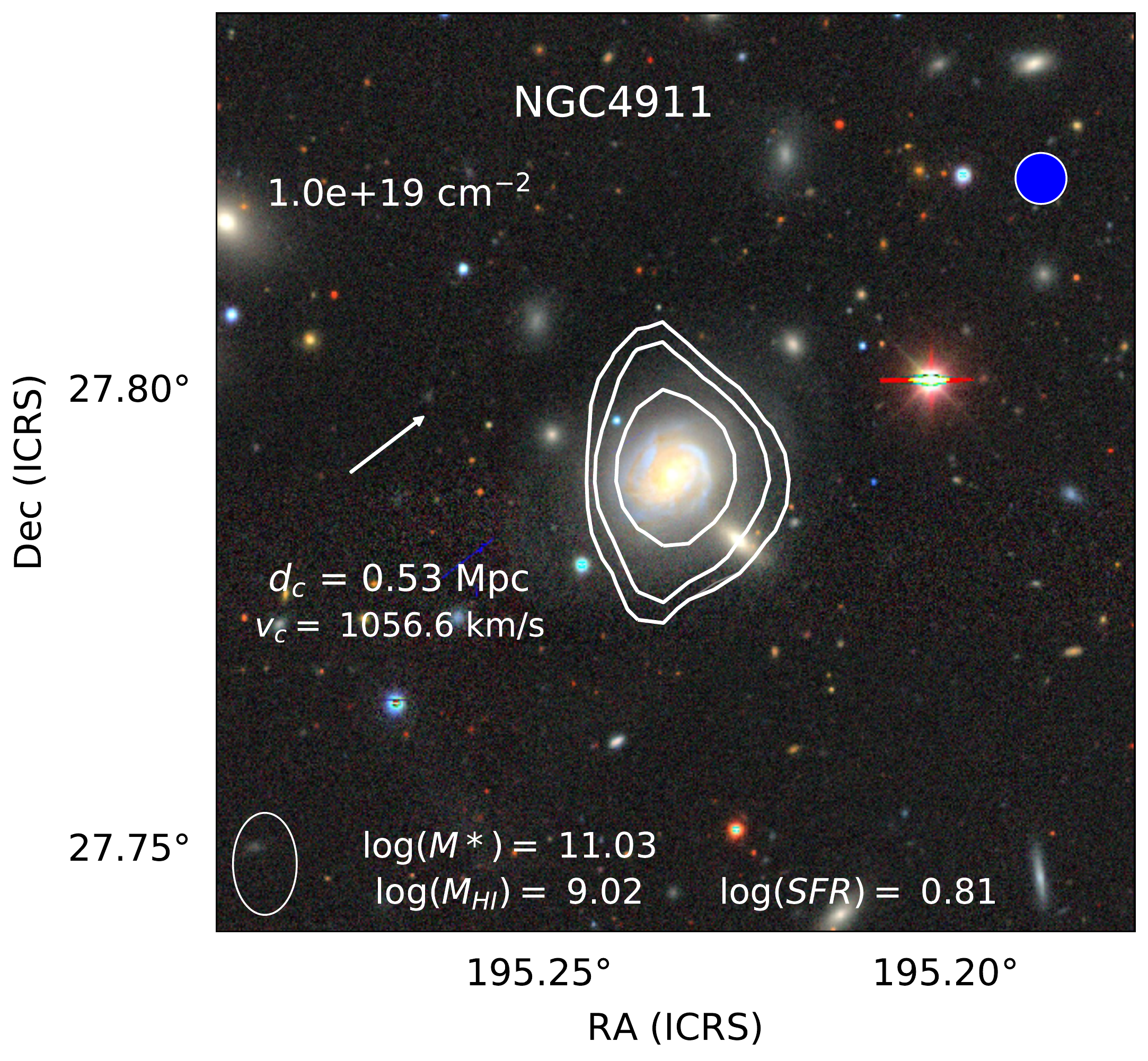} 
  \includegraphics[width=0.45\textwidth]{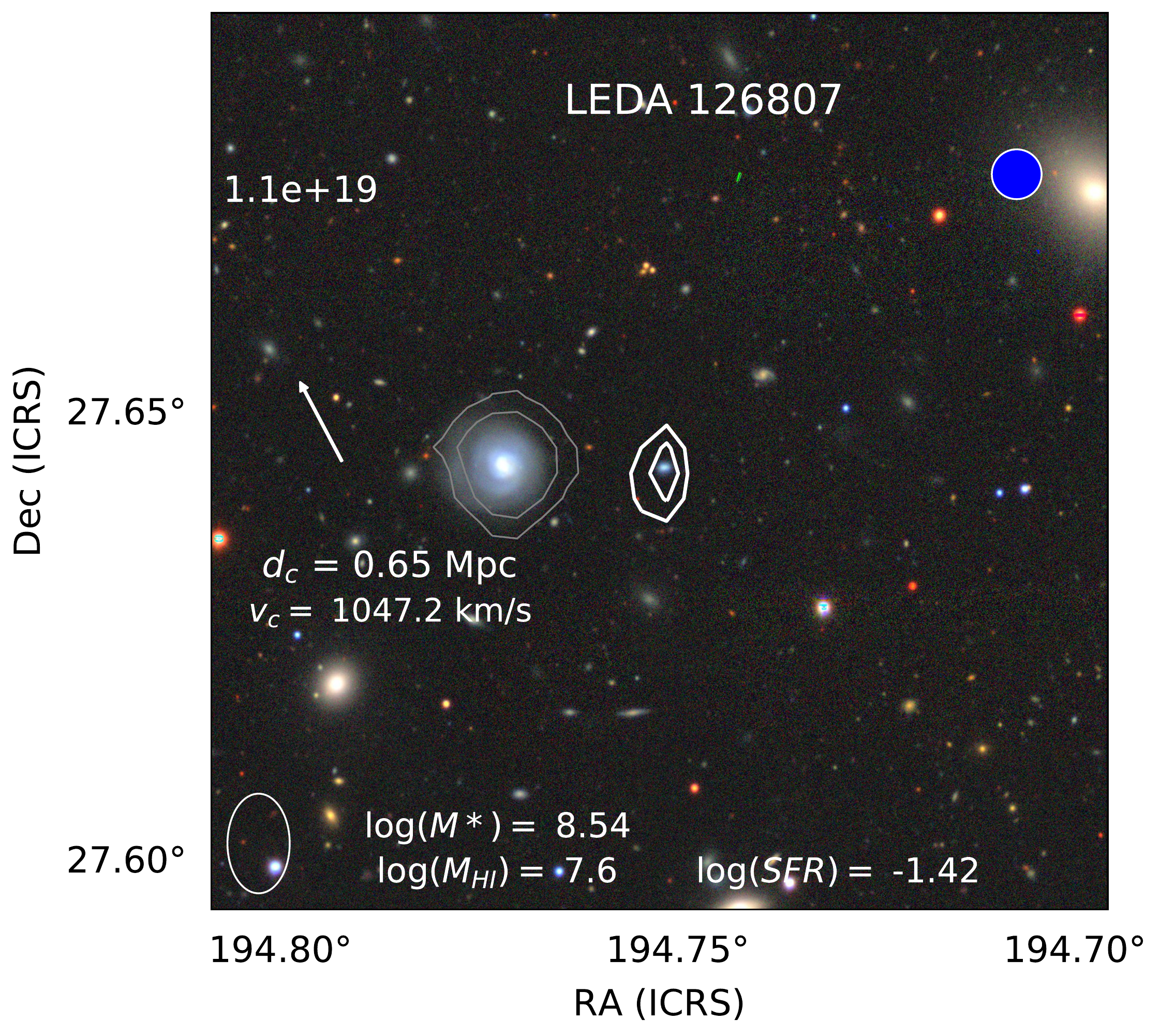}f 
  \addtocounter{figure}{-1}
   \caption{continued.}
              \label{fig::stamp3}%
\end{figure*}

\begin{figure*}[h!] 
   \centering
   \includegraphics[width=0.45\textwidth]{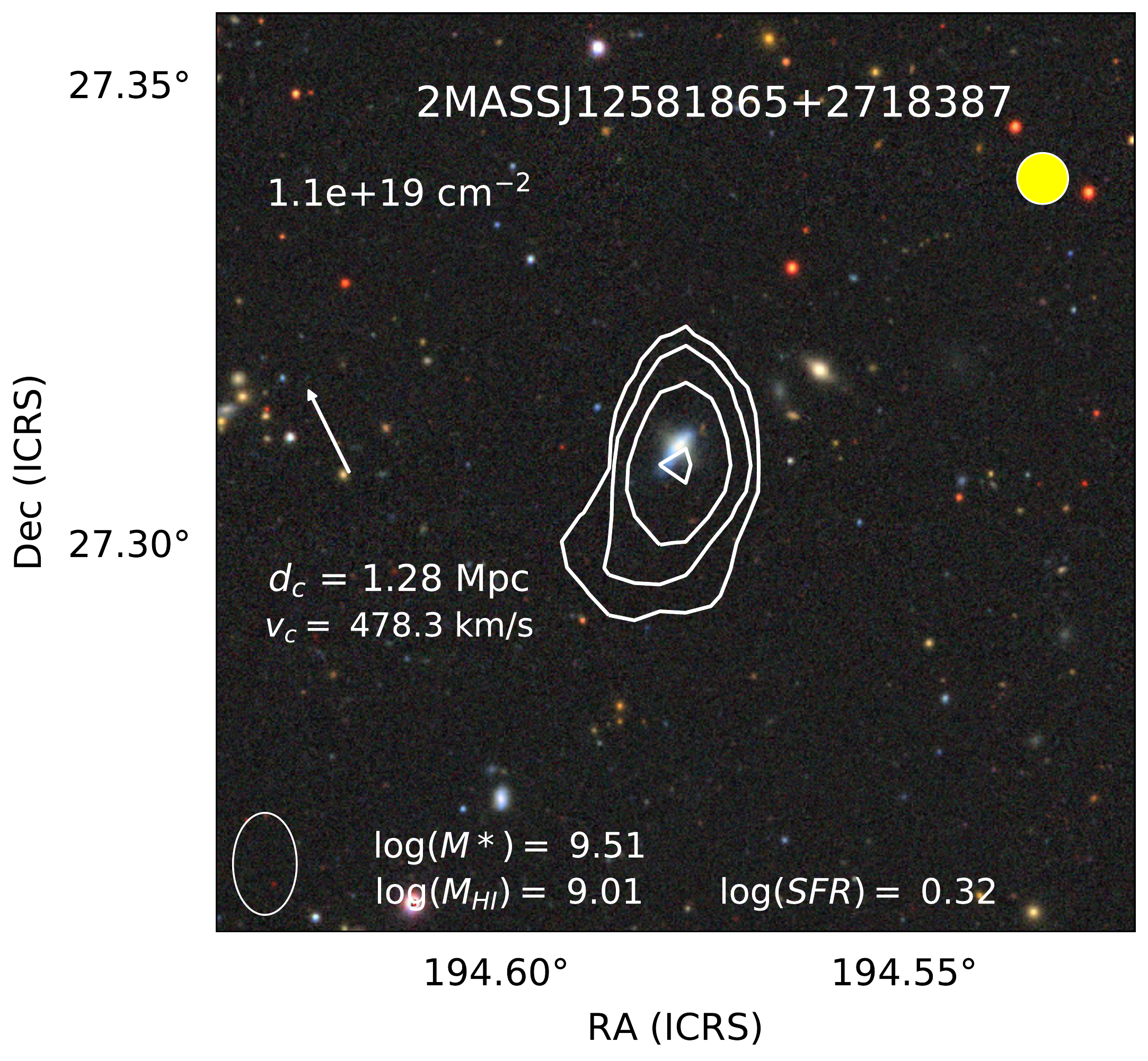} 
   \includegraphics[width=0.45\textwidth]{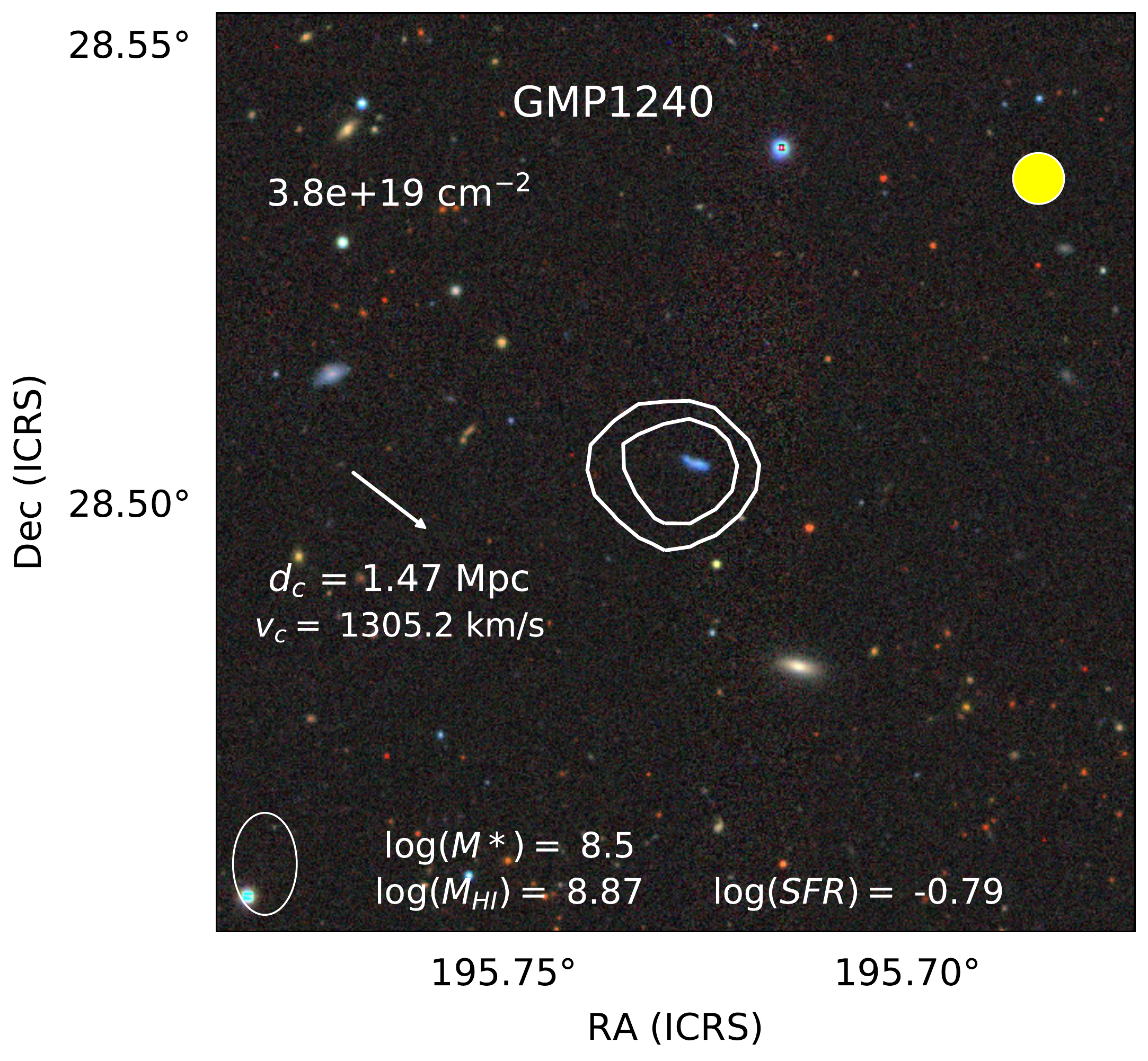} 
   \includegraphics[width=0.45\textwidth]{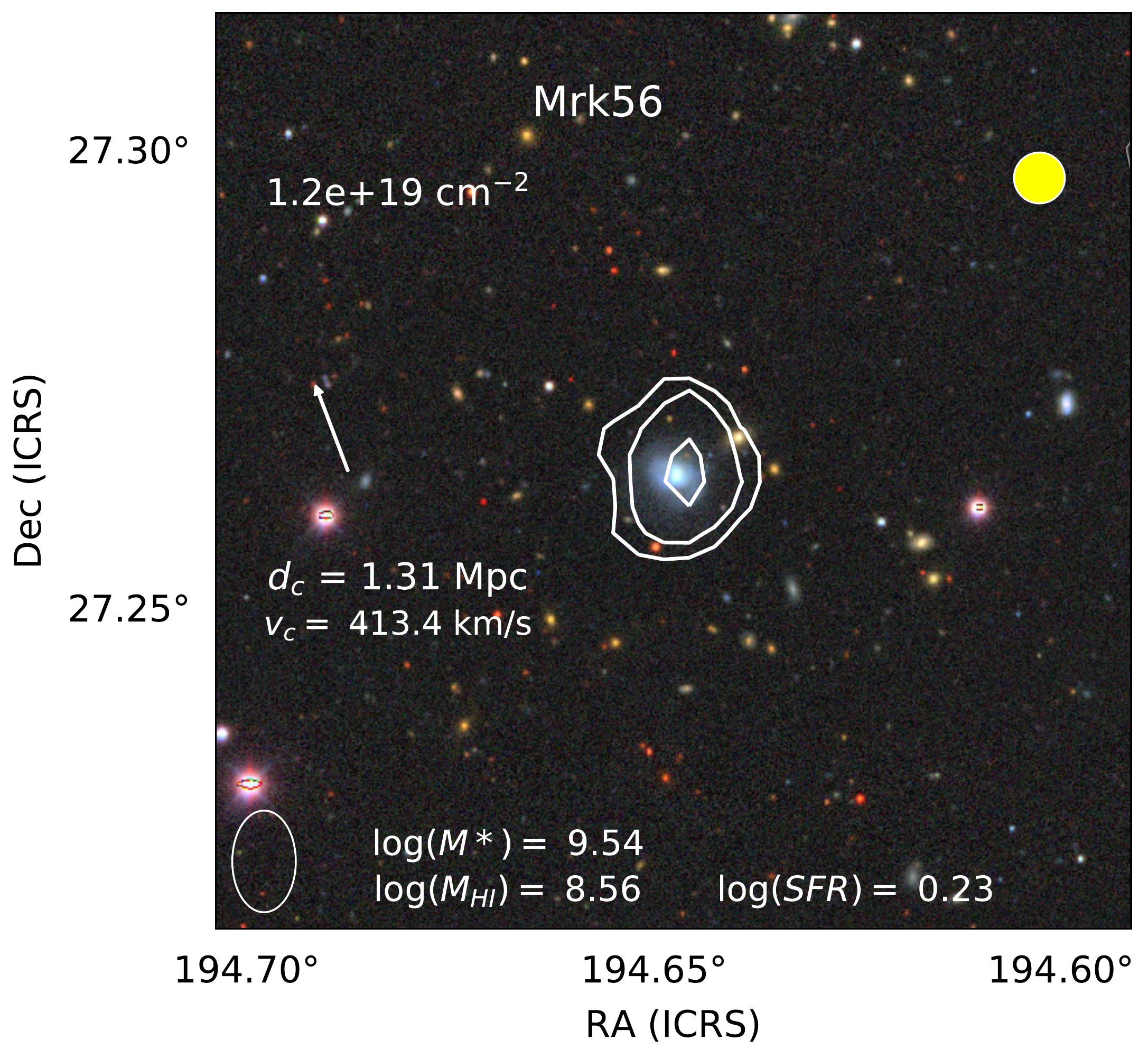} 
   \includegraphics[width=0.45\textwidth]{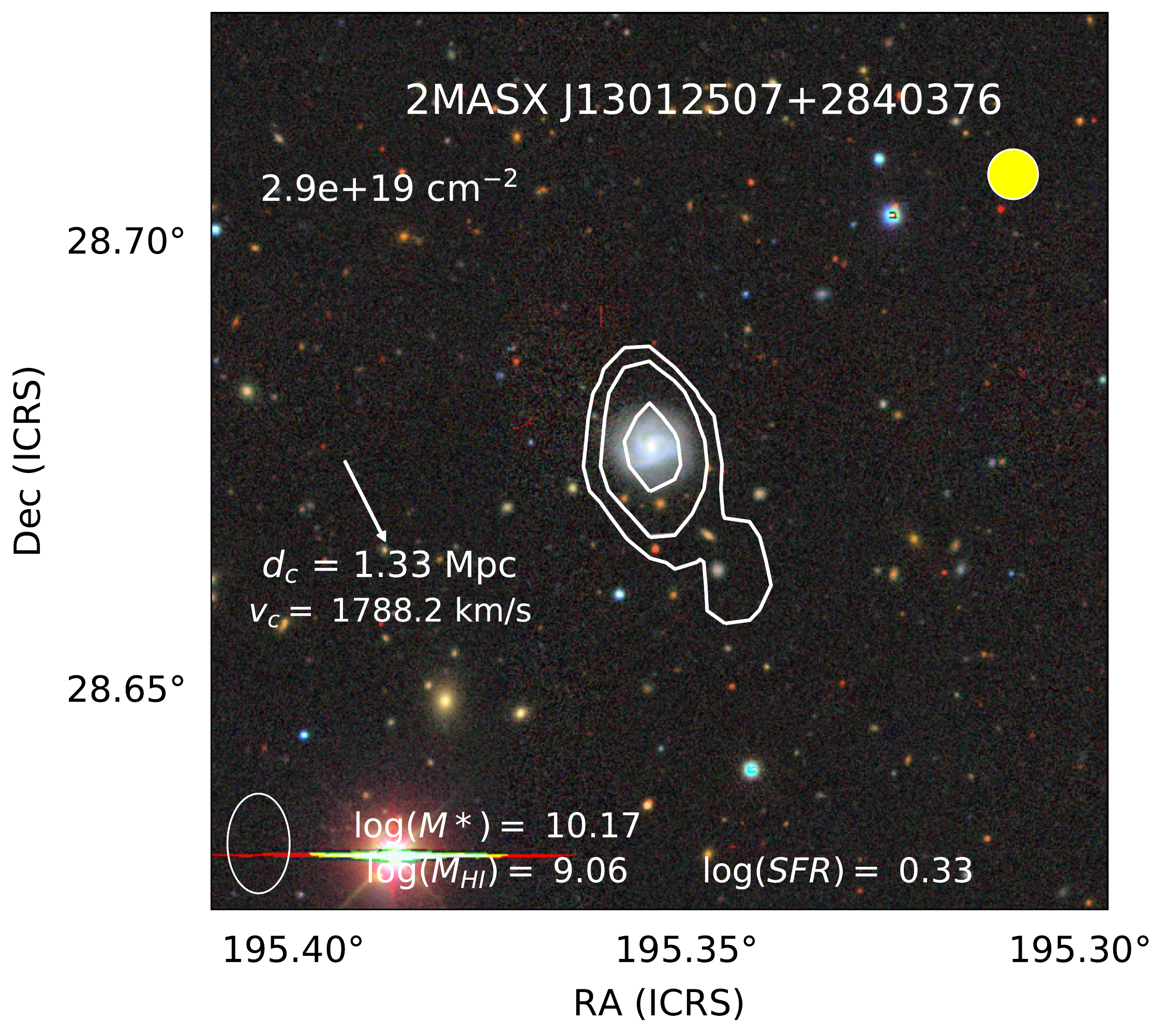} 
   \includegraphics[width=0.45\textwidth]{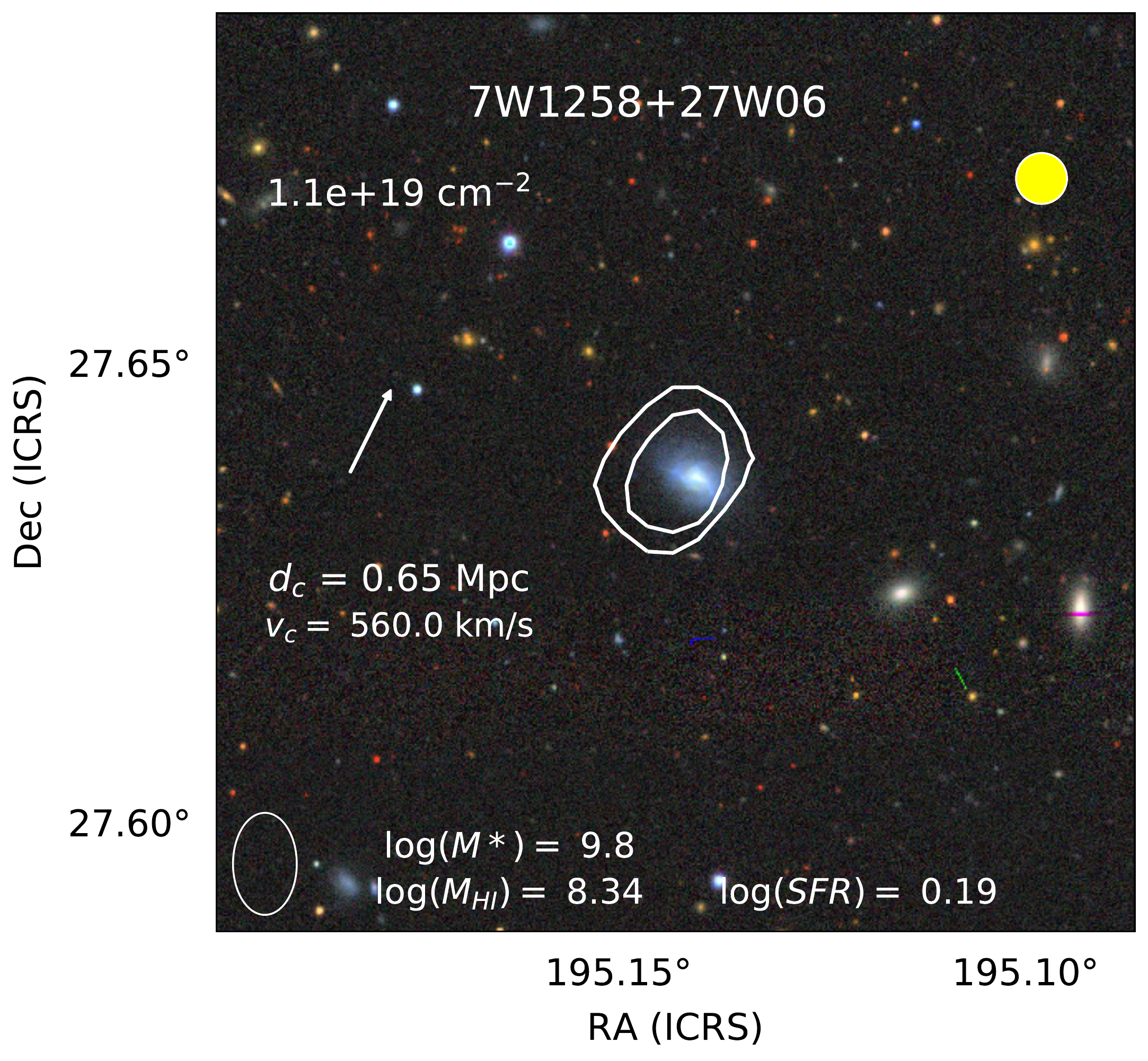} 
   \includegraphics[width=0.45\textwidth]{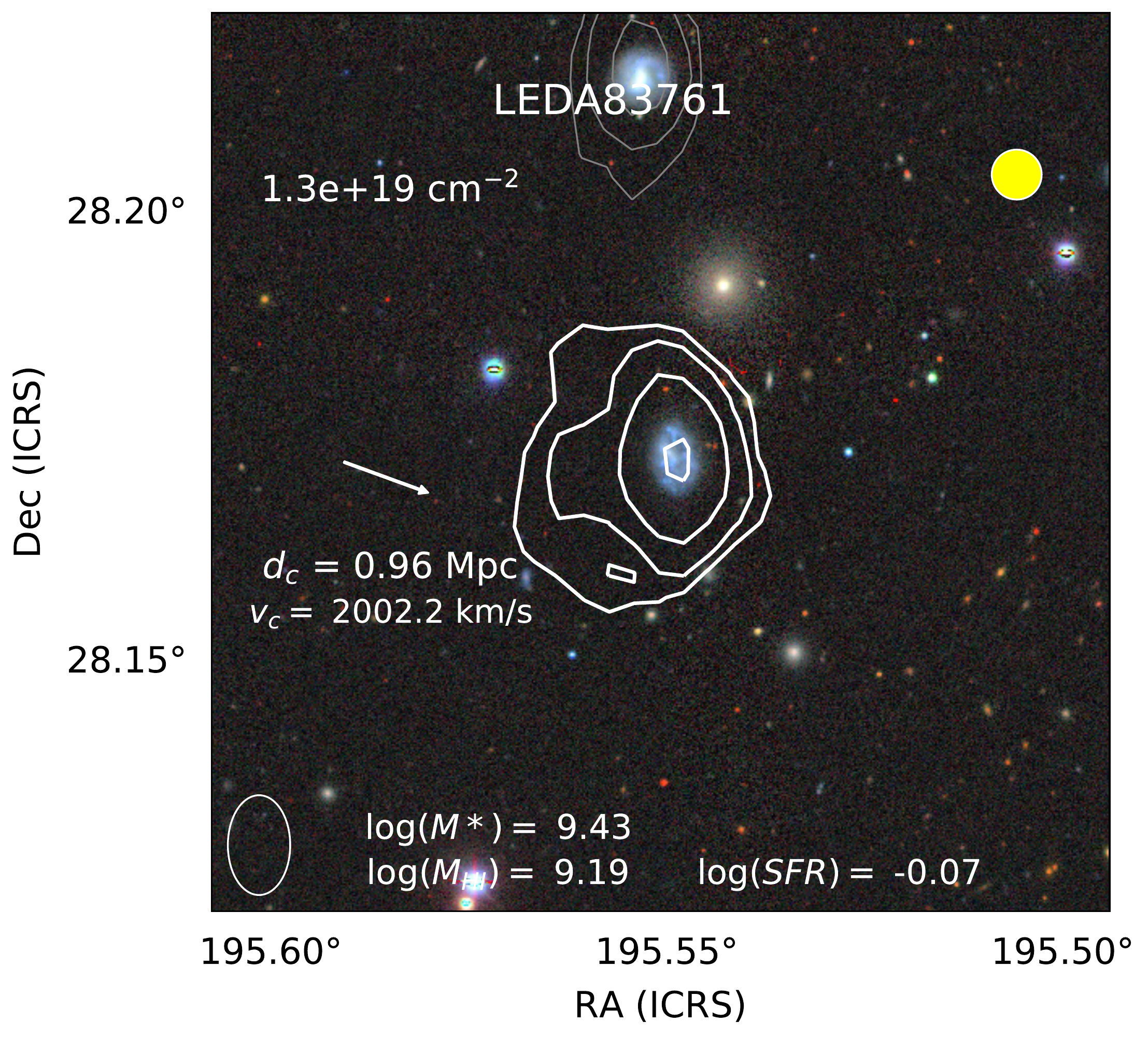} 
   \caption{Same as Fig \protect\ref{fig::stamp1}, but showing H$\,\textsc{i}$ sources with one-sided asymmetry.}
              \label{fig::stamp4}
\end{figure*}

\begin{figure*}[h!] 
   \centering
   \includegraphics[width=0.45\textwidth]{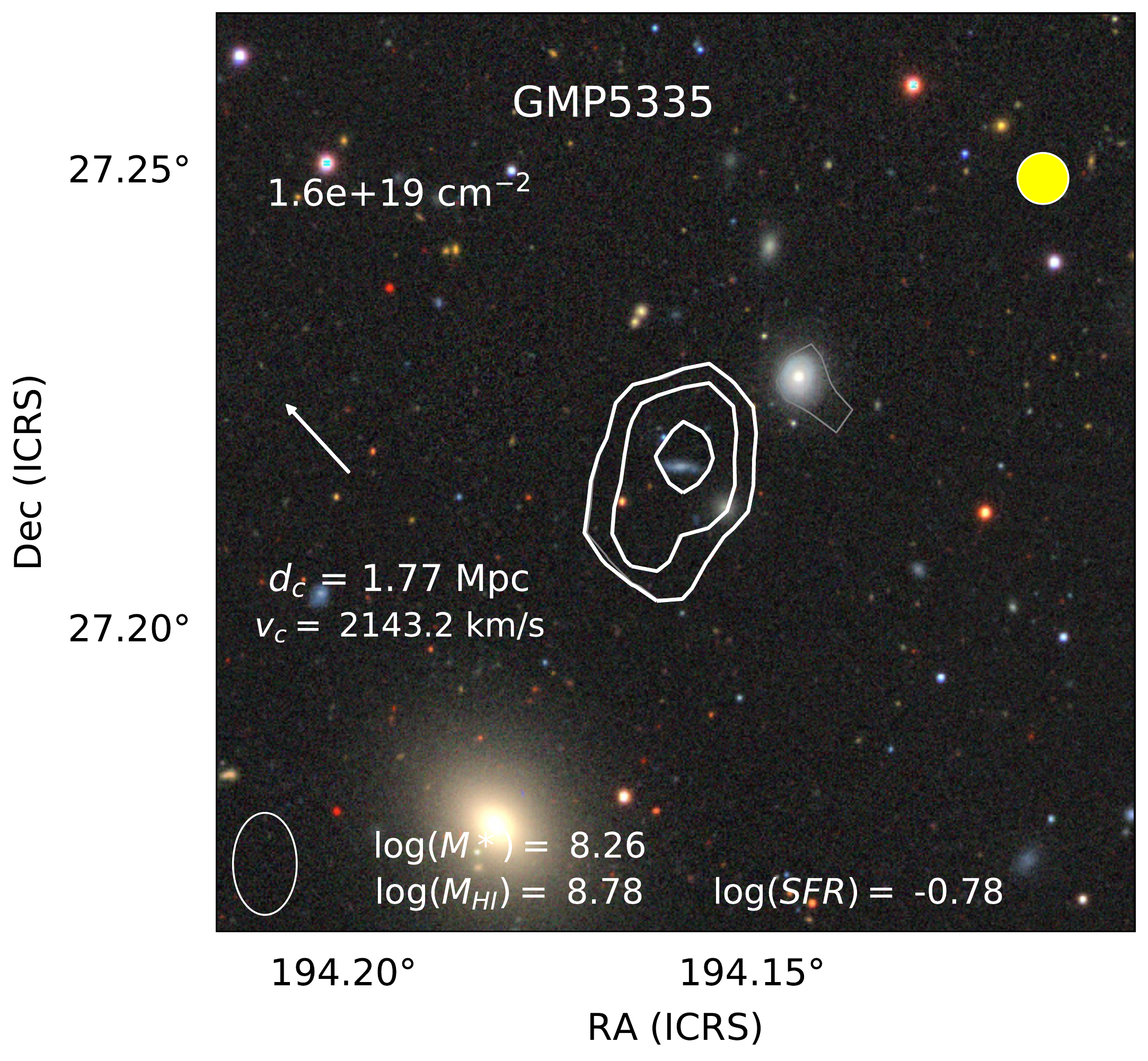} 
   \includegraphics[width=0.45\textwidth]{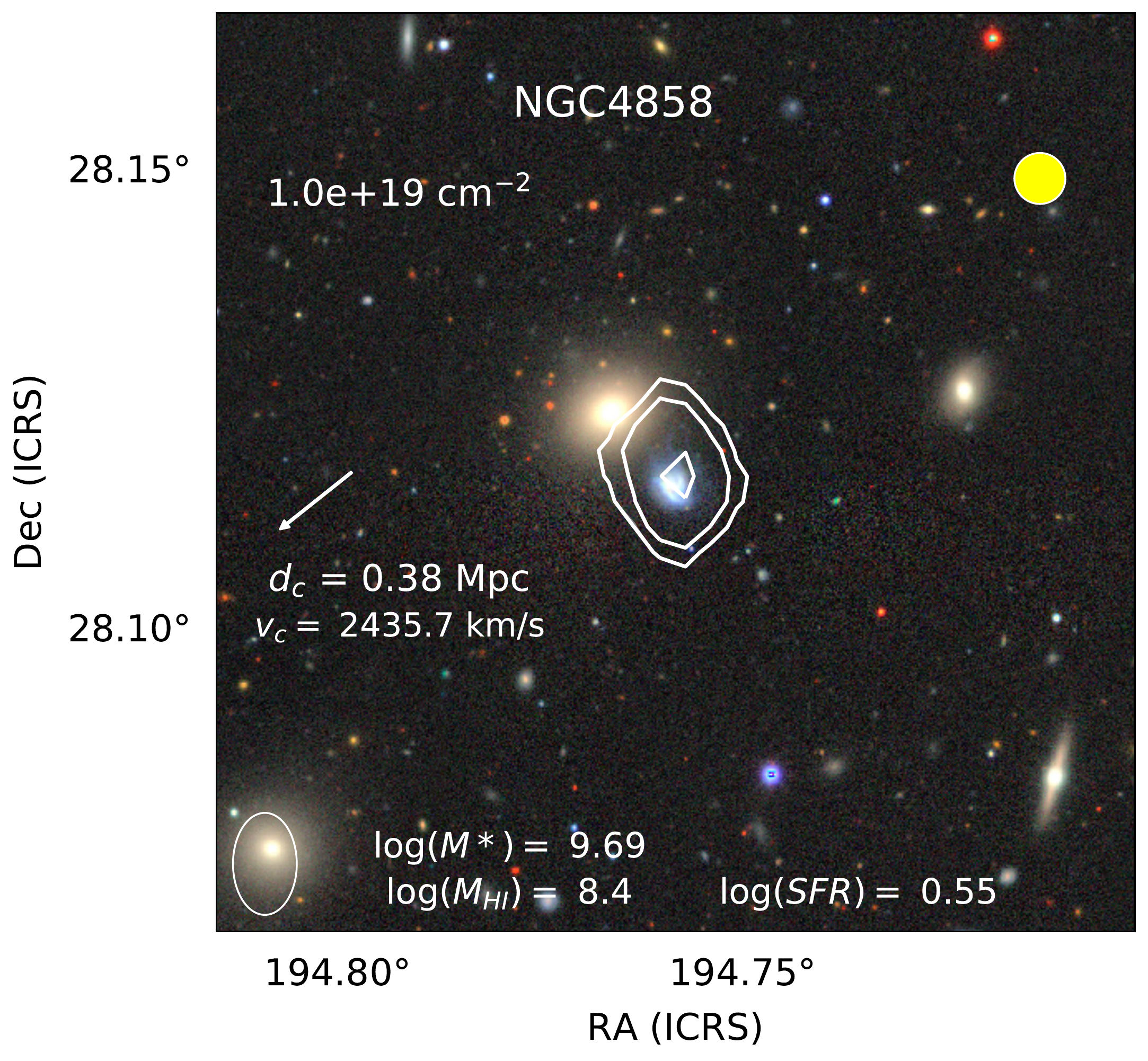} 
   \includegraphics[width=0.45\textwidth]{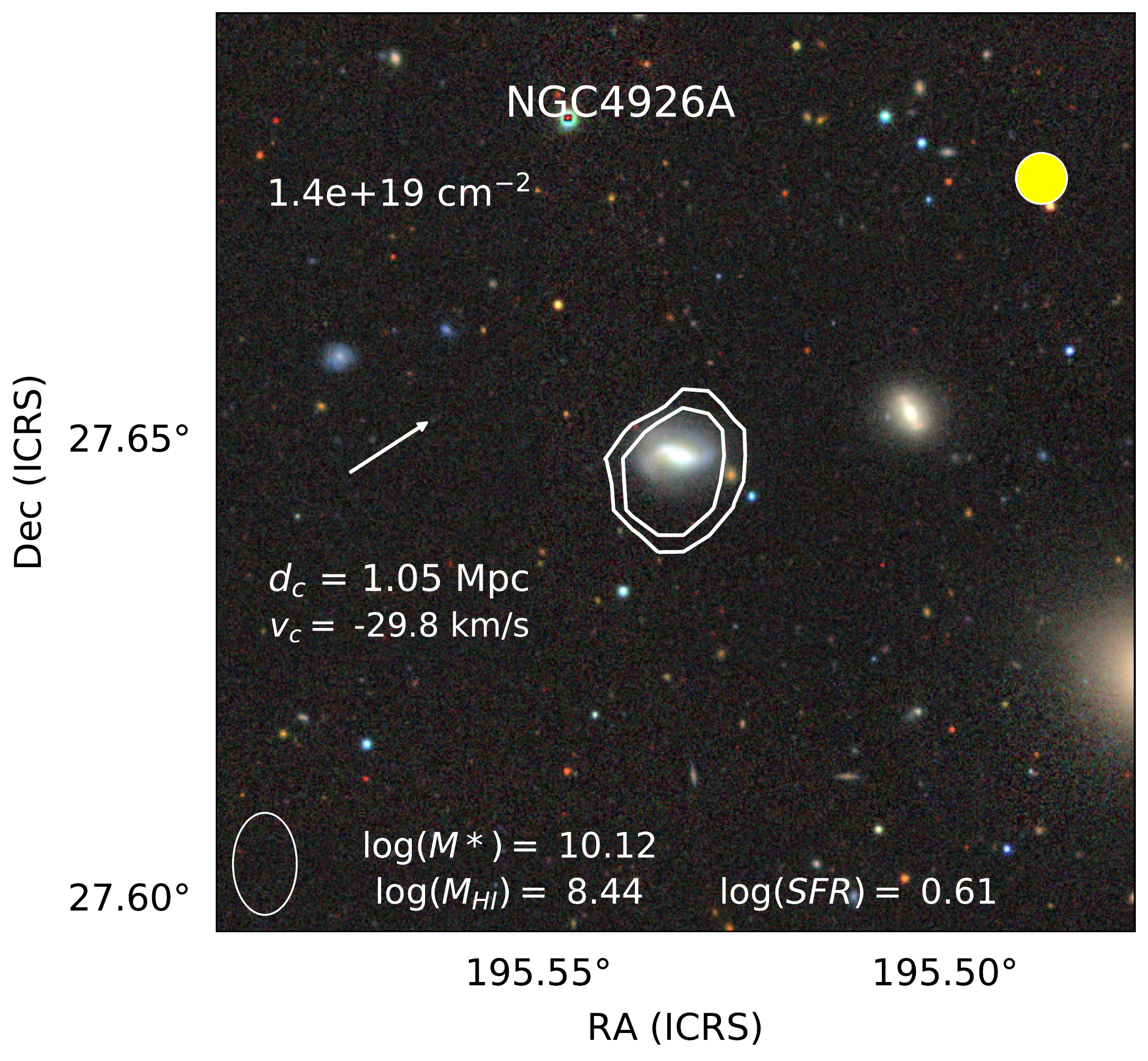} 
   \includegraphics[width=0.45\textwidth]{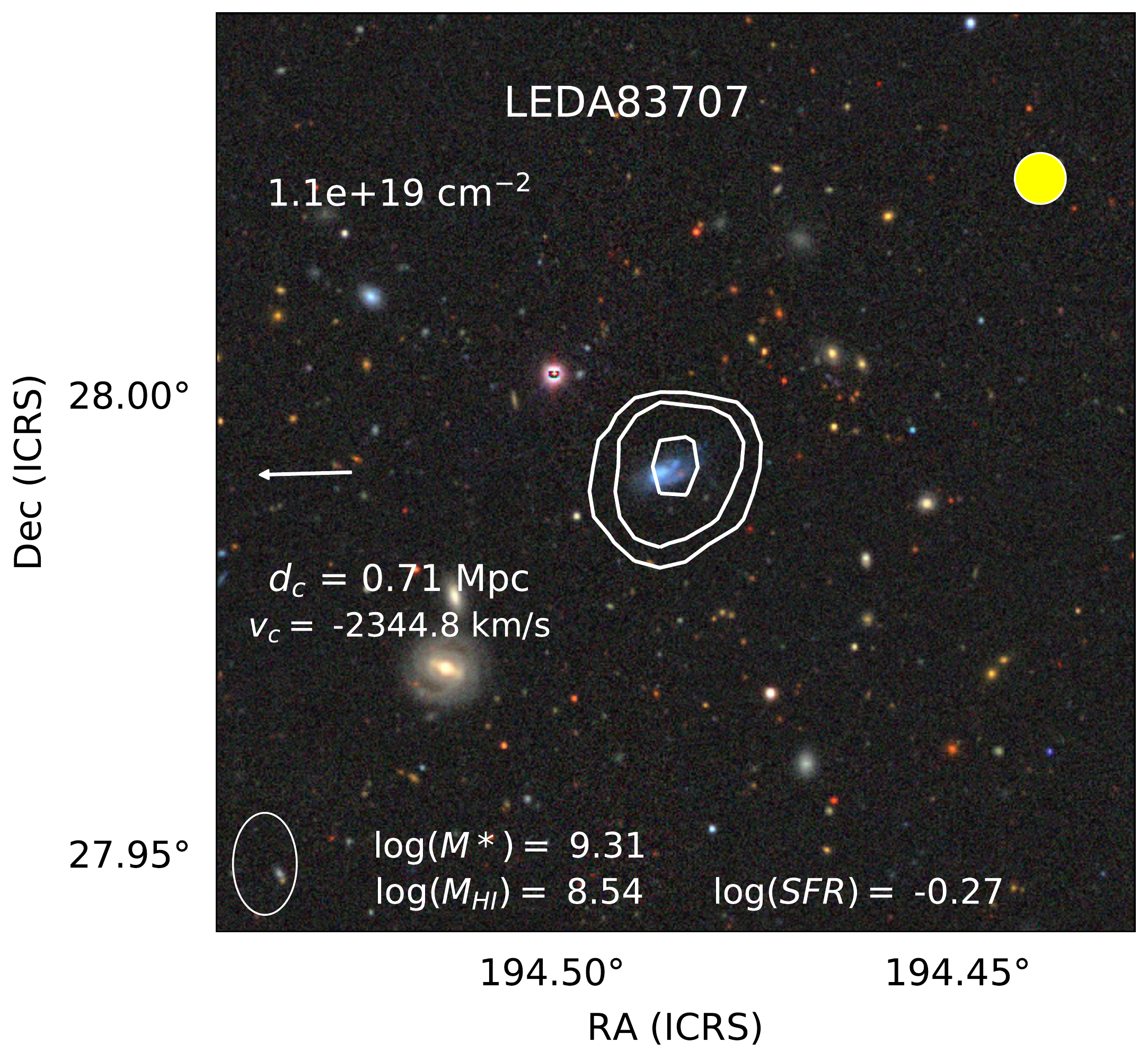} 
   \includegraphics[width=0.45\textwidth]{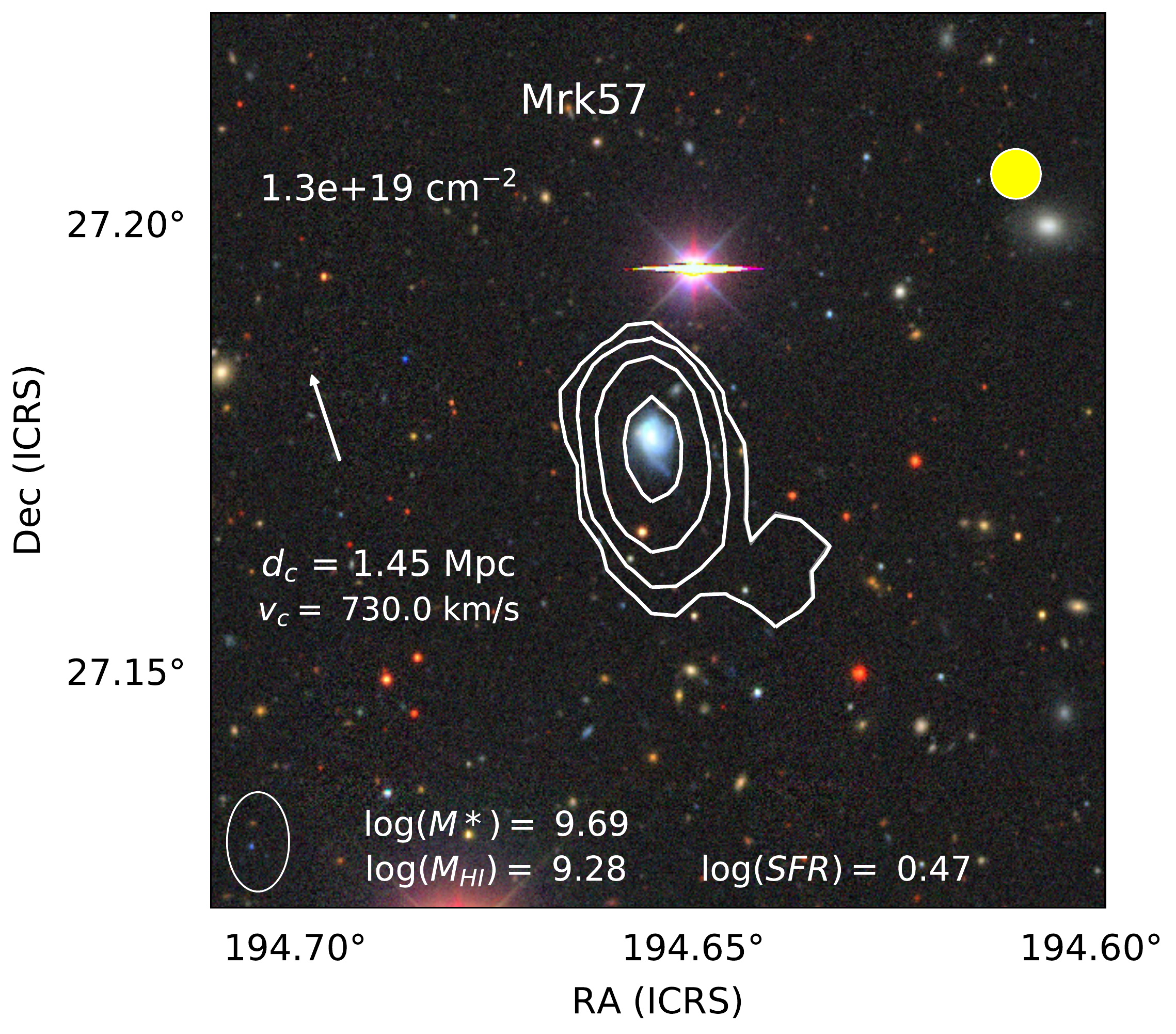} 
   \includegraphics[width=0.45\textwidth]{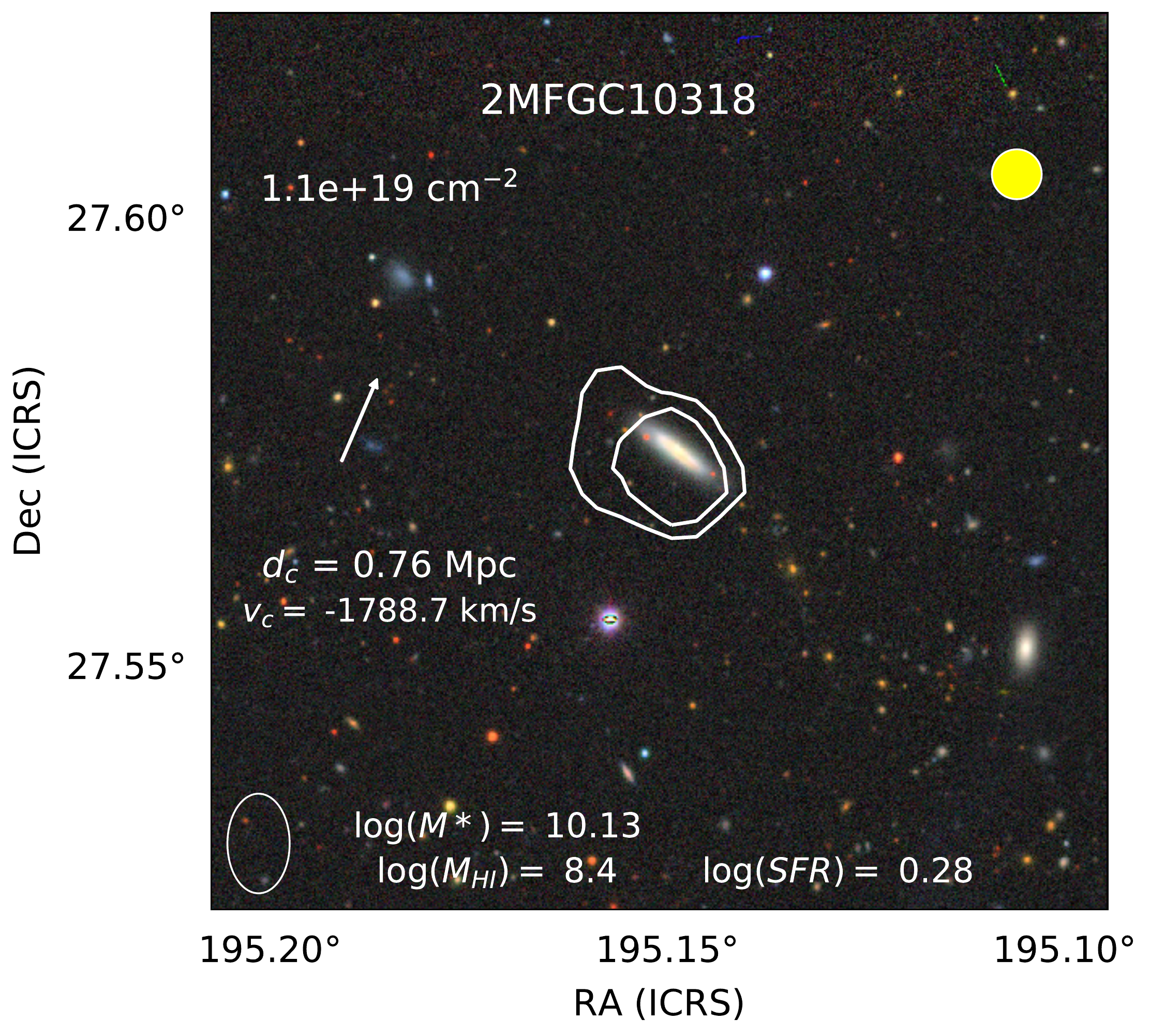} 
   \addtocounter{figure}{-1}
   \caption{continued.}
              \label{fig::stamp5}
\end{figure*}

\begin{figure*}[h!] 
   \centering
   \includegraphics[width=0.45\textwidth]{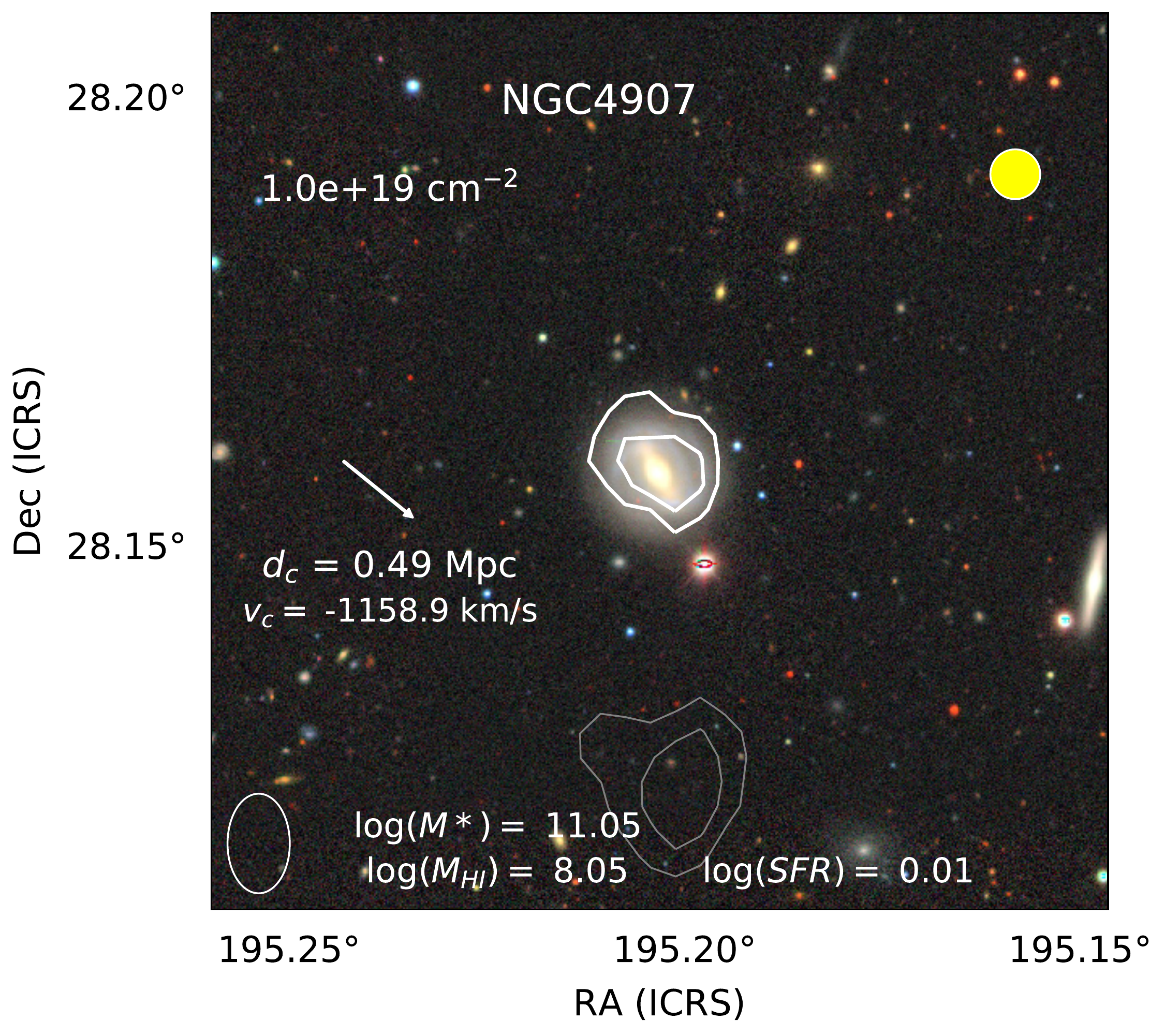} 
   \includegraphics[width=0.45\textwidth]{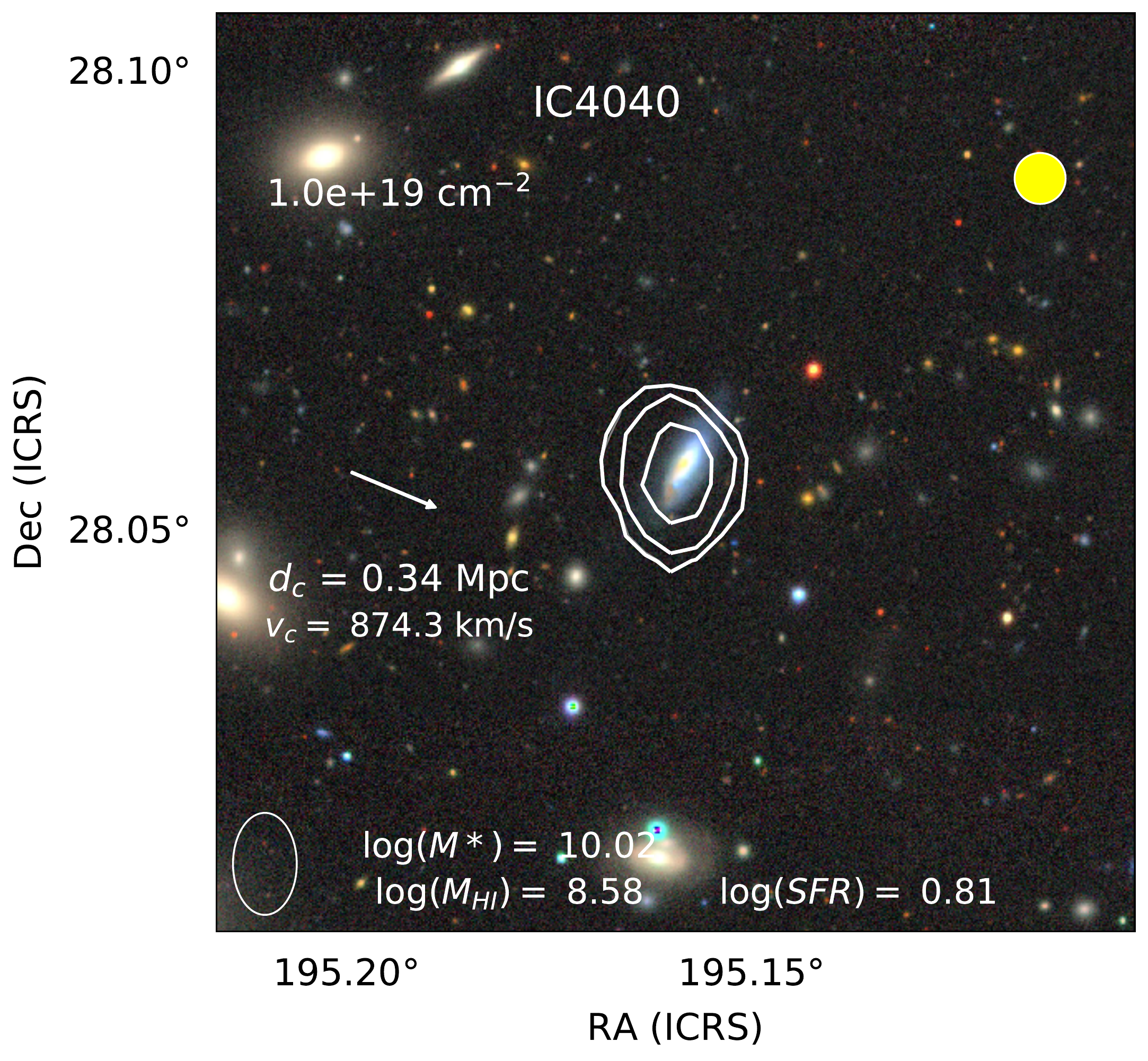} 
   \includegraphics[width=0.45\textwidth]{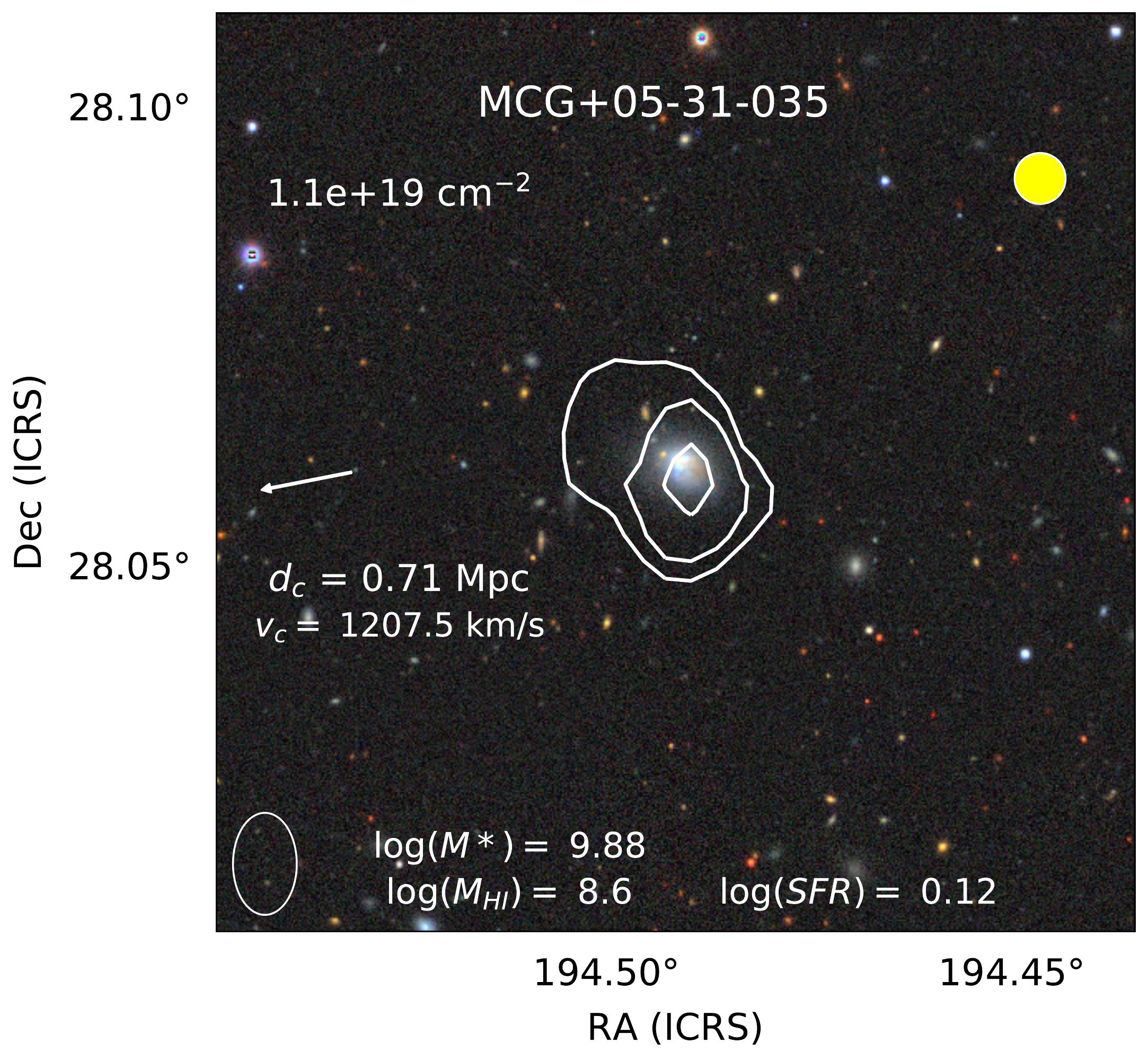} 
    \includegraphics[width=0.45\textwidth]{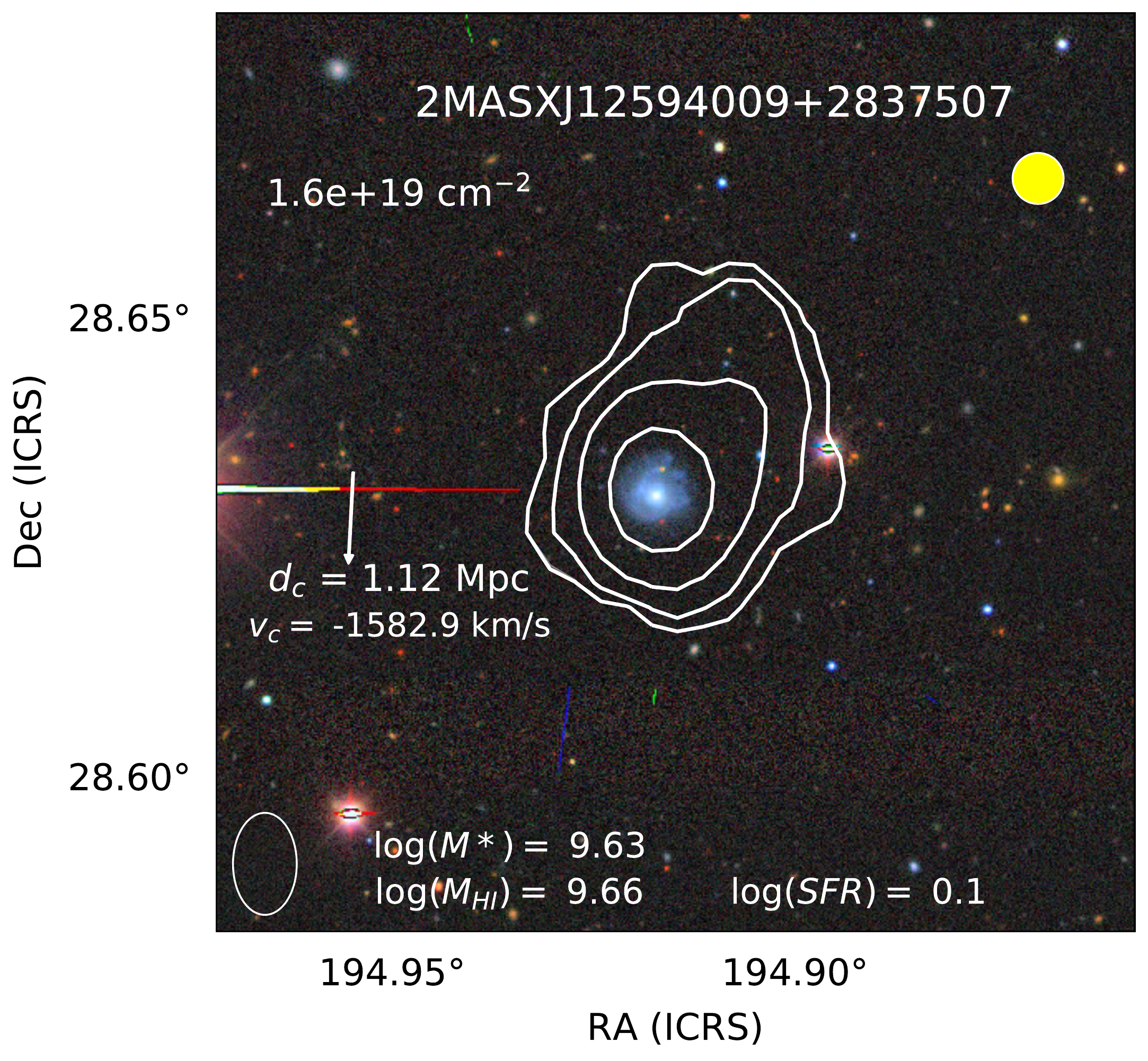} 
   \includegraphics[width=0.45\textwidth]{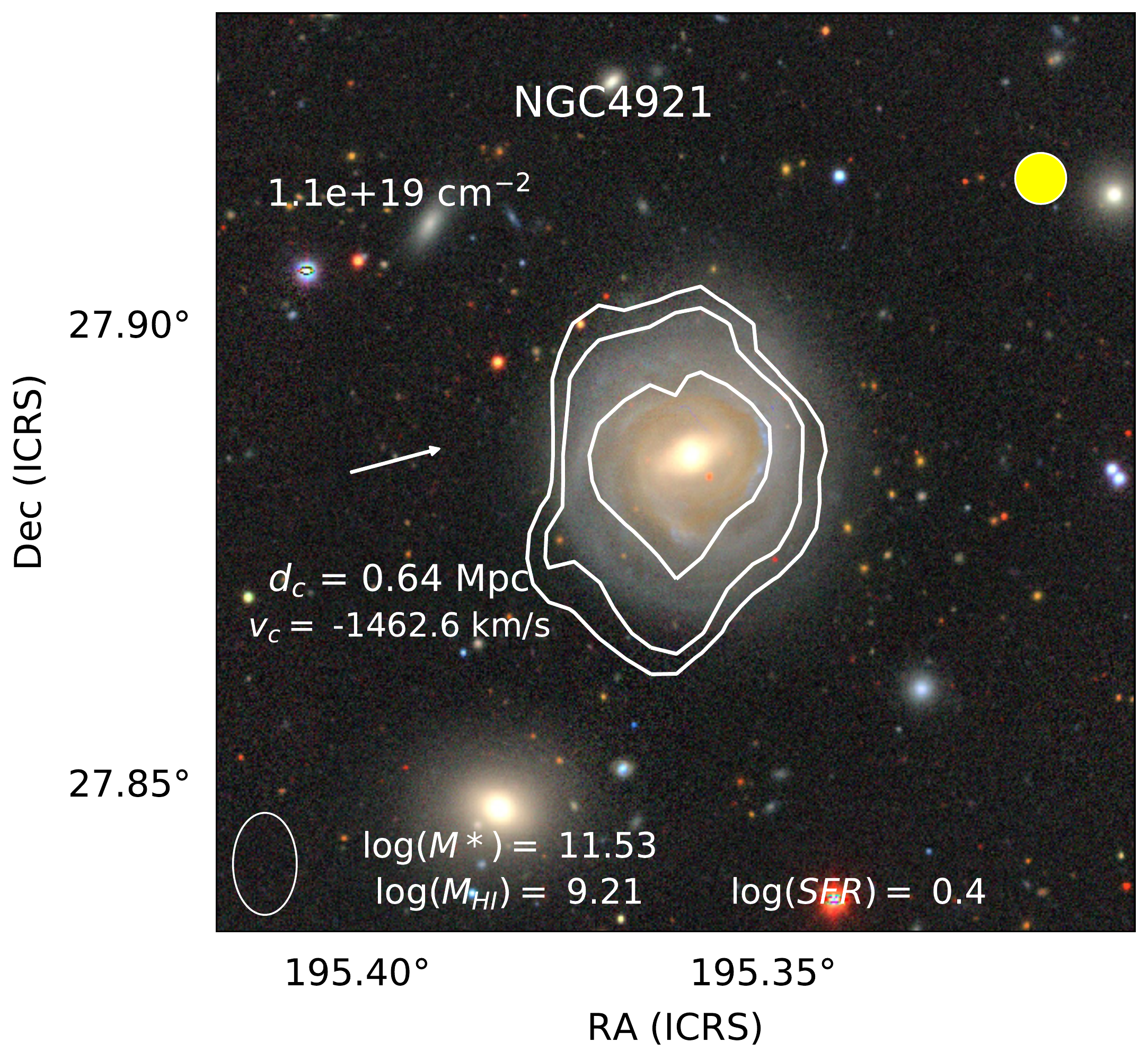} 
   \addtocounter{figure}{-1}
   \caption{continued.}
              \label{fig::stamp7}
\end{figure*}

\begin{figure*}[h!] 
   \centering
   \includegraphics[width=0.45\textwidth]{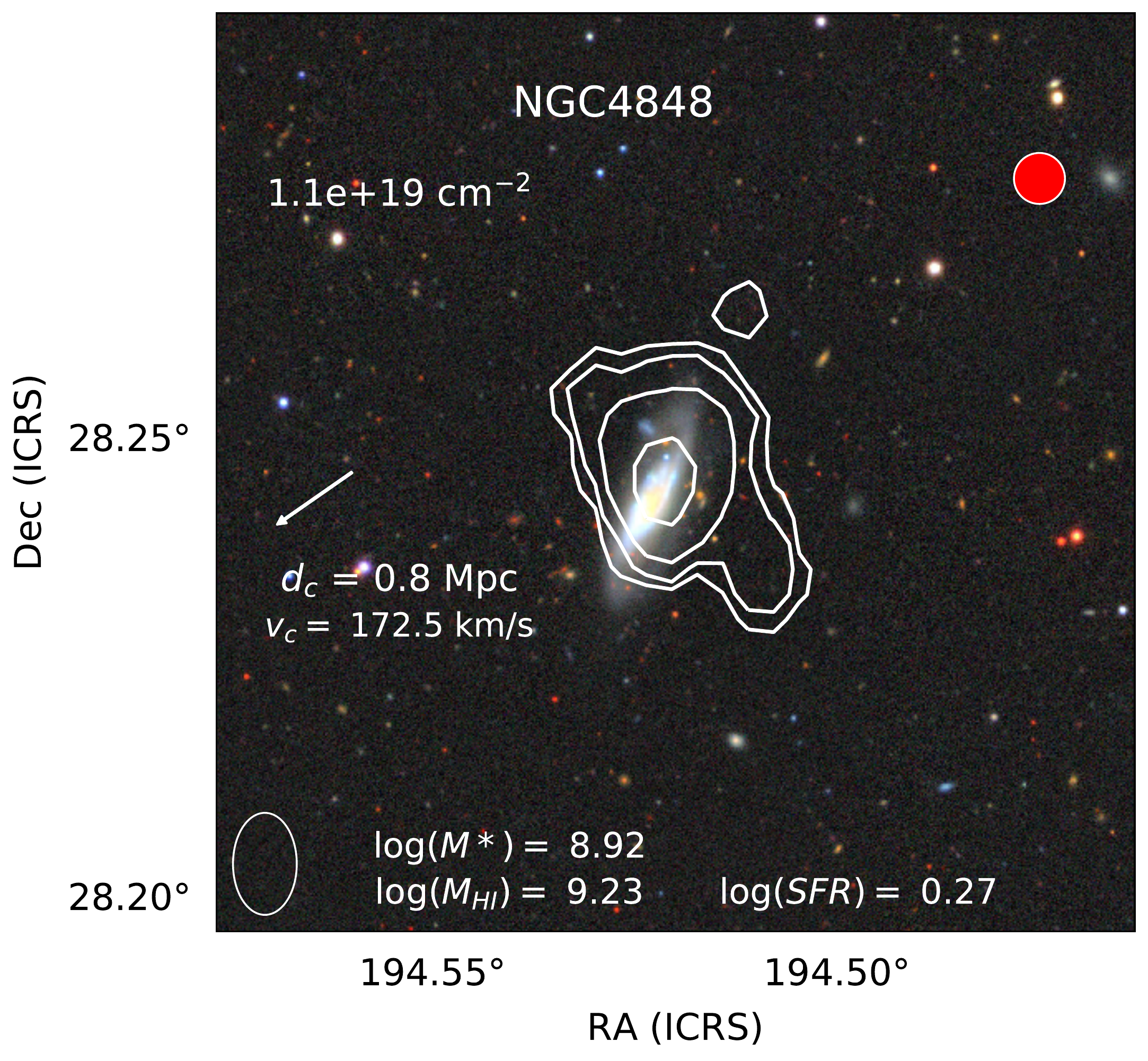} 
   \includegraphics[width=0.45\textwidth]{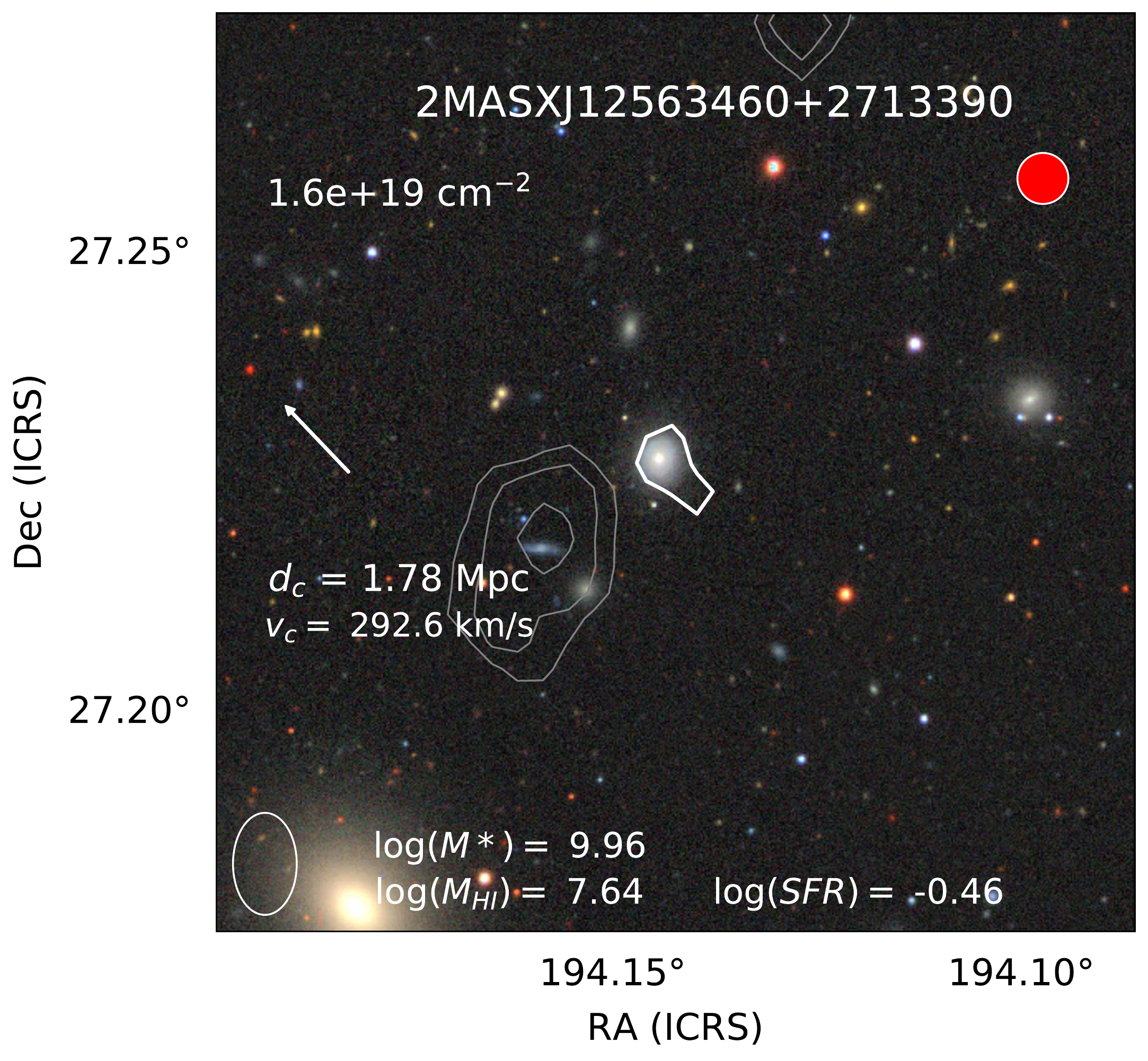} 
   \includegraphics[width=0.45\textwidth]{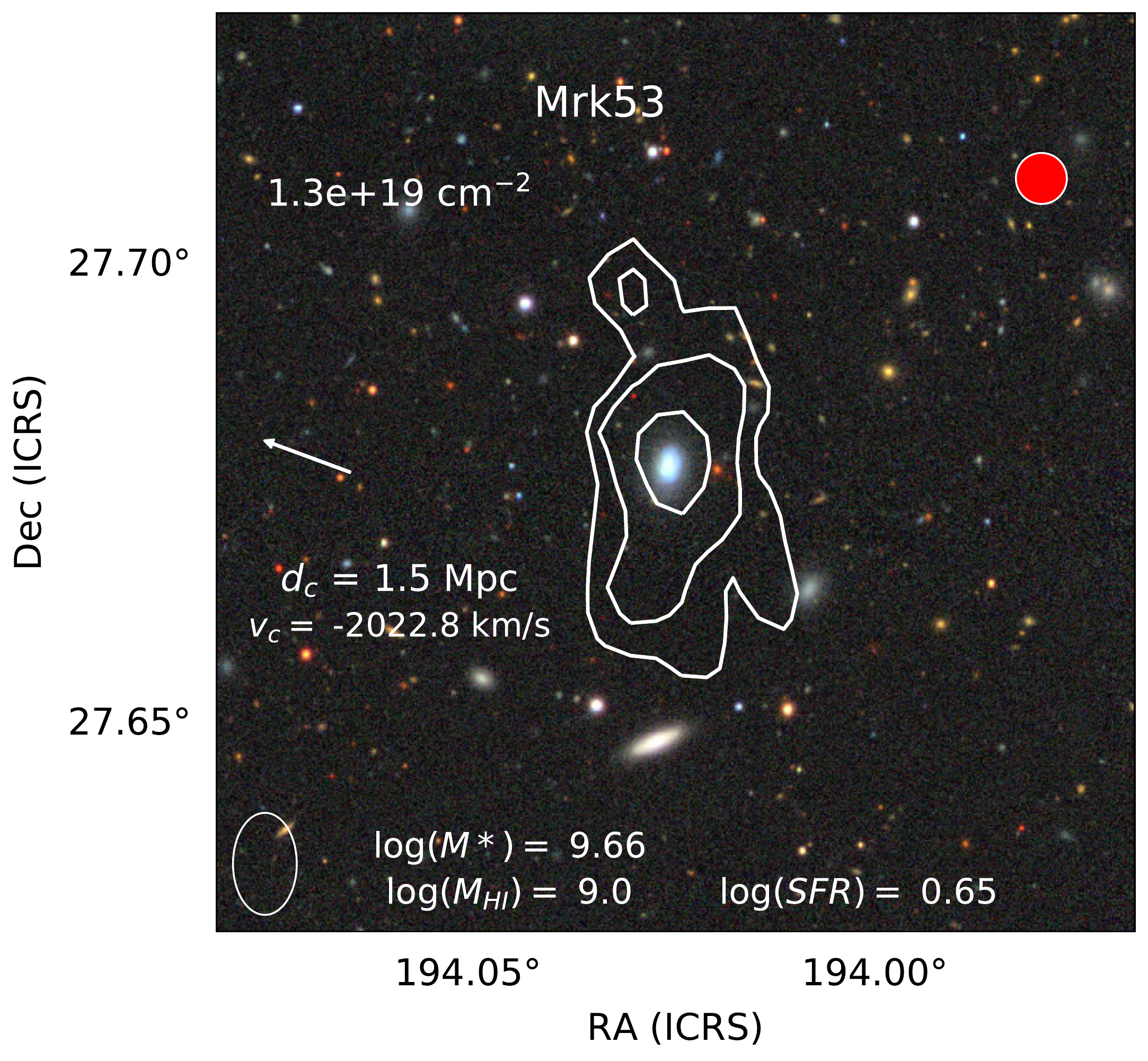} 
   \includegraphics[width=0.45\textwidth]{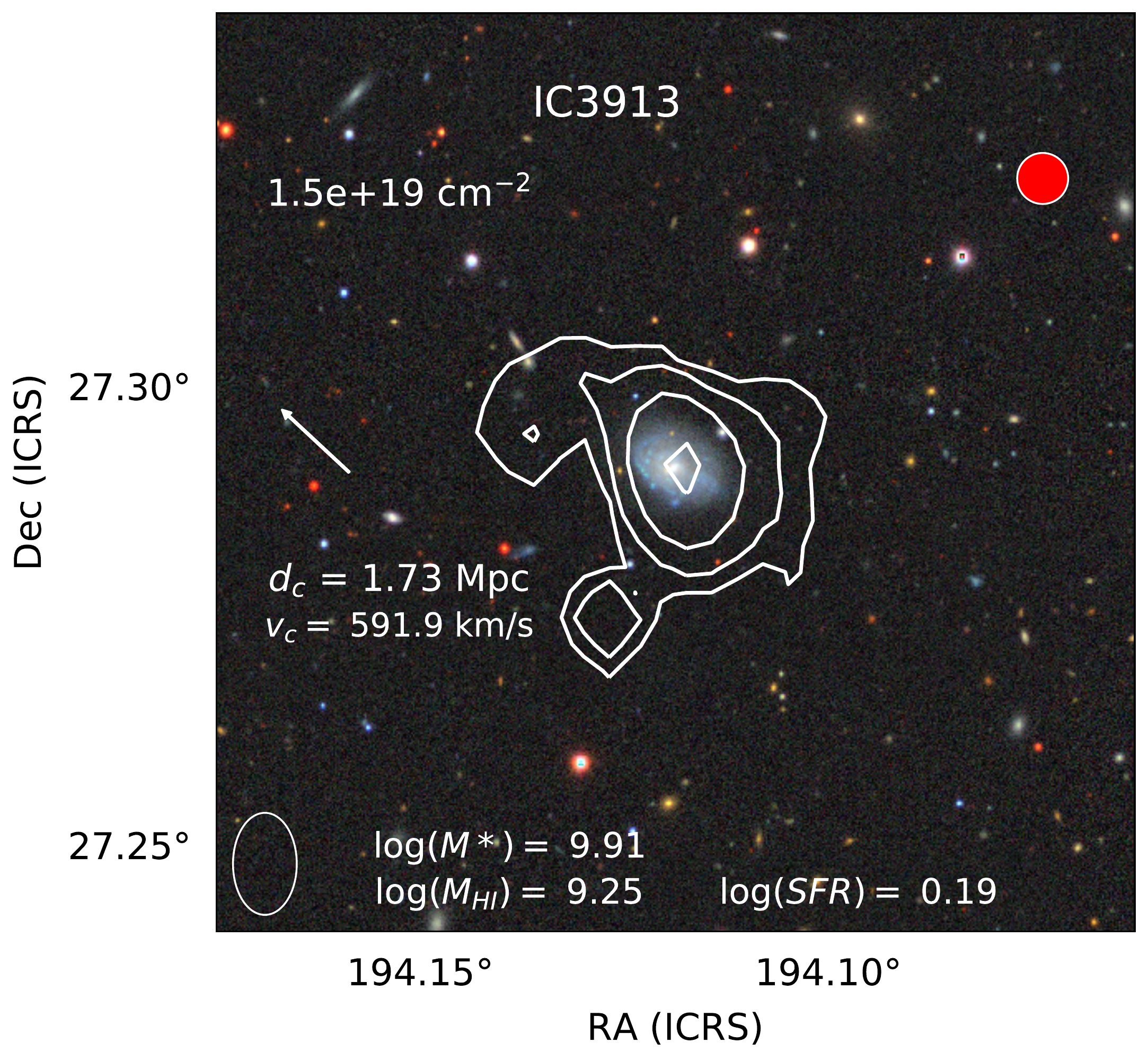} 
   \includegraphics[width=0.45\textwidth]{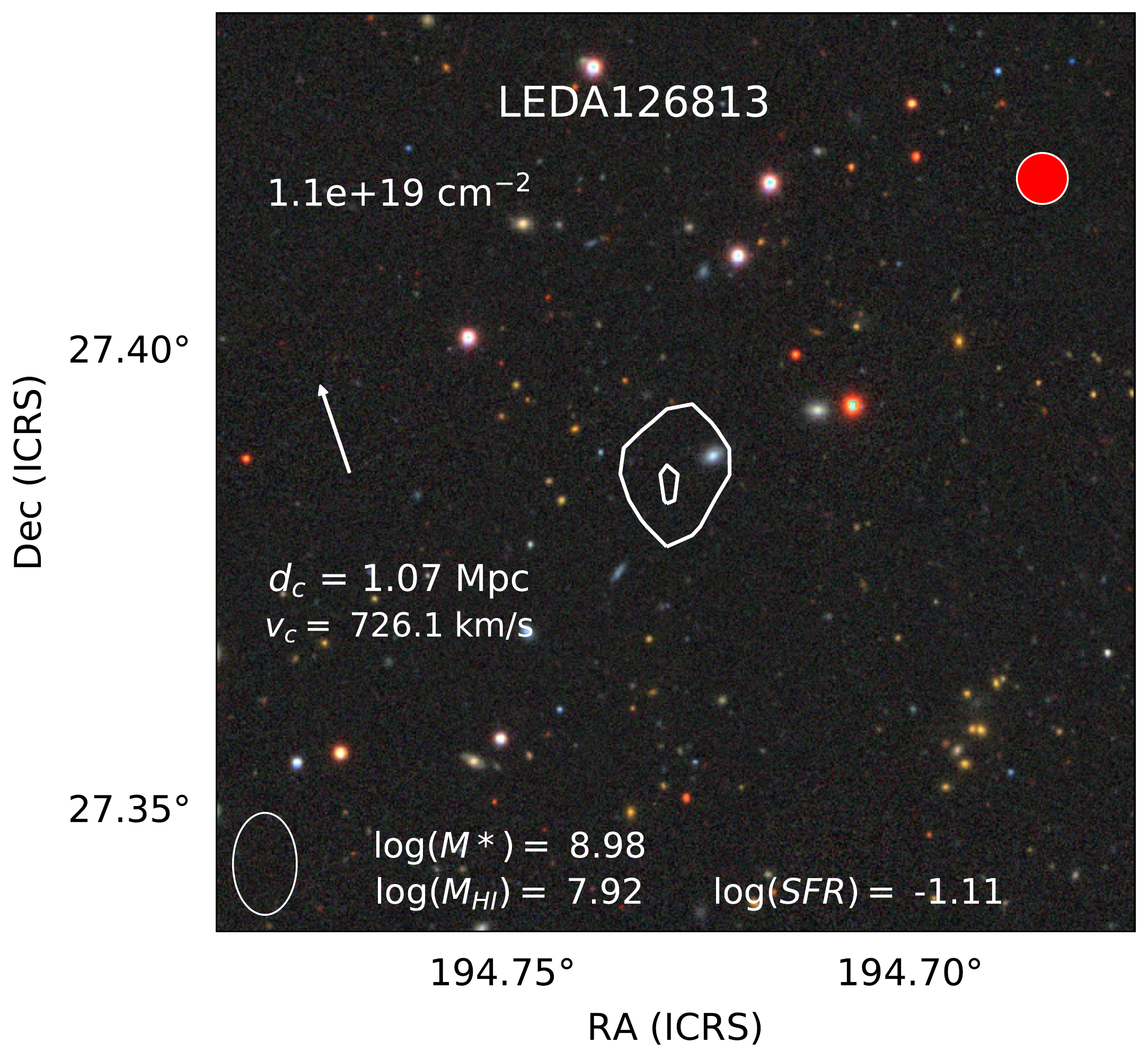} 
   \caption{Same as Fig \protect\ref{fig::stamp1}, but showing H$\,\textsc{i}$ sources with unsettled morphologies.}
              \label{fig::stamp6}
\end{figure*}

\FloatBarrier

\section{Properties of H$\,\textsc{i}$-selected Coma galaxies}

In Table \ref{tab::hi_coma} we show our catalogue containing the properties of our H$\,\textsc{i}$-selected Coma galaxy sample. For details on our H$\,\textsc{i}$ morphological classification, see Sect. \ref{sect::hi_morph}. Our methods of estimating physical properties are described in Sect. \ref{sect::mstar_sfr_calc}.

\begin{onecolumn}
\begin{landscape}
\vspace{-1.5cm}

\begin{longtable}{cccccccccccccc}
\caption{Properties of our H$\,\textsc{i}$-selected galaxies in Coma. $^1$: Angular distance between the centre of our H$\,\textsc{i}$ detections and the position of the optical counterpart. $^2$: projected distance to Coma centre. $^3$: difference of H$\,\textsc{i}$ and optical counterpart line-of-sight velocities. $^4$: our H$\,\textsc{i}$ morphological class (see Sect \protect\ref{sect::hi_morph}.). $^5$: 1 if source was detected to have a tail at UV/H$_\alpha$ wavelengths \protect\citet{smith10,yagi10}, otherwise 0. $^6$: 1 for sources identified as AGN in the literature \protect\citep{mahajan10,gavazzi11,toba14}. $^7$: 1 if source was not detected by either \protect\citet{bravo00,bravo01} or \protect\citet{gavazzi06}, otherwise 0.}\label{tab::hi_coma} 
\endfirsthead
 ID & RA & Dec & off$^1$ & $v_{HI}$ & $d_C$ $^2$ & $v_{\rm off}$ $^3$ & $\log(M_{\star})$ & $\log(\rm SFR)$ & $\log(M_{HI})$ & H$\,\textsc{i}$ type$^4$ & Tail$^5$ & AGN$^6$ & NewDet$^7$ \\
- & [deg] & [deg] & ["] & [$\rm km s^{-1}$] & [Mpc] & [$\rm km s^{-1}$] & [$\log(M_{\odot})$] & [$\log(M_{\odot} \mathrm{yr^{-1}})$] & [$\log(M_{\odot})$] & - & - & - & - \\
\noalign{\smallskip}\hline\noalign{\smallskip}
NGC4926A & 195.533 & 27.647 & 6 & 6903 & 1.05 & 4 & 10.12$\pm$0.1 & 0.61$\pm$0.01 & 8.44 & 1 & 0 & 0 & 0 \\
GMP4023 & 194.691 & 28.693 & 2 & 6825 & 1.28 & -27 & 8.86$\pm$0.65 & -0.66$\pm$0.07 & 8.17 & 0 & 0 & 0 & 1\\
\makecell{2MASX \\ J13013085+2833237} & 195.378 & 28.558 & 3 & 6879 & 1.18 & -3 & 9.22$\pm$0.17 & -0.69$\pm$0.09 & 9.02 & 0 & 0 & 0 & 1\\
GMP2486 & 195.187 & 28.528 & 2 & 6888 & 1.00 & -1 & 7.84$\pm$1.3 & -1.11$\pm$0.18 & 8.23 & 0 & 0 & 0 & 1\\
NGC4848 & 194.52 & 28.247 & 18 & 7106 & 0.80 & -106 & 8.92$\pm$0.27 & 0.27$\pm$0.02 & 9.23 & 2 & 1 & 1 & 0 \\
LEDA1795276 & 194.66 & 27.014 & 3 & 6975 & 1.71 & -42 & 9.27$\pm$0.23 & -0.58$\pm$0.16 & 7.95 & 0 & 0 & 0 & 1 \\
\makecell{2MASX \\ J12563460+2713390} & 194.142 & 27.226 & 8 & 7226 & 1.78 & 36 & 9.96$\pm$0.12 & -0.46$\pm$0.09 & 7.64 & 2 & 0 & 0 & 1 \\
\makecell{2MASS \\ J12581865+2718387} & 194.578 & 27.308 & 10 & 7411 & 1.28 & 0 & 9.51$\pm$0.11 & 0.32$\pm$0.03 & 9.01 & 1 & 0 & 0 & 0 \\
Mrk56 & 194.648 & 27.265 & 2 & 7346 & 1.31 & -13 & 9.54$\pm$0.11 & 0.23$\pm$0.03 & 8.56 & 1 & 0 & 0 & 1 \\
IC3913 & 194.119 & 27.291 & 1 & 7525 & 1.73 & 9 & 9.91$\pm$0.12 & 0.19$\pm$0.03 & 9.25 & 2 & 1 & 0 & 0 \\
AGC725242 & 194.629 & 26.995 & 1 & 7435 & 1.76 & -19 & 8.78$\pm$0.71 & -0.70$\pm$0.08 & 8.44 & 0 & 0 & 0 & 1 \\
7W1258+27W06 & 195.143 & 27.638 & 9 & 7493 & 0.65 & 9 & 9.8$\pm$0.12 & 0.19$\pm$0.03 & 8.34 & 1 & 1 & 0 & 0 \\
MCG+05-31-041 & 194.539 & 28.709 & 3 & 7605 & 1.39 & 0 & 10.18$\pm$0.1 & 0.34$\pm$0.03 & 9.33 & 0 & 0 & 0 & 0 \\
Mrk57 & 194.654 & 27.174 & 10 & 7663 & 1.45 & -2 & 9.69$\pm$0.10 & 0.47$\pm$0.02 & 9.28 & 1 & 0 & 0 & 0 \\
IC4040 & 195.159 & 28.057 & 5 & 7807 & 0.34 & -67 & 10.02$\pm$0.10 & 0.81$\pm$0.01 & 8.58 & 1 & 1 & 0 & 0 \\
LEDA126813 & 194.729 & 27.387 & 16 & 7659 & 1.07 & -114 & 8.98$\pm$0.39 & -1.11$\pm$0.17 & 7.92 & 2 & 0 & 0 & 1 \\
NGC4911 & 195.233 & 27.791 & 2 & 7990 & 0.53 & 17 & 11.03$\pm$0.01 & 0.81$\pm$0.02 & 9.02 & 0 & 1 & 1 & 0 \\
MCG+05-31-035 & 194.492 & 28.061 & 6 & 8141 & 0.71 & 4 & 9.88$\pm$0.13 & 0.12$\pm$0.05 & 8.6 & 1 & 1 & 0 & 1 \\
MCG+05-31-108 & 195.554 & 28.216 & 4 & 8180 & 0.99 & -5 & 9.82$\pm$0.11 & 0.25$\pm$0.03 & 8.71 & 0 & 0 & 0 & 1 \\
GMP1240 & 195.729 & 28.504 & 9 & 8238 & 1.47 & 33 & 8.5$\pm$1.15 & -0.79$\pm$0.08 & 8.87 & 1 & 1 & 0 & 1 \\
LEDA1838241 & 195.313 & 28.523 & 4 & 8434 & 1.07 & 0 & 8.92$\pm$0.49 & -0.77$\pm$0.08 & 9.08 & 0 & 0 & 0 & 1 \\
\makecell{2MASX \\ J13012507+2840376} & 195.353 & 28.676 & 6 & 8721 & 1.33 & -18 & 10.17$\pm$0.10 & 0.33$\pm$0.02 & 9.06 & 1 & 0 & 0 & 0 \\
\makecell{2MASX \\ J13004067+2831116} & 195.17 & 28.52 & 6 & 8898 & 0.98 & -3 & 9.34$\pm$0.16 & 0.11$\pm$0.02 & 9.3 & 0 & 0 & 0 & 0 \\
LEDA83761 & 195.551 & 28.172 & 5 & 8935 & 0.96 & -40 & 9.43$\pm$0.14 & -0.07$\pm$0.03 & 9.19 & 1 & 0 & 0 & 1 \\
GMP5335 & 194.159 & 27.217 & 3 & 9076 & 1.77 & -39 & 8.26$\pm$1.31 & -0.78$\pm$0.17 & 8.78 & 1 & 0 & 0 & 1 \\
NGC4858 & 194.758 & 28.117 & 6 & 9369 & 0.38 & -48 & 9.69$\pm$0.10 & 0.55$\pm$0.02 & 8.4 & 1 & 1 & 0 & 1 \\
GMP3597 & 194.823 & 27.027 & 1 & 10280 & 1.64 & 2 & 8.08$\pm$1.27 & -1.07$\pm$0.18 & 8.46 & 0 & 0 & 0 & 1 \\
LEDA83707 & 194.485 & 27.992 & 6 & 4588 & 0.71 & 24 & 9.31$\pm$0.28 & -0.27$\pm$0.08 & 8.54 & 1 & 1 & 0 & 1 \\
FOCA0728-214 & 195.812 & 27.882 & 6 & 4739 & 1.31 & -22 & 8.55$\pm$1.27 & -1.00$\pm$0.13 & 8.79 & 0 & 0 & 0 & 1 \\
Mrk53 & 194.025 & 27.677 & 4 & 4910 & 1.5 & -31 & 9.66$\pm$0.11 & 0.65$\pm$0.01 & 9 & 2 & 0 & 0 & 0 \\
\makecell{2MASX \\ J12593980+2734357} & 194.916 & 27.577 & 2 & 4981 & 0.69 & -29 & 9.63$\pm$0.14 & -0.34$\pm$0.07 & 7.82 & 0 & 1 & 0 & 1 \\
2MFGC10318 & 195.151 & 27.573 & 8 & 5144 & 0.76 & 47 & 10.13$\pm$0.10 & 0.28$\pm$0.04 & 8.4 & 1 & 0 & 0 & 1 \\
FOCA0728-637 & 194.27 & 28.21 & 3 & 5142 & 1.10 & -7 & 8.8$\pm$0.54 & -0.77$\pm$0.09 & 8.58 & 0 & 0 & 0 & 1 \\
\makecell{2MASX \\ J12594009+2837507} & 194.915 & 28.634 & 12 & 5350 & 1.12 & 3 & 9.63$\pm$0.12 & 0.1$\pm$0.03 & 9.66 & 1 & 0 & 0 & 0 \\
NGC4921 & 195.361 & 27.884 & 9 & 5470 & 0.64 & -1 & 11.53$\pm$0.10 & 0.4$\pm$0.04 & 9.21 & 1 & 0 & 1 & 0 \\
Mrk58 & 194.771 & 27.645 & 3 & 5436 & 0.64 & 2 & 10.1$\pm$0.11 & 0.41$\pm$0.03 & 8.23 & 0 & 1 & 0 & 0 \\
GMP5257 & 194.193 & 27.817 & 3 & 5772 & 1.18 & -13 & 9.53$\pm$0.16 & -0.98$\pm$0.12 & 9.22 & 0 & 0 & 0 & 1 \\
NGC4907 & 195.203 & 28.16 & 5 & 5774 & 0.49 & -33 & 11.05$\pm$0.10 & 0.01$\pm$0.05 & 8.05 & 1 & 0 & 1 & 0 \\
GMP2943 & 195.025 & 28.253 & 3 & 6319 & 0.48 & -42 & 8.92$\pm$0.27 & -0.60$\pm$0.06 & 9.39 & 0 & 0 & 0 & 1 \\
LEDA 126807 & 194.752 & 27.644 & 3 & 7980 & 0.65 & -51 & 8.54$\pm$0.94 & -1.42$\pm$0.05 & 7.60 & 1 & 0 & 0 & 1 \\
\noalign{\smallskip}
    \end{longtable}

\end{landscape}
\end{onecolumn}

\end{appendix}

%
%

\end{document}